\documentclass{aa}
\usepackage{graphicx}
\usepackage{txfonts}
\usepackage{rotating}

\begin{document}
\authorrunning{Schegerer et al.} 

\titlerunning{10\,$\mu$m spectroscopic survey of T\,Tauri systems}

\title{Analysis of the dust evolution in the circumstellar disks of T\,Tauri
stars }

\author{A. Schegerer \inst{1} \and S. Wolf \inst{1} 
\and N.V. Voshchinnikov \inst{2,3} \and F. Przygodda \inst{1} 
\and J.E. Kessler-Silacci \inst{4} }

\offprints{A. Schegerer \email{schegerer@mpia.de}}

\institute{Max Planck Institute for Astronomy, K\"onigstuhl 17, 69117 Heidelberg, Germany  
\and Sobolev Astronomical Institute, St. Petersburg University, Universitetskii prosp. 28, St. Petersburg 198504, Russia 
\and Isaac Newton Institute of Chile in Eastern Europe and Eurasia, St. Petersburg Branch, Russia
\and Department of Astronomy, University of Texas, C-1400, Austin, TX 78712, USA}

\date{Received $<$date$>$; Accepted $<$date$>$}

\abstract{}
{We present a compositional analysis of $\rm 8-13\,\mu m$ spectra of
32 young stellar objects (YSOs). Our sample consists of 5 intermediate-mass
stars and 27 low-mass stars.} 
{While the spectra and first scientific
results have already been published by Przygodda et al. (2003) and
Kessler-Silacci et al. (2004) we perform a more detailed analysis
of the $\rm \sim 10\,\mu m$ silicate feature. In our analysis we assume that 
this emission feature
can be represented by a linear superposition of the wavelength-dependent
opacity $\kappa_{\rm abs}(\lambda)$ describing the optical properties of silicate grains
with different chemical composition, structure, and grain size. The determination
of an adequate fitting equation is another goal of this study. 
Using a restricted number of fitting
parameters we investigate which silicate species are necessary for
the compositional fitting.
Particles with radii of 0.1\,$\mu$m- and 1.5\,$\mu$m consisting of amorphous olivine
and pyroxene, forsterite, enstatite, and quartz have been considered.
Only compact, homogeneous dust grains have
been used in the presented fitting procedures. In this context we
show that acceptable fitting results can also be achieved if
emission properties of porous silicate grains are considered instead.}
{Although some previous studies give reasons for the similarity between the dust in circumstellar 
disks of T\,Tauri stars and Herbig Ae/Be stars, a quantitative comparison has been missing,
so far. Therefore, we conclude with a discussion of the results of
a $\rm 10\,\mu$m spectroscopic survey of van Boekel et al. (2005)
who focus on Herbig Ae/Be stars, the higher mass counterparts of T\,Tauri
stars and draw comparisons to this and other studies. We find
that the results of our study of T\,Tauri systems partly agree 
with previous studies of Herbig Ae/Be stars.}
{}


\maketitle

\section{Introduction}

T\,Tauri stars are known as precursors of low-mass main sequence stars
similar to our sun. In contrast to their more evolved counterparts
one of their characteristics is excess radiation observed
in the infrared and sub-millimeter range. High-resolution interferometric
observations (e.g., Leinert et al. 2004; van Boekel et al. 2004; Millan-Gabet
et al. 2001) combined with radiative transfer models (e.g., Cotera
et al. 2001; Wolf et al. 2003c; Akeson et al. 2005) confirm what has
been assumed for 30 years (Mendoza 1968; Adams et al. 1987): the excess
arises from a circumstellar, passive disk heated by the central star.
Last but not least images in the near-infrared up to the sub-millimeter
wavelength regime have provided impressive illustrations of the circumstellar
structure of young stellar objects (YSOs) (e.g., Padgett et al. 1999;
Allen et al. 2002; Krist et al. 2005). 

Several different absorption and emission features have been detected
towards YSOs, so far (e.g., Molster \& Waters 2003 and Table~1 in Voshchinnikov 2002).
A broad, frequently found emission band is the silicate feature at
$\rm \sim 10\,\mu$m which is attributed to the stretching mode of
the Si-O bond in silicate minerals (e.g., Dorschner et
al. 1988). Hanner et al. (1995) successfully reproduced this feature
and the underlying continuum by a two component fit: 
$\lambda \mathcal{F}_{\lambda} \propto C_{0} \lambda^{n} + C_{1} \varepsilon(\lambda ) \lambda^{m}$,
where $C_{0}$ and $C_{1}$ as well as $m$ and $n$ are fitting constants
and $\varepsilon(\lambda )$ is the normalised, smoothed silicate band
of $\rm \theta $ Ori D (Elias 1978) in the N band
\footnote{N band covers the wavelength range $\rm 8.0\,\mu m - 13.0\,\mu m$
}. The first term represents the underlying continuum from the optically thick, 
inner region of the disk, while the second term
stands for the emission feature which is assumed to have its origin
in the optically thin surface layer (Natta et al. 2000). 
Instead of
using the normalised emission feature of an arbitrarily selected source,
Bouwman et al. (2001) performed compositional fits which are based
on the emission profiles of different silicate compounds, particle
shapes and radii (0.1$\mu$m and 2.0$\mu$m for olivine, 0.1$\mu$m
for forsterite, enstatite and quartz). This approach was motivated by
previous stu\-dies of the optical parameters of dif\-ferent silicate minerals
(e.g., Dorschner et al. 1995). The resulting profile is a linear combination
of the opacities $\kappa_{{\rm abs;} i}(\lambda)$ for the different silicate
species $i$. 

In respect of the observed shift in peak position of the silicate band,
Bouwman et al. (2001) concluded a change in average grain size and dust 
crystallisation in Herbig Ae/Be (HAeBe) stars indicating dust grain evolution. 
In addition, this result was supported by further comparisons with mid-infrared
spectra of the interstellar medium as the parent material for YSOs, and the
mid-infrared spectra of the comet Hale Bopp as a representative of
the evolved planetary systems. 
While the mid-infrared flux of interstellar medium shows 
a maximum at a wavelength of $\sim 9.8 \mu m$, the si\-licate feature of comet 
Hale Bopp is flatter with a maximum at $\sim 11.3 \mu m$. Corresponding results
were found in laboratory for silicate grains of increasing size and 
crystallinity.

Because of instrumental sensitivity constraints most stu\-dies have
focused on HAeBe stars rather than on the fainter T\,Tauri systems up to the present.
Apart from the similarity of ge\-neral properties like geometry and structure, evolutionary processes
like dust coagu\-lation and crystallization are assumed to be accelerated
or at least enabled in HAeBe stars because of the higher energy
input from the star. There are only a few authors who ana\-lysed the
10\,$\mu$m-feature of classical T\,Tauri stars, so far. Natta et al. (2000) presented 
low-resolution spectra of nine classical T\,Tauri stars associated with the 
Chameleon I dark cloud. Honda et al. (2003)
found a larger amount of crystalline silicate like forsterite, enstatite
and quartz in the primary of the well-studied binary Hen\,3-600.
Furthermore, Przygodda et al. (2003, hereafter PR03) and Kessler-Silacci
et al. (2005, hereafter KS05) selected a sample of 16\,T\,Tauri and 5\,HAeBe
stars. They established a linear correlation between feature
strength and feature shape which were explained by grain
growth. The similar correlation has been ascertained for HAeBe stars
(van Boekel et al. 2003, 2005).

In this paper we complete the work presented in PR03 and KS05 by providing
a more detailed ana\-lysis of the silicate emission feature. While
the objects are introduced in Sect. \ref{sec:The-sample}, in Sect.
\ref{sec:The-fitfunction} the fitting function is scrutinised
in detail in respect of the temperature (Sect. \ref{sub:The-single-temperature-black}) 
and the emission properties (Sect. \ref{sub:Emission-profiles}) 
of the dust grains which contribute to the silicate feature.
After studying our fit results (Sect. \ref{sec:Results}) we investigate
the dependence of grain growth and crystallization on stellar parameters
and look for further correlations. Finally, we raise the question,
if the silicate feature could be reproduced by dust species that have
not been considered, so far (Sect. \ref{sec:Fit-results-with}). In
this context, porous dust grains are taken into
account. A discussion of future prospects conclude this study in Sect.
\ref{sec:Conclusion-and-future}.

\section{The sample\label{sec:The-sample}}

In Table \ref{object-properties}, our sample of 27 T\,Tauri is presented. 
We additionally consider 5 HAeBe stars in order to show differences
and similarities between low-mass and intermediate-mass systems. Furthermore, 
we are able to analyse differences between our results and previous studies.
The targets were observed in the N band with the Thermal Infrared Multi Mode Instrument 2 (TIMMI2; Reimann
et al. 1998, 2000) at ESO's observatory La Silla and with the Long
Wavelength Spectrometer (LWS) at the W.\,M.\,Keck\,Observatory (e.g.,
Marler et al. 1995), respectively. Information about the observations
and the subsequent data reduction are given in PR03 and KS05. 

Six of our targets are known binaries (e.g., VW\,Cha, Hen\,3-600\,A, AK\,Sco),
but we have seperate spectra only for two binary objects (AS\,205 and S\,CrA).
Therefore, we present the spectral analysis of (30+2) objects altogether.
Selection effects in terms of stellar age and stellar mass can not
be found (see Table \ref{object-properties}), although most of our
objects are younger (age $\rm < 3\,$Myr) T\,Tauri stars. Assuming similar
stellar masses, the putative more frequent observation of \emph{younger}
T\,Tauri stars can be explained by their stronger infrared brightness
in contrast to \emph{older} T\,Tauri stars ($\rm > 6\,$Myr) where the circumstellar
disk, which is responsible for the infrared excess, has already vanished
(Haisch et al. 2001; Carpenter et al. 2005). The age is certainly
the least constrained para\-meter in Table \ref{object-properties}.

Considering the inclination of the circumstellar disks of the objects
as a further, geometrical selection effect, the ap\-pearance of an \emph{emission}
feature favours smaller inclination angles%
\footnote{with respect of the face-on orientation%
} if the feature originates from the innermost regions of the \emph{optical
thin} surface of the disk. 

\section{The fitting function\label{sec:The-fitfunction}}

The determination of an adequate fitting function of the $\rm 10\,\mu$m
silicate feature and its physical justification is non-trivial. The
difficulty results from the attempt to simulate the complex spectral
energy distribution (SED) with the least number of para\-meters.

In fact, our first approach was 
the determination of a potential grain size distribution 
of dust in a circumstellar disk, only considering the 
10$\mu$m silicate feature. 
This approach failed because unique fitting models could not be obtained.
In this context, we have to mention that Mathis et al. (1977) found a particle size 
distribution for graphite and
several silicates in interstellar dust that can roughly be reproduced
by a power law. It is still an open issue if a \emph{similar} particle
size distribution is existent in the environment of YSOs, too (e.g.,
Tanaka et al. 1996).

Nevertheless, it has been found, that a \emph{simple} function with
only a \emph{small} number of fitting parameters finally guarantees a successful, 
robust fit. Therefore, the balance between physical
justification and fitting robustness is our guiding idea. 

\onecolumn{\small
\begin{sidewaystable}
\begin{center}
\caption{\label{object-properties} Our sample of 27 T\,Tauri stars and 5 HAeBe stars (lower 5 panels) with the corresponding coordinates.}
\begin{tabular}[t]{clllllllllll}
\hline\hline
\# & Object & RA (J2000) & DEC (J2000) & $T$\raisebox{-0.5ex}{$\rm \star$} [K] & $M$\raisebox{-0.5ex}{$\star$} [M\raisebox{-0.5ex}{$\odot$}] & $d$ [pc]& Sp. Type & $L$\raisebox{-0.5ex}{$\star$} [L\raisebox{-0.5ex}{$\odot$}] & age [Myr] & Other common names & Instrument \\[1.0ex] \hline 

1 & GG\,Tau &  04 32 30.3 & +17 31 41 & 3802\raisebox{0.5ex}{\tiny (1)} & 1.3\raisebox{0.5ex}{\tiny (3)} & 140 & K7\raisebox{0.5ex}{\tiny (2)}  & 1.6\raisebox{0.5ex}{\tiny (3)} & 0.7-1.7\raisebox{0.5ex}{\tiny (3)} & & TIMMI2 \\[1.0ex]

2 & AA\,Tau & 04 34 55.6 & +24 28 54 & 4060\raisebox{0.5ex}{\tiny (4)} & 0.7\raisebox{0.5ex}{\tiny (5)} & 140\raisebox{0.5ex}{\tiny (4)} & K7\raisebox{0.5ex}{\tiny (5)} & 1.0\raisebox{0.5ex}{\tiny (5)} & 2.4\raisebox{0.5ex}{\tiny (4)} & & LWS\\[1.0ex]

3 & LkCa\,15 & 04 39 17.8 & +22 21 05 & 4350\raisebox{0.5ex}{\tiny (4)} & 1.0\raisebox{0.5ex}{\tiny (3)} & 140\raisebox{0.5ex}{\tiny (4)} & K5 V\raisebox{0.5ex}{\tiny (4)} & 0.7\raisebox{0.5ex}{\tiny (3,4)} & 2.0\raisebox{0.5ex}{\tiny (4)} & & LWS \\[1.0ex]

4 & DQ\,Tau & 04 46 53 & +17 00 00 & 4000\raisebox{0.5ex}{\tiny (21)} & 0.4\raisebox{0.5ex}{\tiny (22)} & 140\raisebox{0.5ex}{\tiny (21)} & M0\raisebox{0.5ex}{\tiny (7)} & 0.6\raisebox{0.5ex}{\tiny (7,22)} & - & & TIMMI2 \\[1.0ex]

5 & DR\,Tau & 04 47 05.5 & +16 58 42 & 4060\raisebox{0.5ex}{\tiny (3,6)} & 1.1\raisebox{0.5ex}{\tiny (6)} & 140\raisebox{0.5ex}{\tiny (1)} & K4 V & 1.0\raisebox{0.5ex}{\tiny (3,7)} & 2.5-3.8\raisebox{0.5ex}{\tiny (3)} & & TIMMI2 \\[1.0ex]

6 & GM\,Aur & 04 55 10.9 & +30 22 01 & 4060\raisebox{0.5ex}{\tiny (4)} & 1.2\raisebox{0.5ex}{\tiny (3,6)} & 140\raisebox{0.5ex}{\tiny (4)} & K7\raisebox{0.5ex}{\tiny (4)} & 1.0\raisebox{0.5ex}{\tiny (4)} & 1.3-1.8\raisebox{0.5ex}{\tiny (3)} & & LWS \\[1.0ex]

7 & SU\,Aur & 04 55 59.4 & +30 34 02 & 5945\raisebox{0.5ex}{\tiny (8)} & 1.7\raisebox{0.5ex}{\tiny (8)} & 152 & G1\raisebox{0.5ex}{\tiny (8)} & 7.8\raisebox{0.5ex}{\tiny (8)} & 8.7\raisebox{0.5ex}{\tiny (8)} & HD\,282624 & TIMMI2 \\[1.0ex]

8 & GW\,Ori & 05 29 08.4 & +11 52 13 & 6030\raisebox{0.5ex}{\tiny (8)} & 3.7\raisebox{0.5ex}{\tiny (8)} & 305 & K3 V\raisebox{0.5ex}{\tiny (8)} & 61.8\raisebox{0.5ex}{\tiny (8)} & 1.0\raisebox{0.5ex}{\tiny (8)} & HD\,244138 & TIMMI2 \\[1.0ex]

9 & FU\,Ori & 05 45 22.6 & +09 04 12 & 5000\raisebox{0.5ex}{\tiny (23)} & 1.0\raisebox{0.5ex}{\tiny (23)} & - & G3 I & - & - & & TIMMI2 \\[1.0ex]

10 & BBW\,76 & 07 50 35.5 & -33 06 24 & 5500\raisebox{0.5ex}{\tiny (24)} & - & 180\raisebox{0.5ex}{\tiny (24)} & G0-G2 I\raisebox{0.5ex}{\tiny (24)} & - & - & & TIMMI2 \\[1.0ex]

11 & CR\,Cha & 10 59 07.0 & -77 01 40 & 4900 & 1.2\raisebox{0.5ex}{\tiny (9)} & 144 & K2 & 3.3\raisebox{0.5ex}{\tiny (9)} & 1.0\raisebox{0.5ex}{\tiny (8)} & & TIMMI2 \\[1.0ex]

12 & TW\,Hya & 11 01 51.9 & -34 42 17 & 4000\raisebox{0.5ex}{\tiny (10)} & 0.6\raisebox{0.5ex}{\tiny (10)} & 55\raisebox{0.5ex}{\tiny (10)} & K8 Ve & 0.2\raisebox{0.5ex}{\tiny (10)} & 10\raisebox{0.5ex}{\tiny (10)} & & TIMMI2 \\[1.0ex]

13 & VW\,Cha & 11 08 01.8 & -77 42 29 & 4350\raisebox{0.5ex}{\tiny (9)} & 0.6\raisebox{0.5ex}{\tiny (9)} & - & K2/K5\raisebox{0.5ex}{\tiny (9)} & 2.9\raisebox{0.5ex}{\tiny (9,26)} & 1.0\raisebox{0.5ex}{\tiny (9)} & & TIMMI2 \\[1.0ex]

14 & Glass\,I & 11 08 15.4 & -77 33 53 & 4600\raisebox{0.5ex}{\tiny (9)} & 0.9\raisebox{0.5ex}{\tiny (9)} & - & K4\raisebox{0.5ex}{\tiny (11)} & 1.6\raisebox{0.5ex}{\tiny (9)} & 1.0\raisebox{0.5ex}{\tiny (9)} & & TIMMI2 \\[1.0ex]

15 & VZ\,Cha & 11 09 23.8 & -76 23 21 & 4200\raisebox{0.5ex}{\tiny (25)} & 0.7\raisebox{0.5ex}{\tiny (25)} & - & K7\raisebox{0.5ex}{\tiny (25)} & 0.5\raisebox{0.5ex}{\tiny (9,25)} & 3.0\raisebox{0.5ex}{\tiny (9)} & & TIMMI2 \\[1.0ex]

16 & WW\,Cha & 11 10 00.7 & -76 34 59 & 4350\raisebox{0.5ex}{\tiny (11)} & 0.6\raisebox{0.5ex}{\tiny (11)} & - & K5 & 2.7\raisebox{0.5ex}{\tiny (11)} & 0.3\raisebox{0.5ex}{\tiny (11)} & & TIMMI2\\[1.0ex]

17 & Hen\,3-600\,A & 11 10 28.9 & -37 32 05 & 3200\raisebox{0.5ex}{\tiny (5)} & 0.3\raisebox{0.5ex}{\tiny (12)} & 50\raisebox{0.5ex}{\tiny (4)} & M4\raisebox{0.5ex}{\tiny (4)} & 0.2\raisebox{0.5ex}{\tiny (12)} & 10\raisebox{0.5ex}{\tiny (4)} & & LWS \\[1.0ex]

18 & IRAS\,14050-4109 & 14 08 10.3 & -41 23 53 & 4405\raisebox{0.5ex}{\tiny (4)} & - & 140\raisebox{0.5ex}{\tiny (4)} & K5\raisebox{0.5ex}{\tiny (4)} & 0.6\raisebox{0.5ex}{\tiny (5)} & 10\raisebox{0.5ex}{\tiny (5)} & & LWS \\[1.0ex]

19 & Sz\,82 & 15 56 42.3 & -37 56 06 & 3800\raisebox{0.5ex}{\tiny (15)} & 0.3\raisebox{0.5ex}{\tiny (26)} & - & M0\raisebox{0.5ex}{\tiny (26)} & 1.6\raisebox{0.5ex}{\tiny (26)} & 1.1\raisebox{0.5ex}{\tiny (26)} & IM\,Lup & TIMMI2 \\[1.0ex]

20 & RU\,Lup & 15 56 42.3 & -37 49 16 & 4000\raisebox{0.5ex}{\tiny (13)} & 0.8\raisebox{0.5ex}{\tiny (13)} & 140\raisebox{0.5ex}{\tiny (14)} & G5 V & 0.5\raisebox{0.5ex}{\tiny (13)} & 0.04\raisebox{0.5ex}{\tiny (15)} & HD\,142560 & TIMMI2 \\[1.0ex]

21 & AS\,205\,N & \raisebox{-0.5ex}{16 11 31.4} & \raisebox{-0.5ex}{-18 38 25} & 4450\raisebox{0.5ex}{\tiny (16)} & 1.5\raisebox{0.5ex}{\tiny (16)} & 160\raisebox{0.5ex}{\tiny (16)} & K5\raisebox{0.5ex}{\tiny (16)} & 7.1\raisebox{0.5ex}{\tiny (16)} & 0.1\raisebox{0.5ex}{\tiny (16)} & & TIMMI2 \\[1.0ex]

22 & AS\,205\,S & & & 3450\raisebox{0.5ex}{\tiny (16)} & 0.3\raisebox{0.5ex}{\tiny (16)} & 160\raisebox{0.5ex}{\tiny (16)} & M3\raisebox{0.5ex}{\tiny (16)} & 2.2\raisebox{0.5ex}{\tiny (16)} & $\rm <$0.1\raisebox{0.5ex}{\tiny (16)} & & TIMMI2 \\[1.0ex]

23 & HBC\,639 & 16 26 23.4 & -24 21 02 & 5250\raisebox{0.5ex}{\tiny (10)} & 2.3\raisebox{0.5ex}{\tiny (10)} & 160\raisebox{0.5ex}{\tiny (10)} & K3\raisebox{0.5ex}{\tiny (10)} & 8.8\raisebox{0.5ex}{\tiny (10)} & 1.5\raisebox{0.5ex}{\tiny (10)} & DoAr\,24 & TIMMI2 \\[1.0ex]

24 & Haro\,1-16 & 16 31 33.5 & -36 53 19 & 4365\raisebox{0.5ex}{\tiny (15)} & - & - & K2-3\raisebox{0.5ex}{\tiny (17)} & 2.0\raisebox{0.5ex}{\tiny (15)} & 0.5\raisebox{0.5ex}{\tiny (15)} & DoAr\,44 & TIMMI2 \\[1.0ex]

25 & AK\,Sco & 16 54 44.9 & -36 53 19 & 6500\raisebox{0.5ex}{\tiny (18)} & - & 152 \raisebox{0.5ex}{\tiny (18)} & F5 V & 0.9\raisebox{0.5ex}{\tiny (19)} & 6.0\raisebox{0.5ex}{\tiny (19)} & HD\,152404 & TIMMI2 \\[1.0ex]

26 & S\,CrA\,N & \raisebox{-0.5ex}{19 01 08.6} & \raisebox{-0.5ex}{-36 57 20} & 4800\raisebox{0.5ex}{\tiny (16)} & 1.5\raisebox{0.5ex}{\tiny (16)} & 130\raisebox{0.5ex}{\tiny (16)} & K3\raisebox{0.5ex}{\tiny (16)} & 2.3\raisebox{0.5ex}{\tiny (16)} & 3.0\raisebox{0.5ex}{\tiny (16)} & & TIMMI2 \\[1.0ex]

27 & S\,CrA\,S & & & 3800\raisebox{0.5ex}{\tiny (16)} & 0.6\raisebox{0.5ex}{\tiny (16)} & 130\raisebox{0.5ex}{\tiny (16)} & M0 & 0.76\raisebox{0.5ex}{\tiny (16)} & 1.0\raisebox{0.5ex}{\tiny (16)} & & TIMMI2 \\[1.0ex] \hline

28 & MWC\,480 & 04 58 46.3 & +29 50 37 & 8670\raisebox{0.5ex}{\tiny (4)} & - & 145\raisebox{0.5ex}{\tiny (4)} & A3\raisebox{0.5ex}{\tiny (4)} & 22\raisebox{0.5ex}{\tiny (4)} & 4.6 & HD\,31648 & LWS \\[1.0ex] 

29 & HD\,163296 & 17 56 21.3 & +29 50 37 & 9332\raisebox{0.5ex}{\tiny (4)} & 2.0\raisebox{0.5ex}{\tiny (20)} & 122\raisebox{0.5ex}{\tiny (4)} & A1 V\raisebox{0.5ex}{\tiny (4)} & 24.0\raisebox{0.5ex}{\tiny (20)} & 4.0\raisebox{0.5ex}{\tiny (4)} & & LWS \\[1.0ex]

30 & HD\,179218 & 19 11 11.3 & 15 47 16 & 10471\raisebox{0.5ex}{\tiny (4)} & 2.9\raisebox{0.5ex}{\tiny (20)} & 240\raisebox{0.5ex}{\tiny (4)} & B9\raisebox{0.5ex}{\tiny (4)} & 100\raisebox{0.5ex}{\tiny (20)} & 0.1\raisebox{0.5ex}{\tiny (4)} & MWC\,614 & LWS \\[1.0ex]

31 & WW\,Vul & 19 25 58.8 & +21 12 31 & 8600\raisebox{0.5ex}{\tiny (4)} & - & 550\raisebox{0.5ex}{\tiny (4)} & A3/B9 V\raisebox{0.5ex}{\tiny (4)} & - & - & HD\,344361 & LWS \\[1.0ex]

32 & HD\,184761 & 19 34 59.0 & +27 13 31 & 7500\raisebox{0.5ex}{\tiny (4)} & - & 65\raisebox{0.5ex}{\tiny (4)} & A8V\raisebox{0.5ex}{\tiny (4)} & 7.4\raisebox{0.5ex}{\tiny (4)} & - & & LWS \\[1.0ex] \hline 

\end{tabular}
\end{center}
{\scriptsize 
Note: Following symbols are used: stellar temperature ($T_{\rm \star}$), stellar mass
 ($M_{\rm \star}$) derived from corresponding stellar evolutionary tracks, distance ($d$), 
 stellar luminosity ($ L_{\rm \star}$), and stellar age
 (mostly derived from their location in the HR diagram).
 References - (1) Thi et al.\,2001; (2) Hartmann et al.\,1998;
 (3) Greaves\,2004; (4) KS05; (5) Metchev et al.\,2004; (6) Kitamura et al.\,2002;
 (7) Muzerolle et al.\,2003; (8) Calvet et al.\,2004; (9) Meeus et al.\,2003;
 (10) Natta et al.\,2004; (11) Natta et al.\,2002;
 (12) Geoffray \& Monin\,2001; (13) Stempels \& Piskunov\,2003;
 (14) Brooks \& Costa\,2003; (15) N\"urnberger et al.\,1997; (16) Prato et al.\,2003;
 (17) Chen et al.\,1995; (18) G\"unther et al.\,2004; (19) Manset et al.\,2005; 
(20) van Boekel et al.\,2005; (21) Mathieu et al.\,1997; (22) Hersant et al.\,(2005);
 (23) Lachaume et al.\,2003; (24) Sch\"utz et al.\,2005; (25) Natta et al.\,2000;
 (26) G\'omez \& Mardonez\,2003.
} 
\end{sidewaystable}}
\twocolumn

In order to compare our results with previous ones, we consider an
ansatz which Hanner et al. (1995) and BO05 used: 
\begin{eqnarray} 
\mathcal{F}_{\nu}=B_{\nu}(T) \left(C_{0} + \sum_{i=1}^{n} C_{i} \kappa_{{\rm abs;} i} \right), 
\label{fitfunktion} 
\end{eqnarray} 
where $C_{0}$ and $C_{i}$ are the fitting parameters, $\kappa_{{\rm abs;} i}$
represents the wavelength-dependent opacity for dust component $i$, and $B_{\nu}(T)$
is the Planck function corresponding to the temperature $T$ 
(see Eq.~\ref{eq:planck}).
In the Sects.~\ref{sub:The-single-temperature-black} and \ref{sub:Emission-profiles}
we will include more detailed discussions of Eq.~(\ref{fitfunktion}). 

Two fitting procedures are considered. First, we examine a fitting
procedure that is based on an iterative Levenberg-Marquardt least-squares
minimization (Press et al. 1986). But, this numerical fitting method
does not guarantee the best fit, i.e. the finding of the \emph{global}
minimum of the fit quality para\-meter $\chi^2$ (reduced chi-squared,
see Sect. \ref{sec:Results}), particularly for data with lower signal-to-noise
ratios. Repeated modifications of the initial fitting parameters and
the subsequent ana\-lysis of $\chi^2$ allows an effective search for
the \emph{global} minimum of $\chi^2$.

Second, we consider the non-negative, linear least-squares fit procedure
used by Lawson \& Hanson (1974) containing the ana\-lytical solution
of the matrix equation $\tilde{C}_{i} \tilde{\kappa}_{{\rm abs;} i} = \mathcal{F}_{\rm \nu}$
with $\tilde{\kappa}_{{\rm abs;} i}=B_{\rm \nu}\kappa_{{\rm abs;} i}$. 
The matrix equation is solved for each integer temperature value between 
a minimum and maximum temperature. The temperature with the lowest corresponding 
$\chi^2$ is the temperature $T$ that we look for. We assume a silicate sublimation 
temperature of 1500\,K (Duschl et al. 1996) and a negligible contribution of the silicate 
emission to the mid-infrared spectrum if the temperature of the corresponding silicate
compound is lower than 50\,K (see Sect. \ref{sub:The-single-temperature-black}). 
Because of robustness and high rapidity the non-negative, linear least-squares 
fit procedure is finally used for our study.

\subsection{The single-temperature black body approximation \label{sub:The-single-temperature-black}}

In Eq.~\ref{fitfunktion} the Planck function has the following
form ($x=(h\nu)/(kT)$): 

\begin{equation}
B_{\nu}(x)=\frac{15}{\pi^{4}}\,\frac{x^{3}}{\exp(x)-1}\label{eq:planck}
\end{equation}
with the Planck constant $h$, the Boltzmann constant $k$, and the
frequency $\nu$. This function is normalised in such a way that its
integral $\int_{0}^{\infty}B_{\nu}(T)d\nu$ is unity.

In fact, dust grains with different temperatures contribute to the SED in 
the N band. But an extensive study of the silicate feature and the underlying continuum 
can only be done with radiative transfer simulations in a dust disk model. 
However, there have been several different efforts in order to determine the 
different silicate compounds with simple, semi-analytical tools. 
Most models are based on the idea that the silicate feature arises from
single-temperature dust grains in an optically thin surface layer. 
Consequently, the emitted flux is given by the 
formula ${\cal F}_{\rm \nu}=\tau_{\rm \nu} B_{\rm \nu}(T)$ with the optical depth 
$\tau_{\nu} = \sum_{i=1}^{n} C_{i} \kappa_{{\rm abs;} i} $. Although this approach does 
not consider the dependence of dust temperature on grain size and composition, 
the mean dust temperature is assumed to be dominated by the absorption and 
emission characteristics of carbon grains assuming thermal equilibrium 
(Kr\"ugel \& Walmsley 1984).
Przygodda (2004) obtained a further simplification of the model by the assumption that 
the temperature of most dust grains, which effectively contribute to the 
emission feature, is $T \approx 300\,K$.  As a maximum flux 
$[B_{\nu}(T)]_{\rm max}$ at $\rm \lambda=10\,\mu$m corresponds to a temperature 
of $\rm \sim$ 290\,K the Planck function $B_{\nu}(T \approx 300\,K)$ is approximately 
constant in N band and the shape of the feature depends only on the emission properties 
of dust grains: ${\cal F}_{\rm \nu} \approx \tau_{\rm \nu}$. 
In order not to preselect a certain dust temperature, the temperature $T$ 
can also be introduced as another free fitting parameter as it was done in 
BO05 and as we do in this study. Also in this approach the underlying continuum of the 
$\rm 10\,\mu m$-feature is assumed to be dominated by dust grains of
the same temperature regime.  We have to mention that this approach does not consider the possibility
that large reservoirs of cool ($\ll$300\,K) and/or hot ($\gg$300\,K) dust grains provide the
main contribution to the SED at $\rm 10\,\mu$m. For this reason a further 
analysis requires the modeling of a more representative, circumstellar disk. Such a representative 
model of a circumstellar disk is gene\-rated by means of a Monte
Carlo code for radiative transfer simulations (MC3D: Wolf et al. 1999;
Wolf 2003a,b). The density distribution is calculated assuming hydrostatic
equilibrium in vertical direction, i.e. perpendicular to the disk
midplane (Schegerer et al., in preparation). 

For this study, stellar properties are chosen that were recently found
by modeling the SED of the low-mass star RY\,Tau (Schegerer et al., in prep.;
Akeson et al. 2005). 
The inner and outer radius of the disk amounts to 0.3 AU and 100AU, respectively.
As absorbing, scattering and emitting material we use a mixture of
carbon and silicate dust in a 1:2 ratio (Weingartner \& Draine 2001;
Draine \& Lee 1984; Mathis et al. 1977)%
\footnote{See also http://www.astro.princeton.edu/\textasciitilde{}draine.%
}. This dust mixture has been used in a large variety of simulations
of circumstellar dust configurations (e.g., Wolf et al. 2003d). The extinction,
absorption and scattering cross sections are averaged over their abundance. 
Such an approach minimises the required computer power and memory as the temperature
and density distribution has not to be determined and stored for each
dust type (Wolf 2003c).
\begin{figure*}[!t]
\hfill{}\includegraphics[%
  scale=0.3]{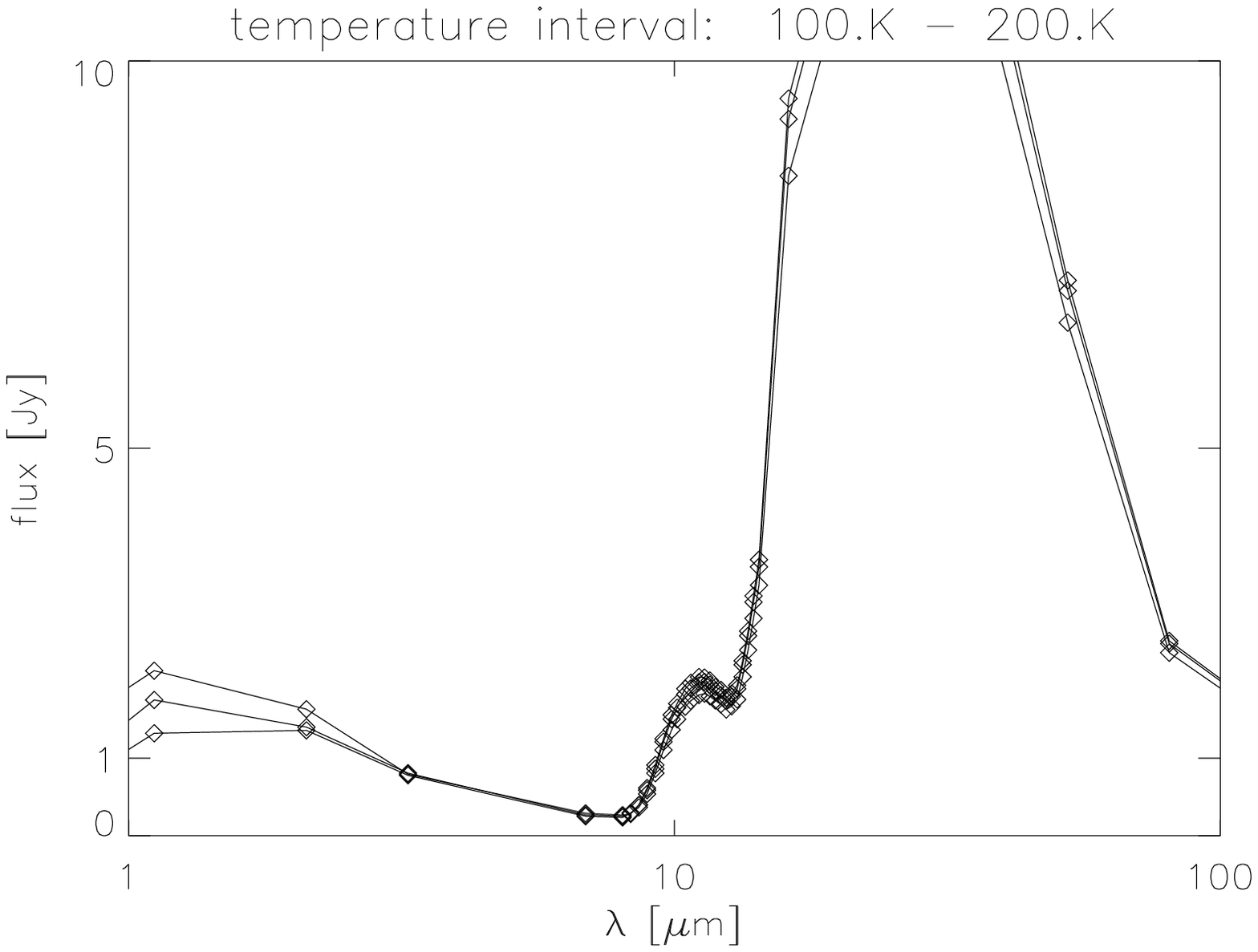}\includegraphics[%
  scale=0.3]{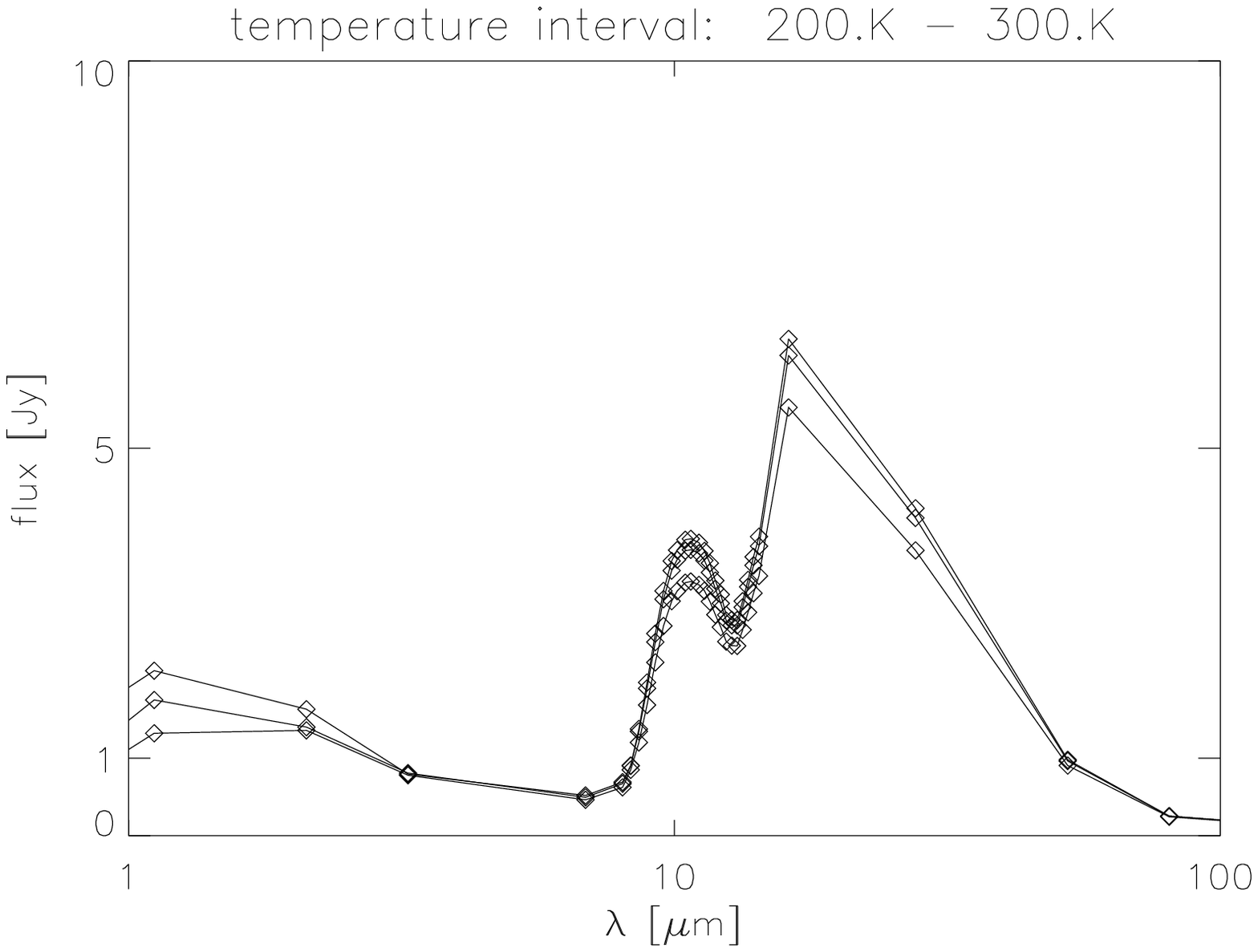}\includegraphics[%
  scale=0.3]{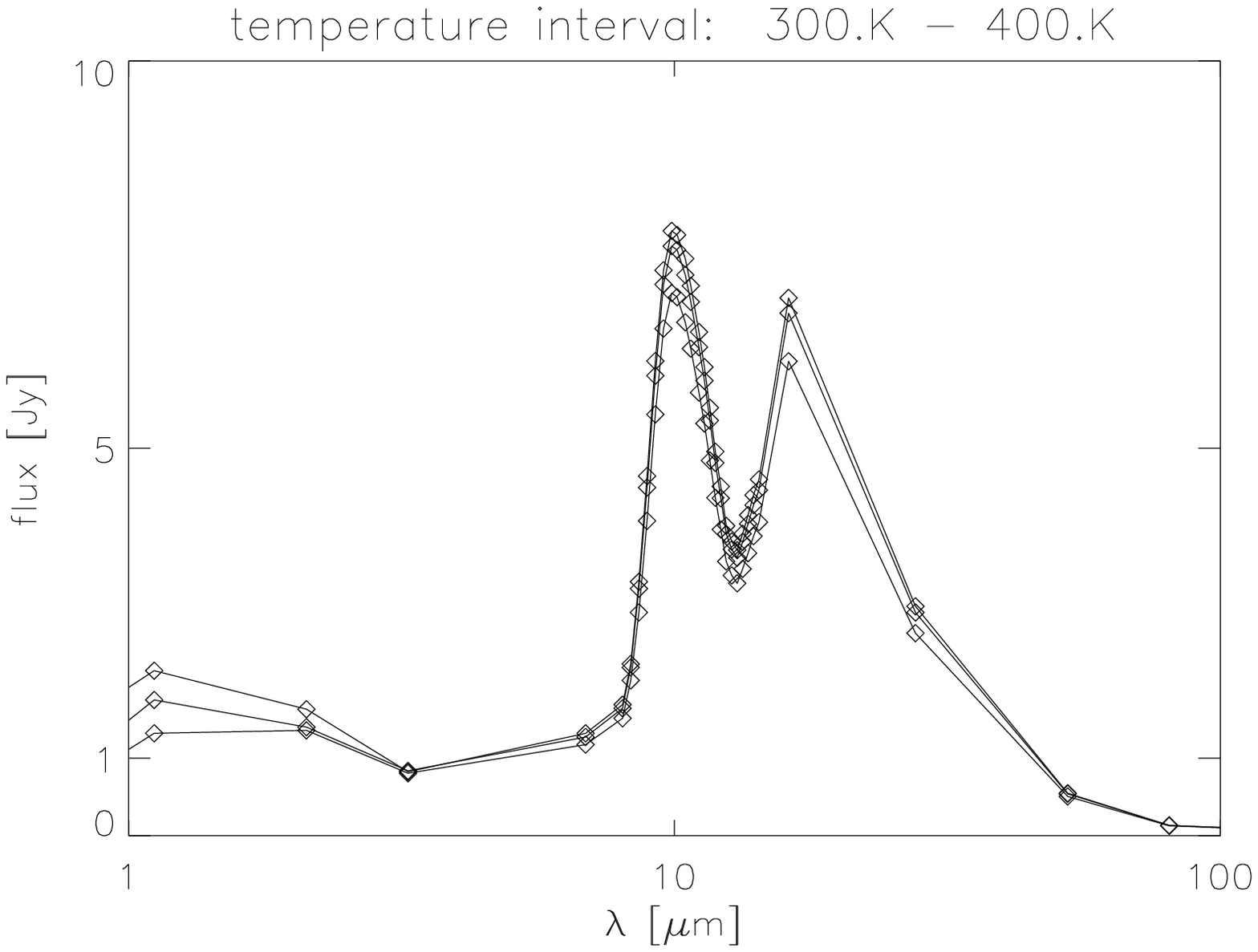}\hfill{}

\hfill{}\includegraphics[%
  scale=0.3]{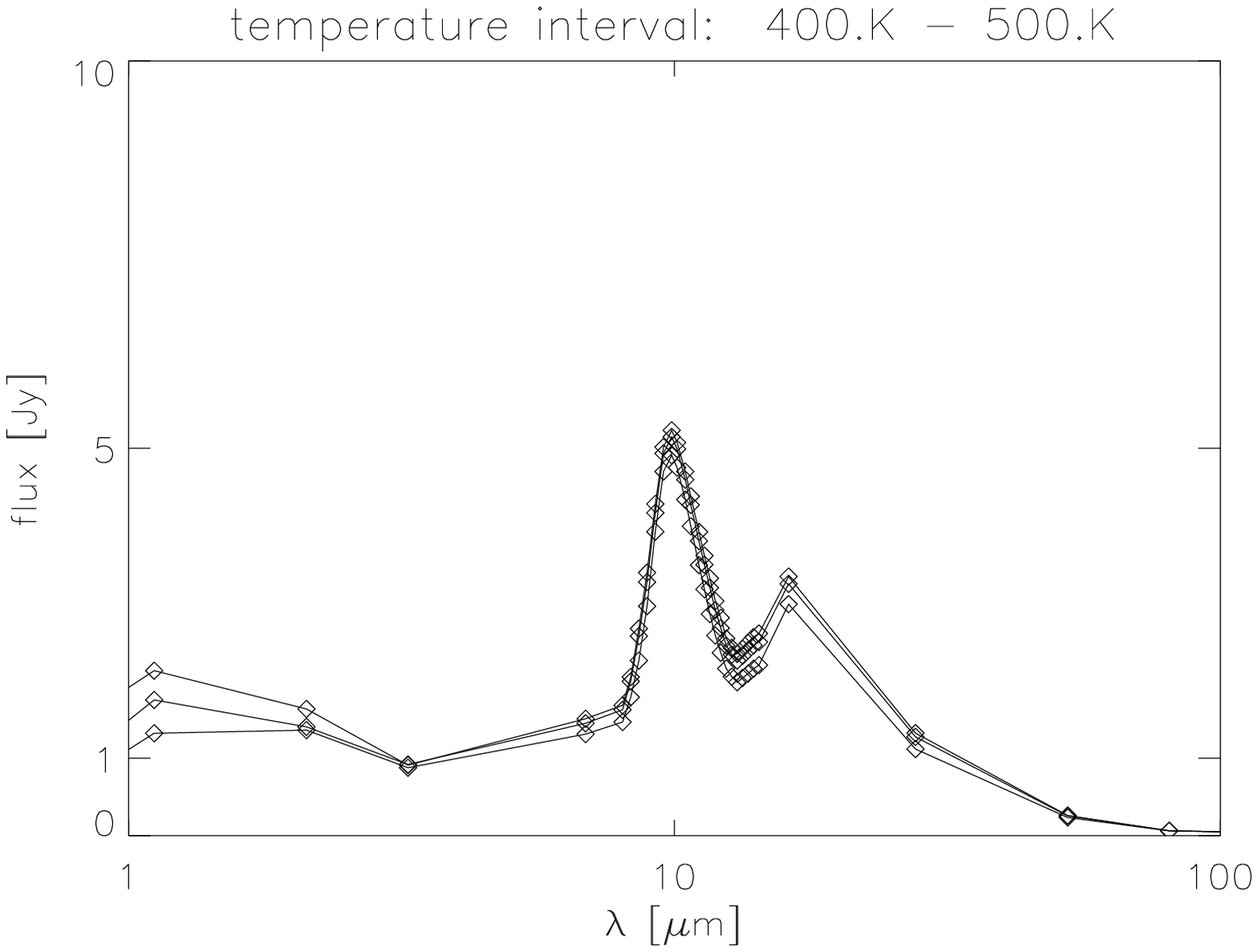}\includegraphics[%
  scale=0.3]{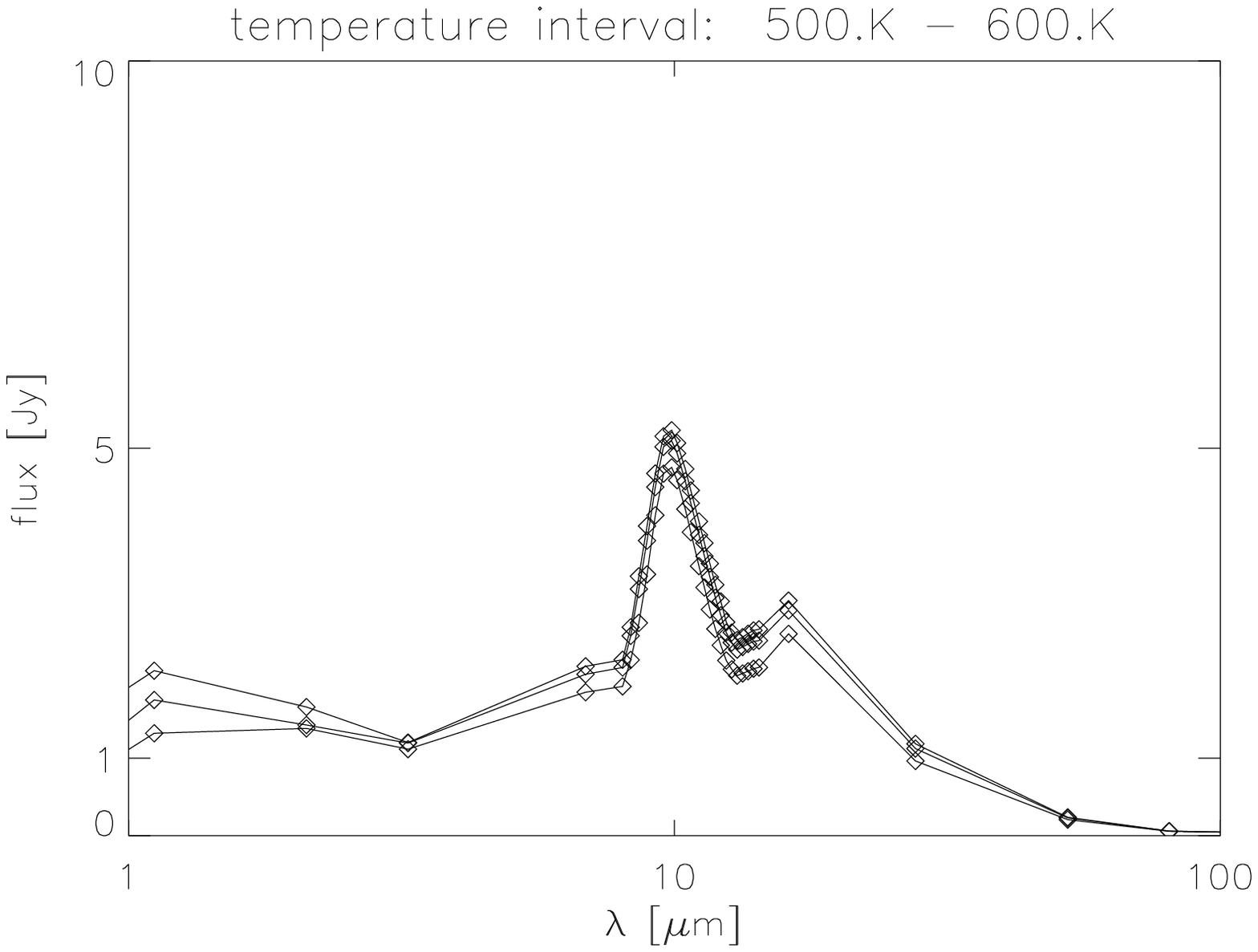}\includegraphics[%
  scale=0.3]{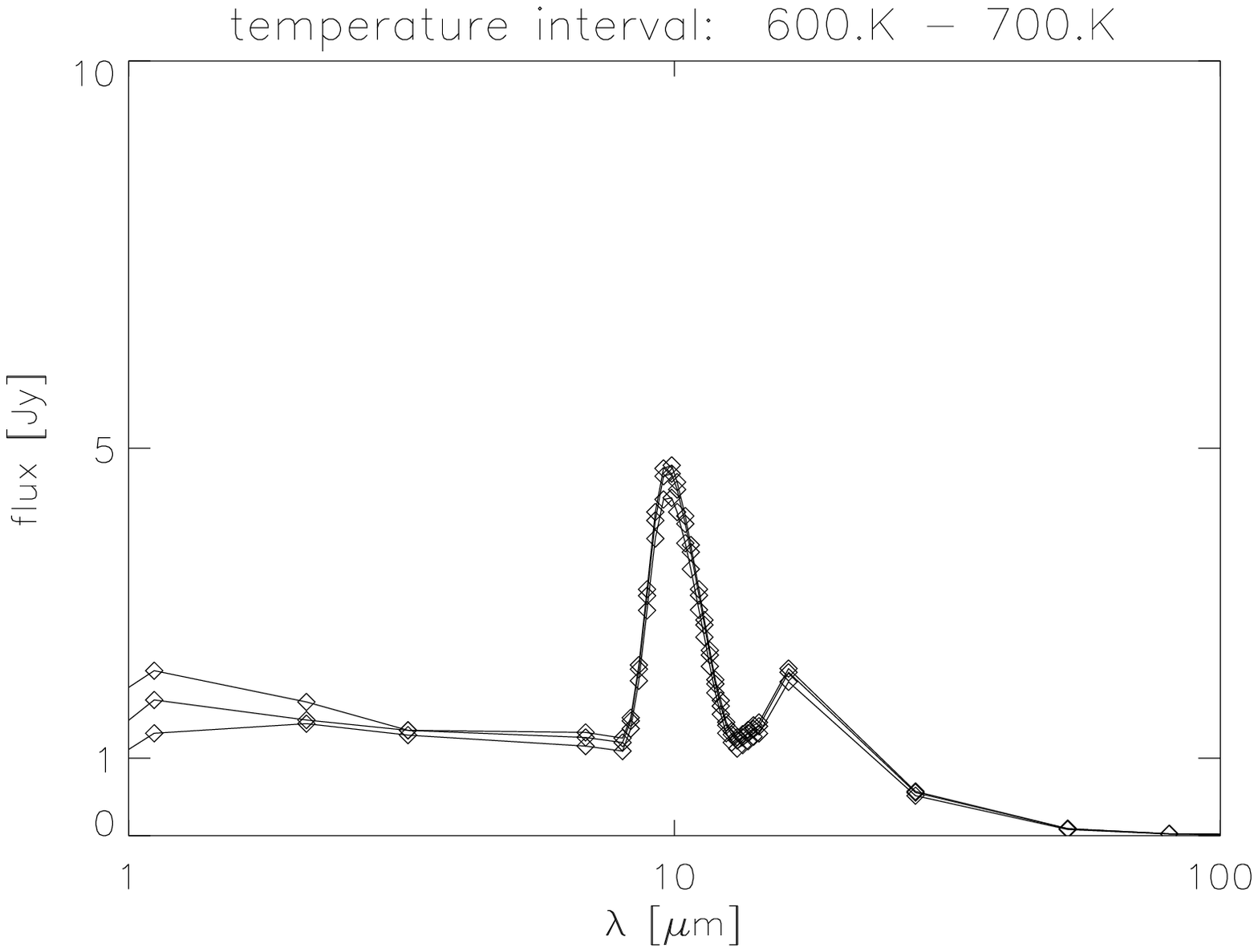}\hfill{}

\hfill{}\includegraphics[%
  scale=0.3]{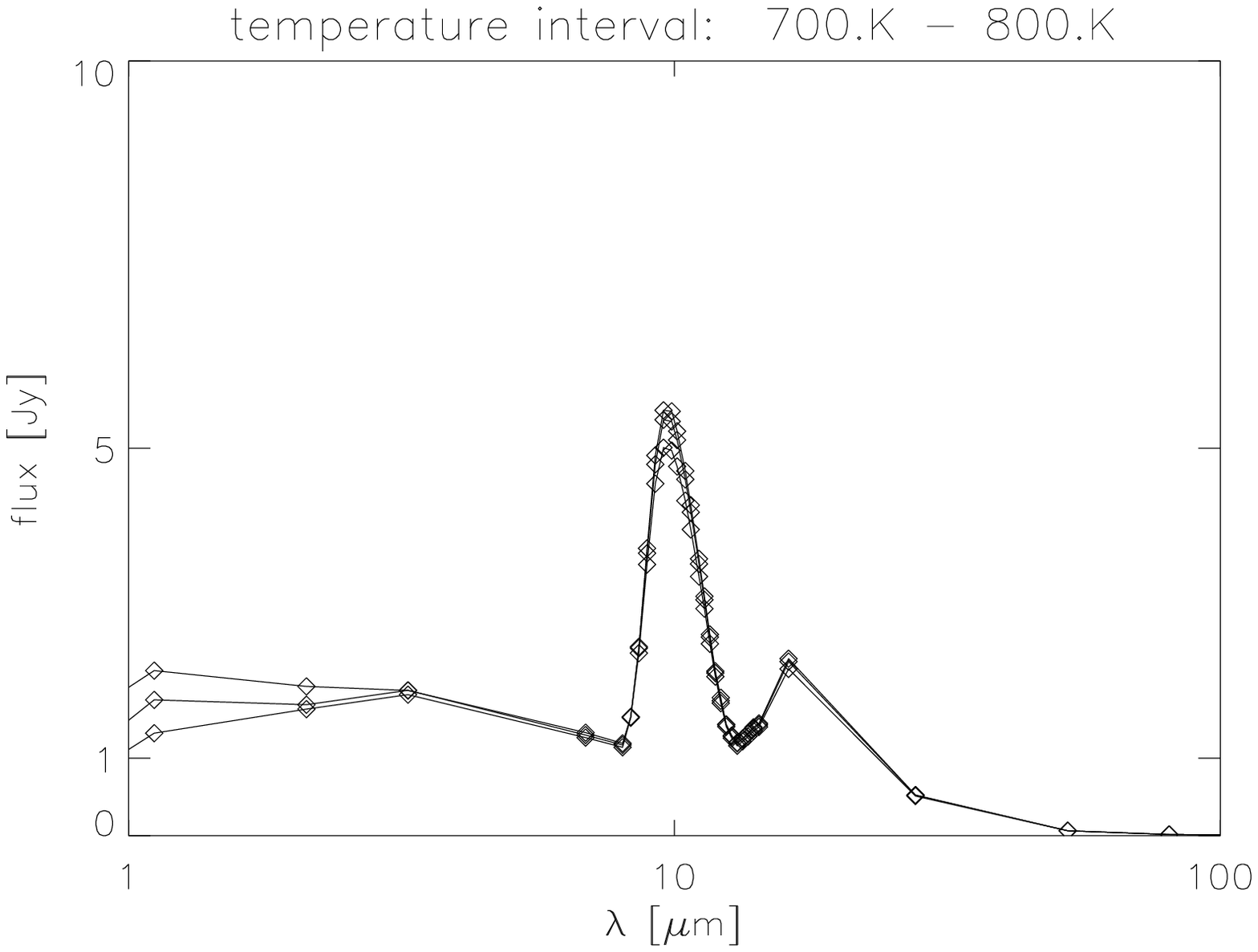}\includegraphics[%
  scale=0.3]{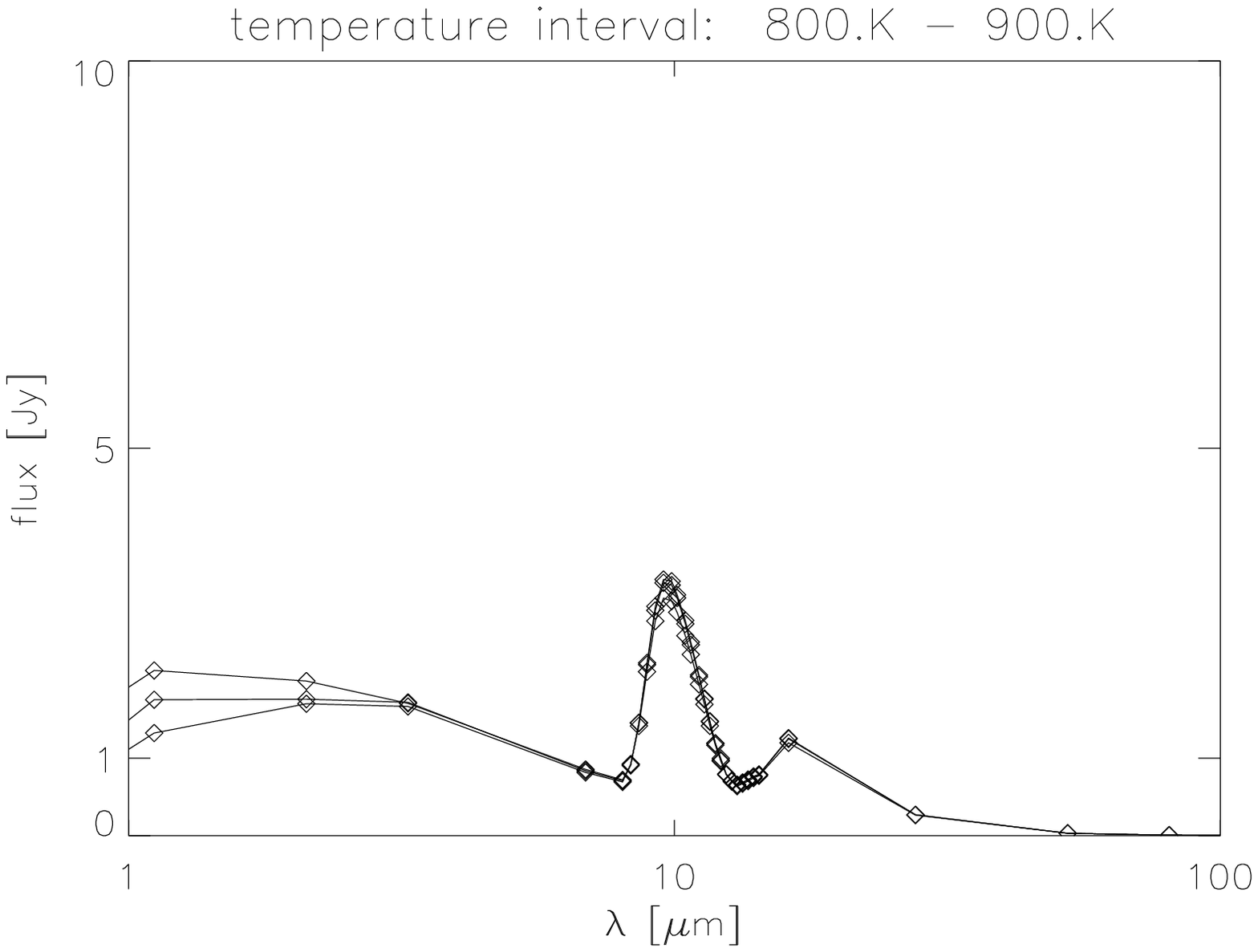}\includegraphics[%
  scale=0.3]{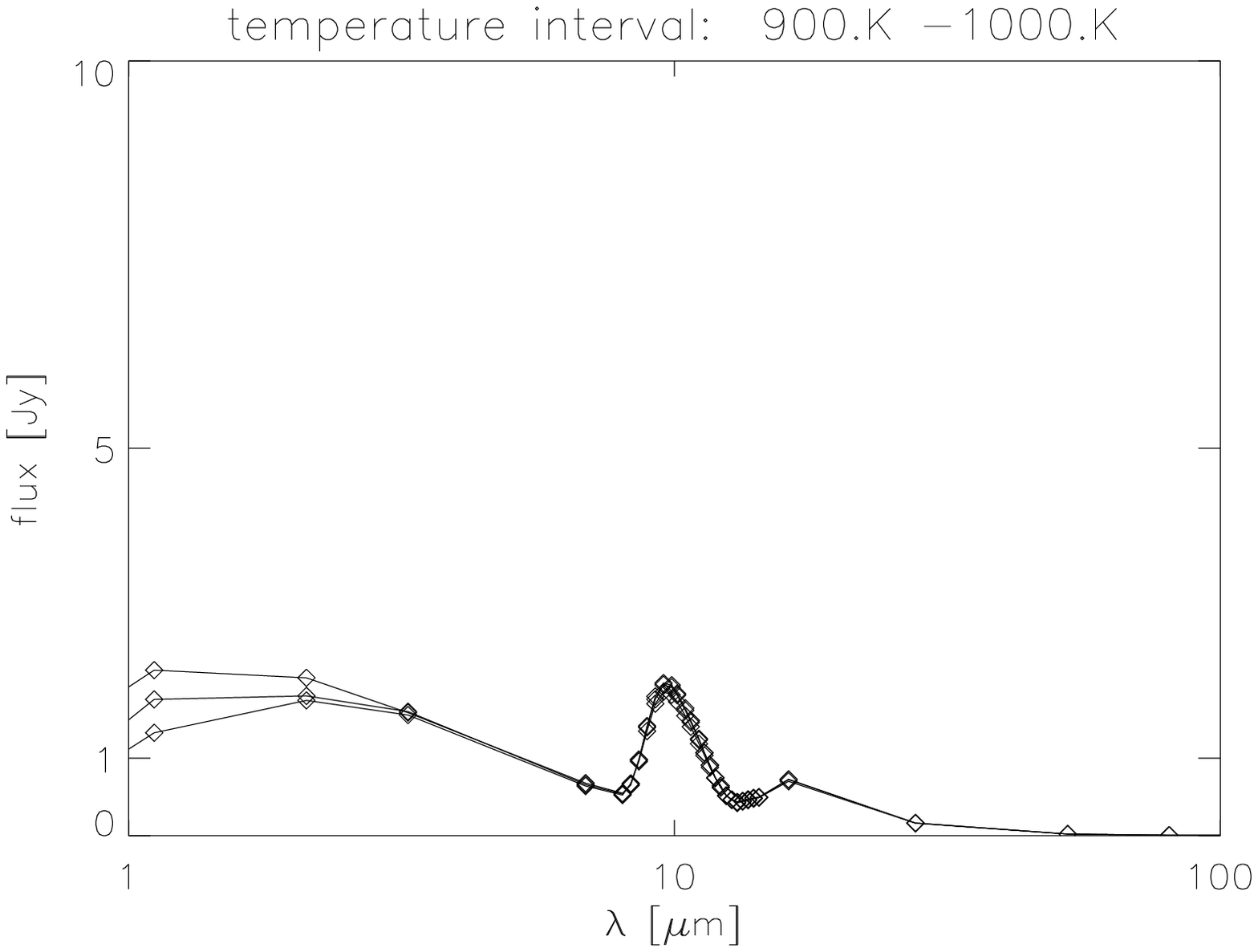}\hfill{}

\caption{\label{cap:The-simulated-SED}The simulated SEDs of a T\,Tauri system
assuming a distance of 140pc. The inner and outer radius of the disk 
amounts to $\rm 0.3\,AU$ and $\rm 100\,AU$, respectively (see Sect. 
\ref{sub:The-single-temperature-black}). The spectra are obtained from systems
with identical geometry and identical stellar properties but in
each case only dust grains that have a temperature in the indicated
temperature interval contribute to the SED. Each figure shows the SED
for inclinations $\rm 0\,\degr$, $\rm 15\,\degr$ and $\rm 45\,\degr$ with respect of 
face-on orientation. Each SED is calculated for 34 wavelengths ($\Diamond$). 
The integration of the flux between $\rm 5.0\,\mu$m and $\rm 15.0\,\mu$m yields
1.84, 3.9, 7.5, 5.1, 5.5, 4.3, 4.8, 2.8, 1.8$\rm \times 10^{14} Jy \times Hz$
from the upper left to the lower right figure. The feature at $\rm \sim 20 \mu m$
is ascribed to silicate, too. }
\end{figure*}
Figure \ref{cap:The-simulated-SED} shows spectra that are obtained
from systems with identical geometry and identical stellar properties
but in each case only dust grains having a temperature in the indicated temperature interval 
are taken into account. Although the $\rm 10\,\mu$m flux is maximised for a 
temperature $T$ of about 375\,K$^($\footnote{This value is obtained by a 
more detailed analysis with more narrow temperature intervals.}$^)$, 
dust grains with a temperature between 300\,K and 400\,K contribute
only moderately ($\rm \sim 20 \%$) to the spectra compared with grains
in adjacent temperature intervals. Therefore, the N band of the SED can not 
be assigned to a certain temperature interval or a certain area of the disk. In 
reality, all grains with different temperatures contribute to this feature.

\begin{figure}[!h]
\hfill{}\includegraphics[%
  scale=0.5]{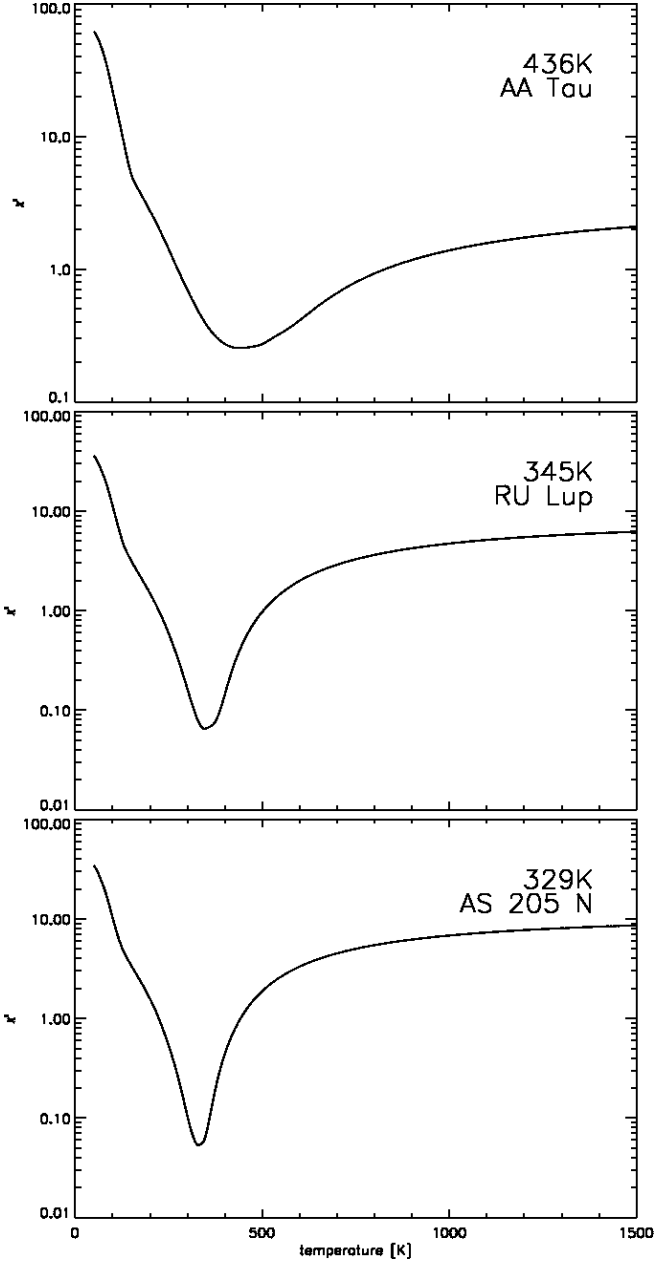}\hfill{}

\caption{\label{fig:temp3}$\chi^{2}$ of the fits versus the pre-specified
temperature $T$. The obtained temperature $T$ for the single fit and the
object name are indicated in each plot.}
\end{figure}

However, the temperature that we obtain in our single-temperature approximation
is an average temperature of all dust grains weighted by their spectral
contribution to the observed silicate feature. The average itself
is assured by the monotonic gradient of the spectral contribution
of dust grains with temperatures in the interval between 100\,K and
600\,K. Figure \ref{fig:temp3} shows the fitting quality parameter $\chi^{2}$ 
(see Eq.~\ref{chi}) of some fits versus the pre-specified temperature $T$ for 
the Planck function (\ref{eq:planck}). The existence of a global mini\-mum
of $(\chi(T))^{2}$ justifies the simplifying single-temperature-model
of the underlying spectral continuum in N band. 

Future radiative transfer simulations will give the 
necessary assistance for the decision if our simplified model for silicate emission in 
circumstellar disk is physically reasonable. Furthermore, interferometric observation 
with the Mid-Infrared Interferometer (MIDI) could help to understand the potential, 
physical meaning of the determined, fitting tempera\-ture $T$ (Schegerer et al., in prep.).  

\subsection{Emission profiles \label{sub:Emission-profiles}}

The emission profiles that we use for the fitting procedure are calculated
for homogeneous, compact, spherical grains 
with Mie theory starting from the complex refractive indices $n_{i}$
for silicate material $i$ (Dorschner et al. 1995; Servoin \& Piriou 1973;
J\"ager et al. 1998b; Spitzer \& Kleinmann 1960). The result is the dimensionless
absorption efficiency $Q_{{\rm abs;} i}$.
This efficiency is used for calculations of the opa\-city%
\footnote{also known as mass absorption coefficient; e.g., Li 2005}%
$\kappa_{{\rm abs;} i}=Q_{{\rm abs;} i} \pi r^2 / (4/3 \pi r^3 \rho)$, where $r$ is the
particle radius and $\rho$ the material density.

\begin{figure}[!h]
\hfill{}\includegraphics[%
  scale=0.39]{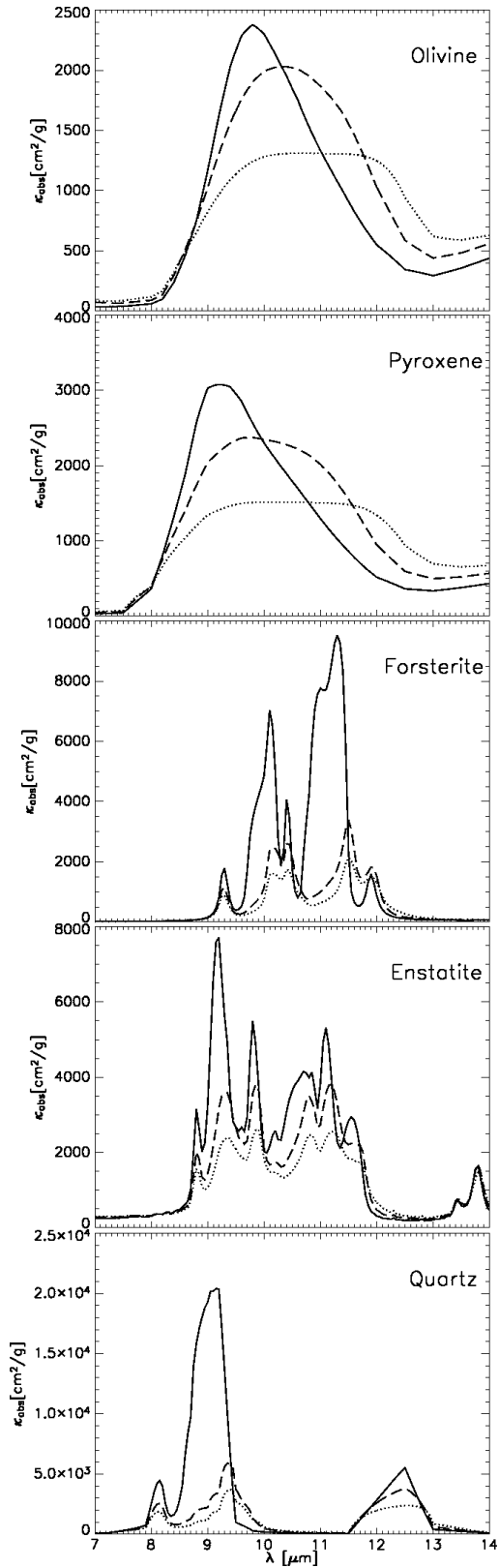}\hfill{}
\caption{\label{fig:emission profiles}
Opacities of grains with radii 0.1\,$\rm \mu$m (solid lines),
1.5\,$\rm \mu$m (dashed lines) and 2.5\,$\rm \mu$m (dotted lines). 
With increasing grain size the maxima shift to
longer wavelengths and the emission profile of olivine
and pyroxene flattens. }
\end{figure}

Fig.~\ref{fig:emission profiles} shows the opacities for amorphous olivine (MgFeSiO$_{\rm 4}$)
and pyroxene (MgFe[SiO$_{\rm 3}]_{\rm 2}$) while
the lower pa\-nels present crystalline forsterite (Mg$_{\rm 2}$SiO$_{\rm 4}$),
enstatite (MgSiO$_{\rm 3}$) and quartz (SiO$_{\rm 2}$), each with
radii 0.1\,$\rm \mu$m, 1.5\,$\rm \mu$m, and 2.5\,$\rm \mu$m.
As Bouwman et al. (2001) pointed out, one 0.1$\rm \mu$m- and
one 2.0$\rm \mu$m-sized dust component for each chemical compound
seem to be sufficiently representative to model the 10\,$\rm \mu$m-silicate-feature.
Here, we use dust grains with a radius of 0.1$\rm \mu m$ and 1.5\,$\rm \mu$m in order
to draw a better comparison to the results of BO05. 
The largest, 2.5\,$\mu$m sized particles show the broad plateau between
9\,$\mu$m and 12\,$\mu$m (see Fig.~\ref{fig:emission profiles},
two upper panels) which does not appear to be existent in any of the observations
modeled here. The dif\-ference in profiles between
1.5\,$\mu$m and 2.5\,$\mu$m sized particles from forsterite,
enstatite and quartz is not significant (Fig.~\ref{fig:emission profiles},
three lower panels).
Therefore, the 2.5\,$\mu$m grains are not considered in the
fitting procedure. Dust grains species 
which do not show an emission feature in the 8-13\,$\rm \mu m$ range, such 
as amorphous, compact carbon grains (e.g., J\"ager et al. 1998a), are not considered. 
However, considering non-homogeneous dust grains (see Sect.~\ref{sec:Fit-results-with}), which consist of a
mixture of various materials, carbon may also modify the silicate emission 
(see Natta et al. 2001).
   
The emission profiles of the remaining dust components are linearly 
independent from the opacities of the other dust components as BO05 showed. This linear
independence and the resulting uniqueness of the solution is not ensured
if too many emission profiles are included. 

Previous investigations (e.g., Natta et al. 2000) do not include certain dust species
like enstatite, quartz and polycyclic aromatic hydrocarbons (PAH)
in circumstellar disks around lower-mass stars. In general, this assumption
seems not to be justified. Honda et al. (2003) found large mass fractions
of enstatite and quartz in the T\,Tauri object Hen\,3-600\,A.
Finally, the robustness of our fitting procedure allows to include
these dust species with one exception: we do not consider PAHs. The PAH features are strongly 
variable in peak position
and profile from source to source and, there\-fore, can not be represented
by one unique emission profile. Additionally, there are
hints that the PAH features seem to be composed of several subfeatures
(Peeters et al. 2002).

\section{Results\label{sec:Results}}

In the following, we focus on the fitting results 
(see Table~\ref{fitresults}) and draw comparisons to previous publications.

\begin{figure*}[!ht]
\hfill{}\includegraphics[%
  clip,
  scale=0.34]{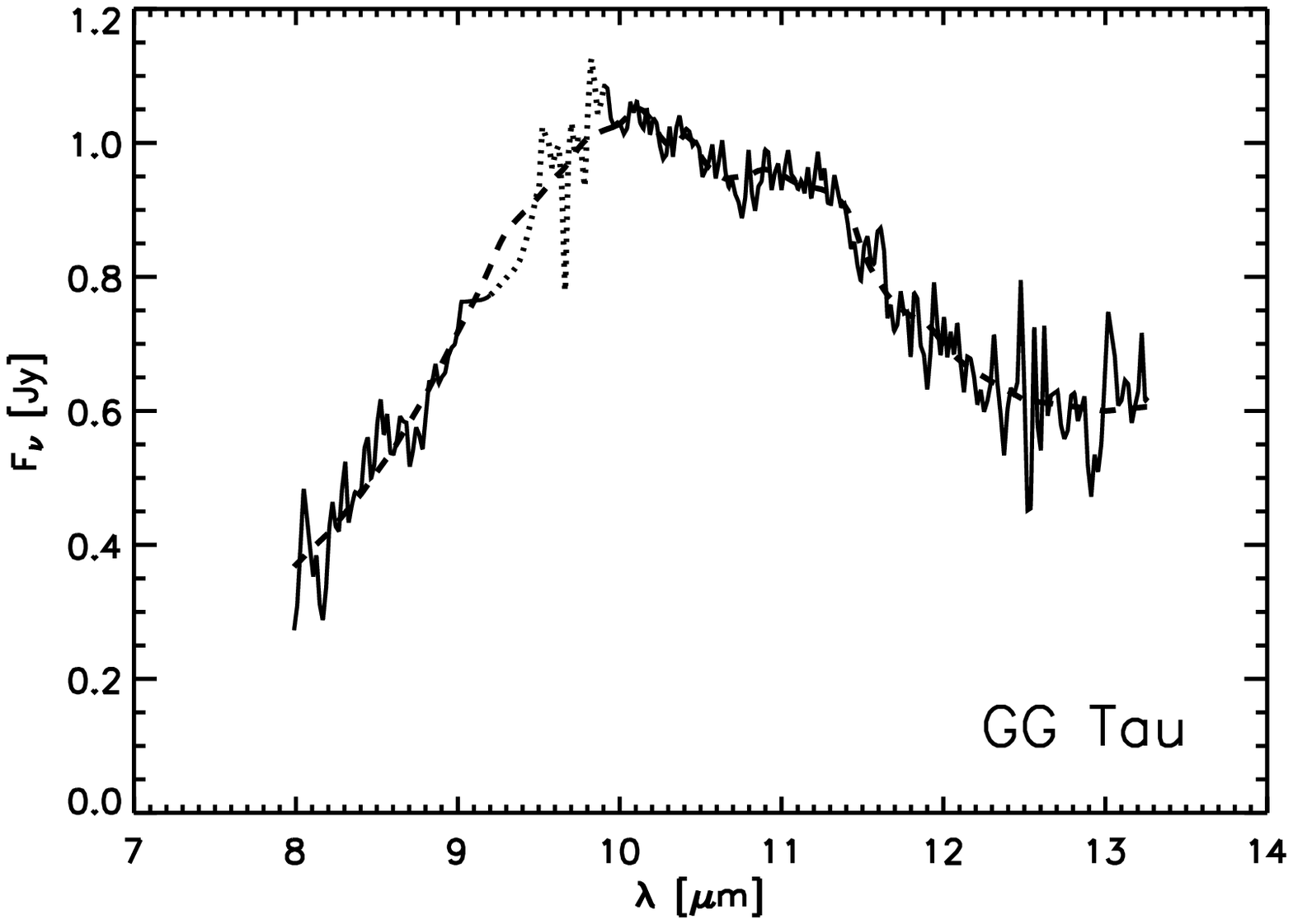}\includegraphics[%
  clip,
  scale=0.34]{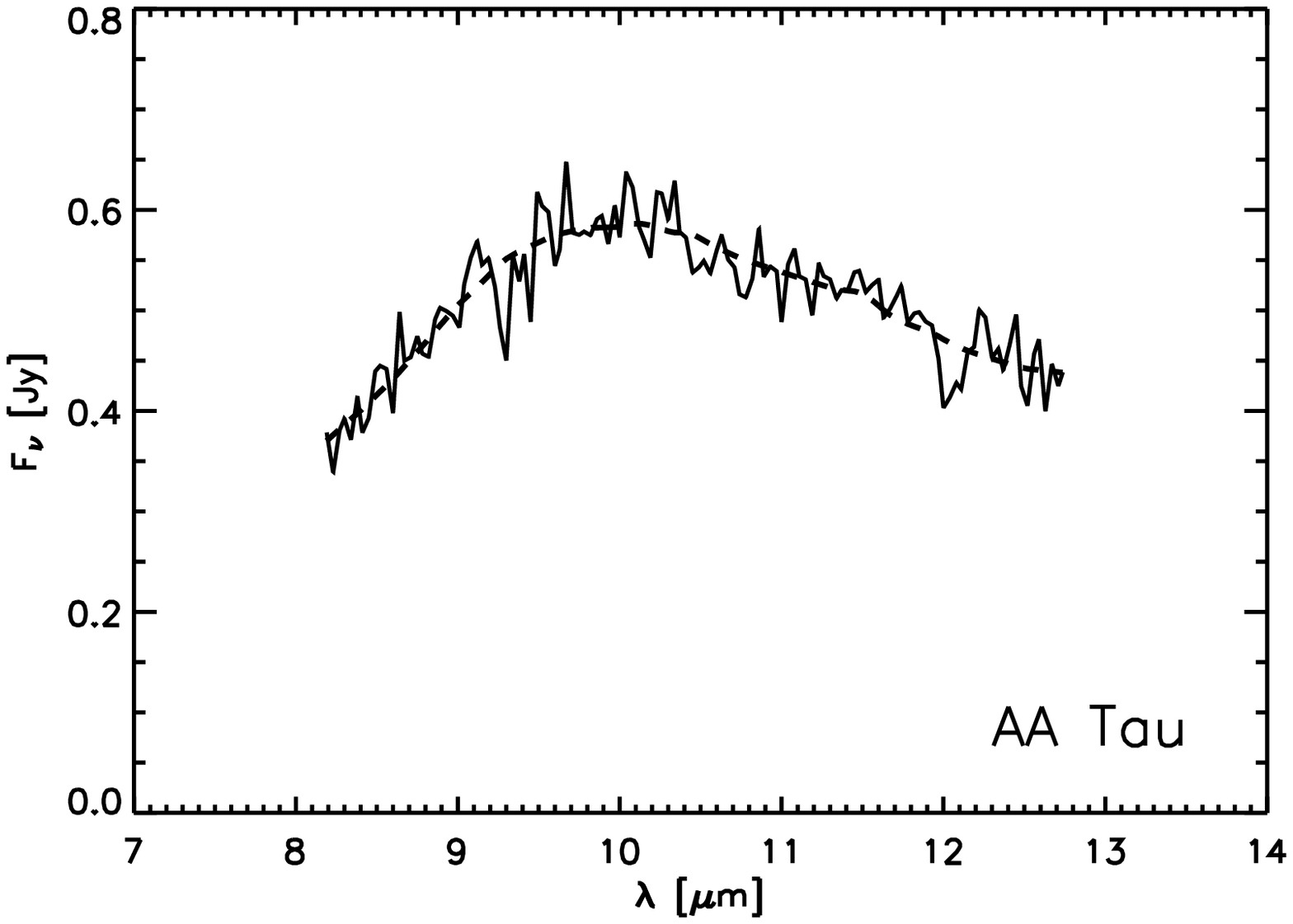}\includegraphics[%
  clip,
  scale=0.34]{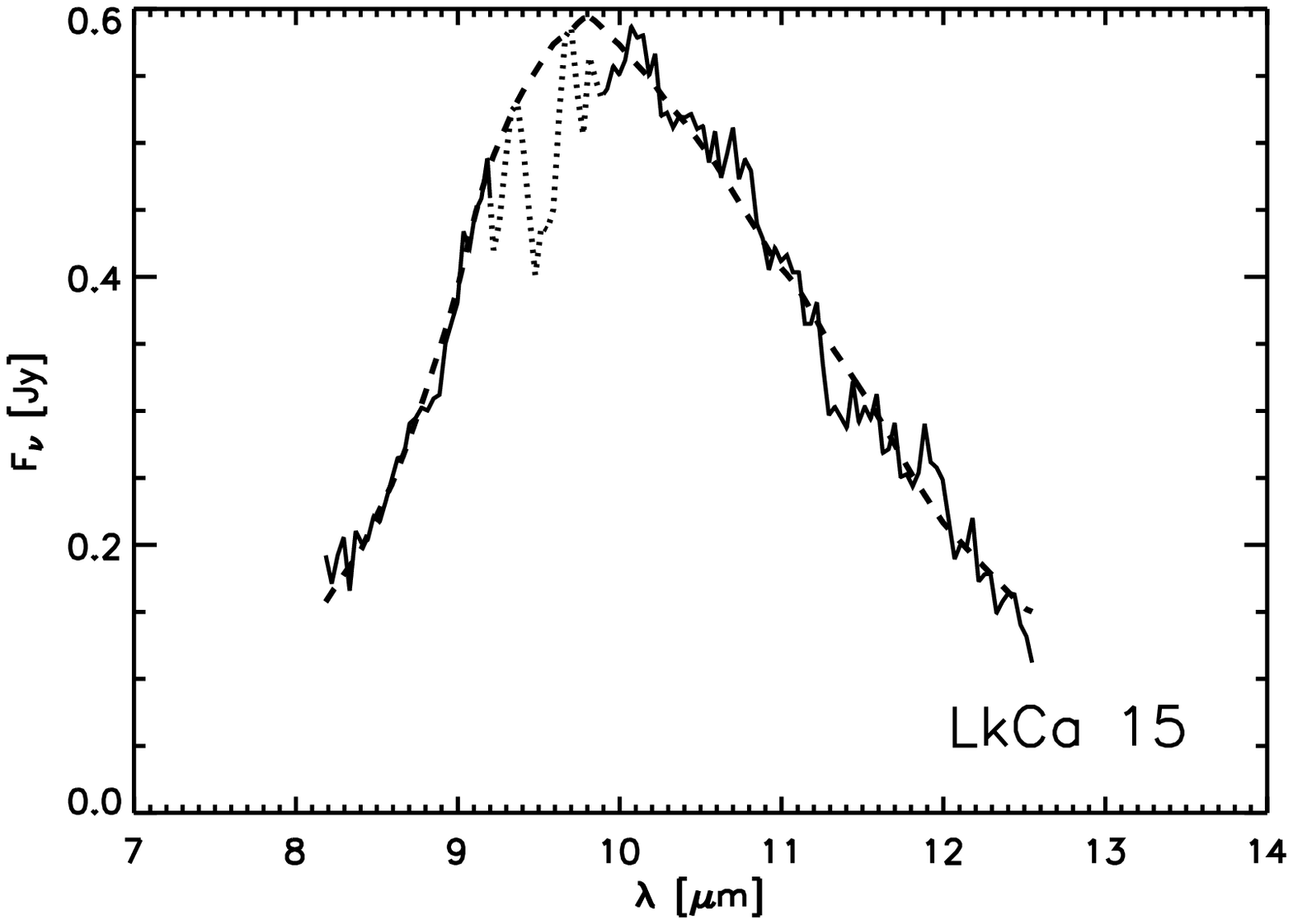}\hfill{}

\hfill{}\includegraphics[%
  clip,
  scale=0.34]{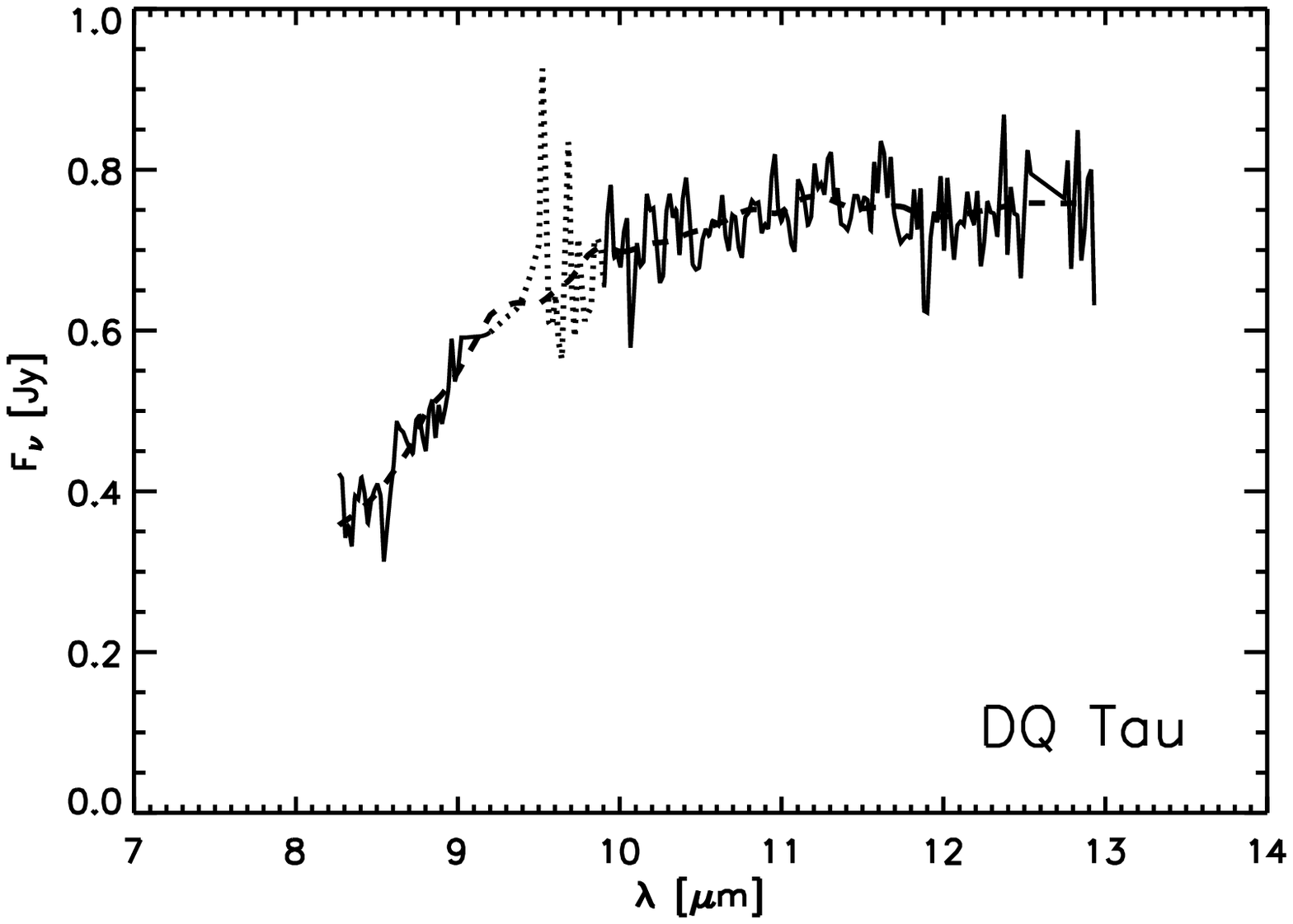}\includegraphics[%
  clip,
  scale=0.34]{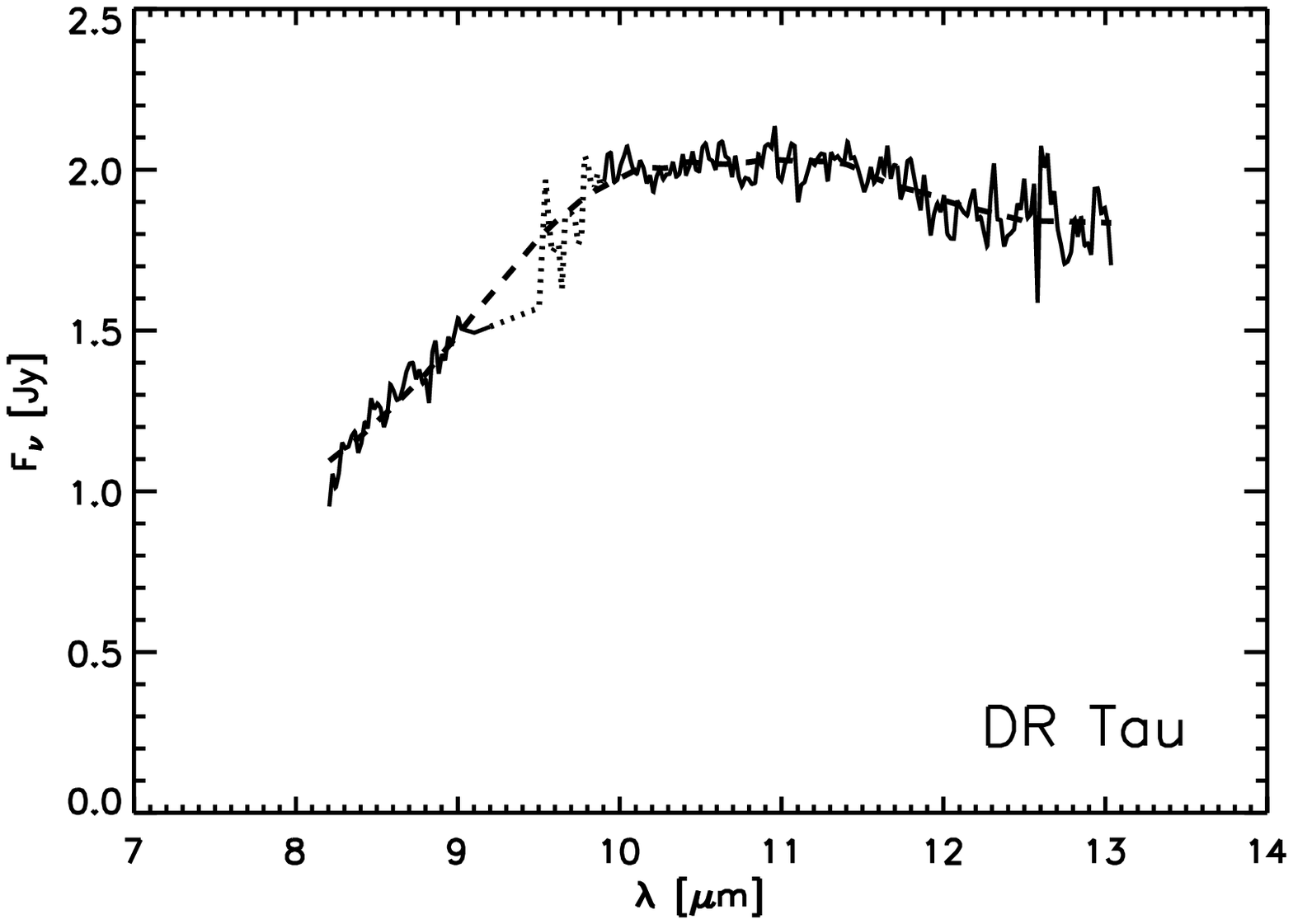}\includegraphics[%
  clip,
  scale=0.34]{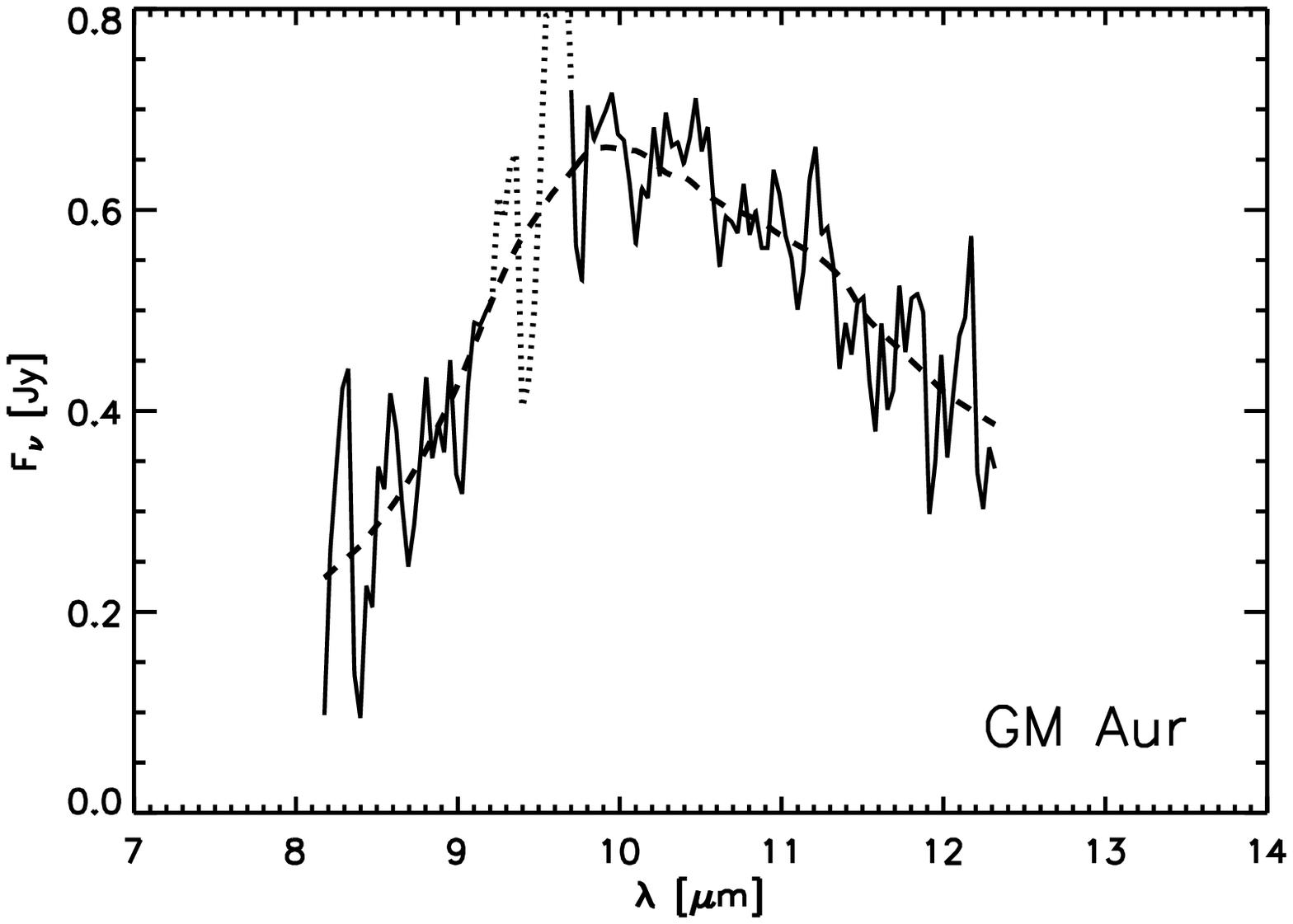}\hfill{}

\hfill{}\includegraphics[%
  clip,
  scale=0.34]{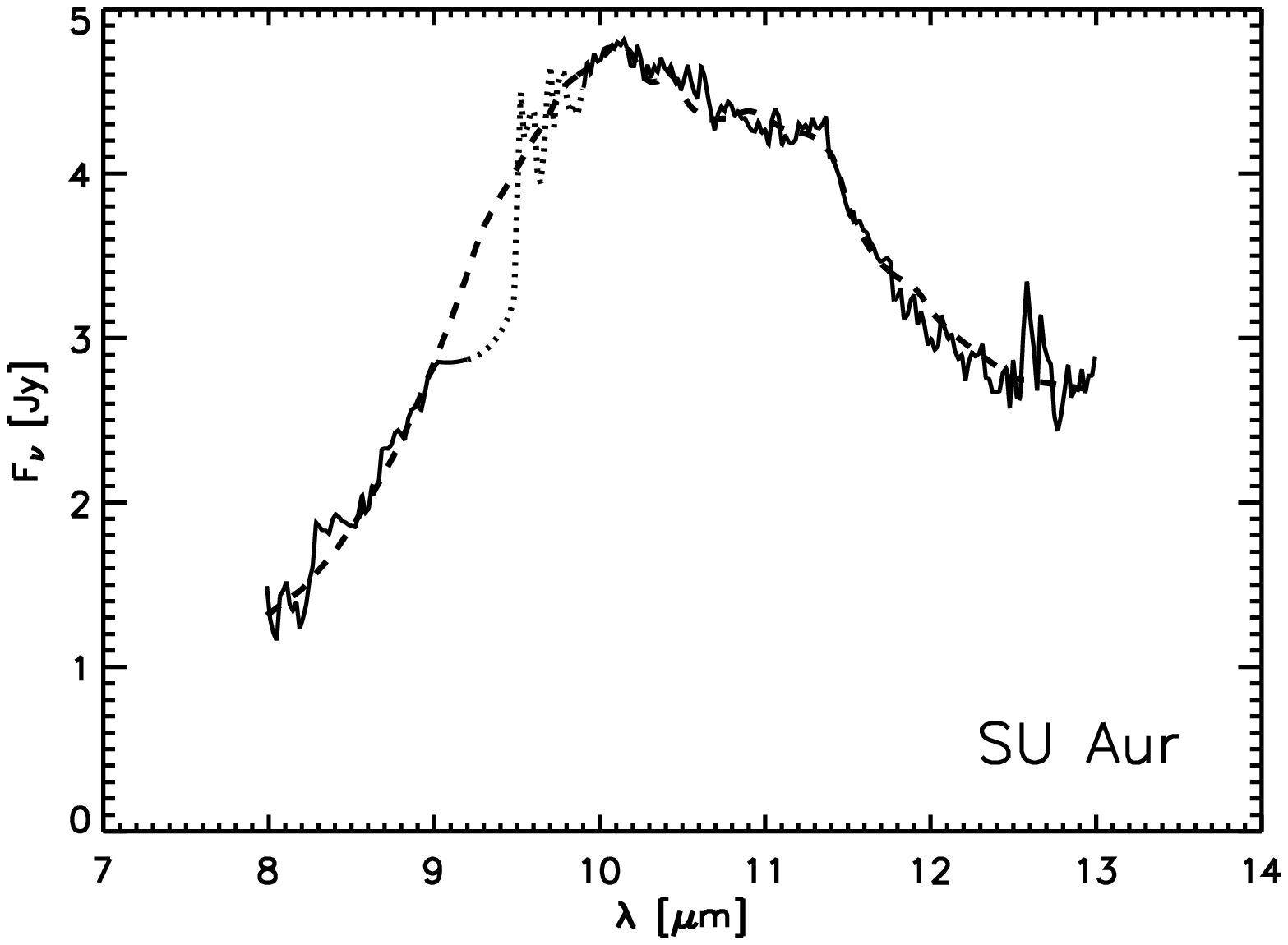}\includegraphics[%
  clip,
  scale=0.34]{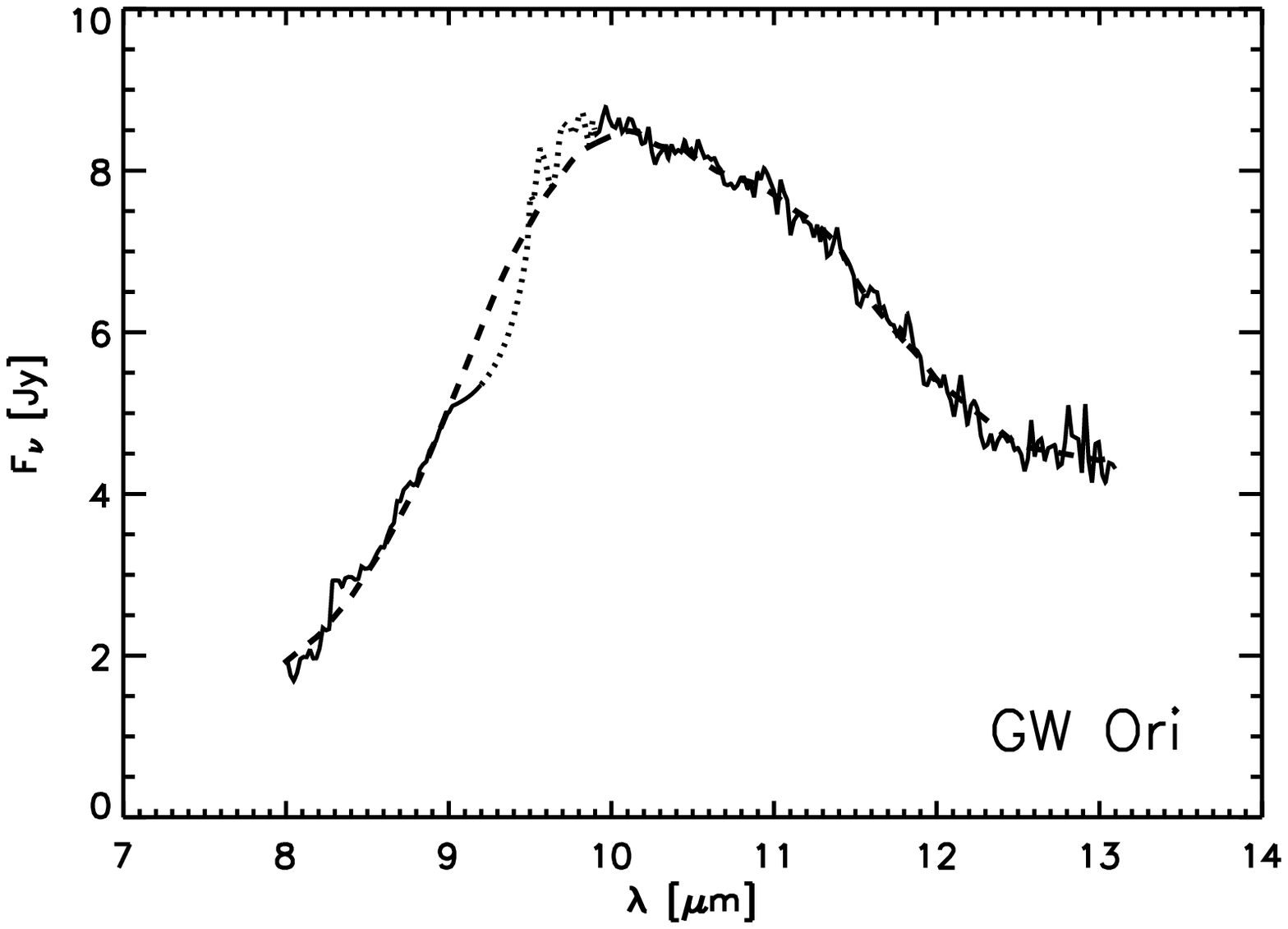}\includegraphics[%
  clip,
  scale=0.34]{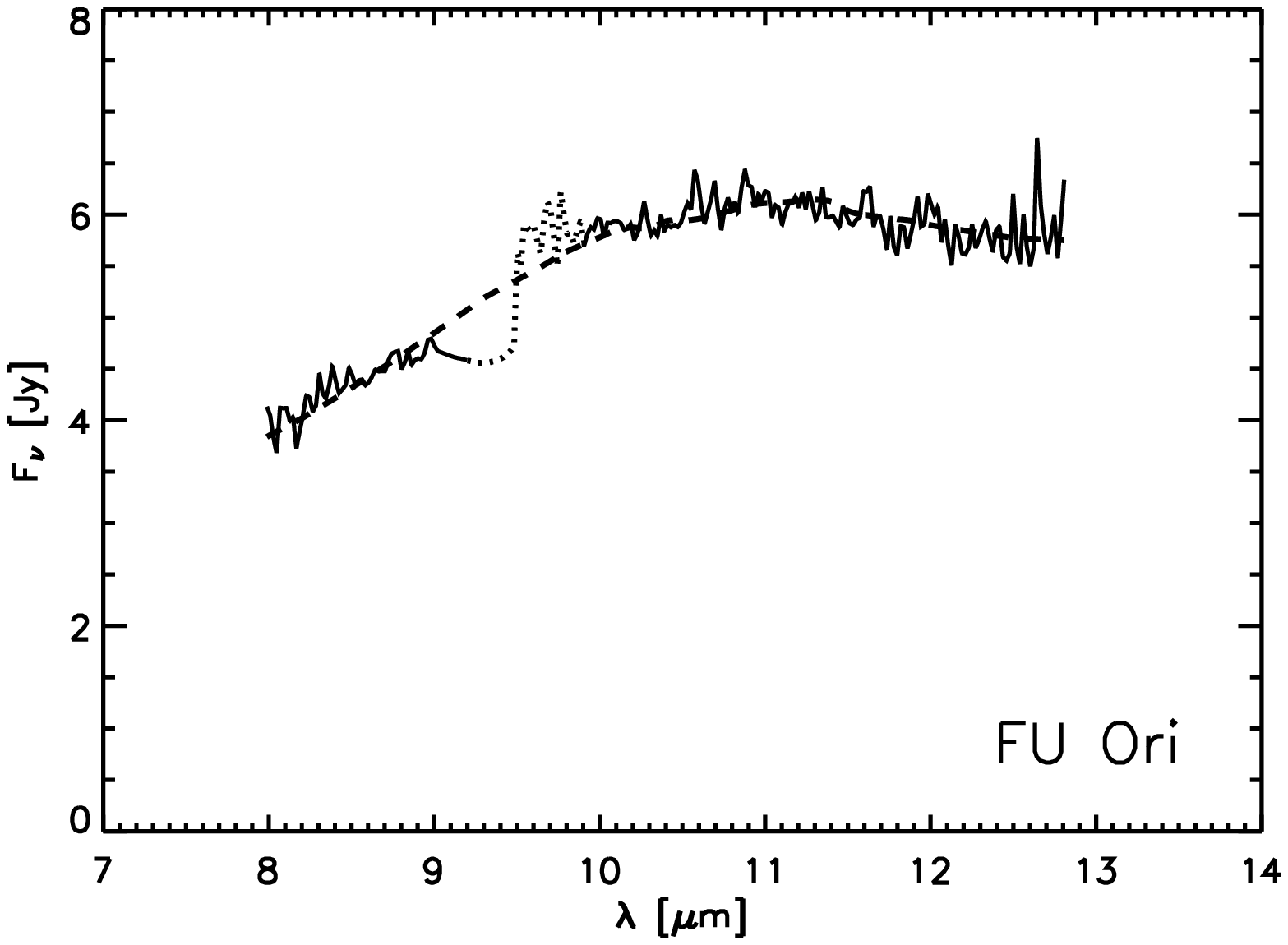}\hfill{}

\hfill{}\includegraphics[%
  clip,
  scale=0.34]{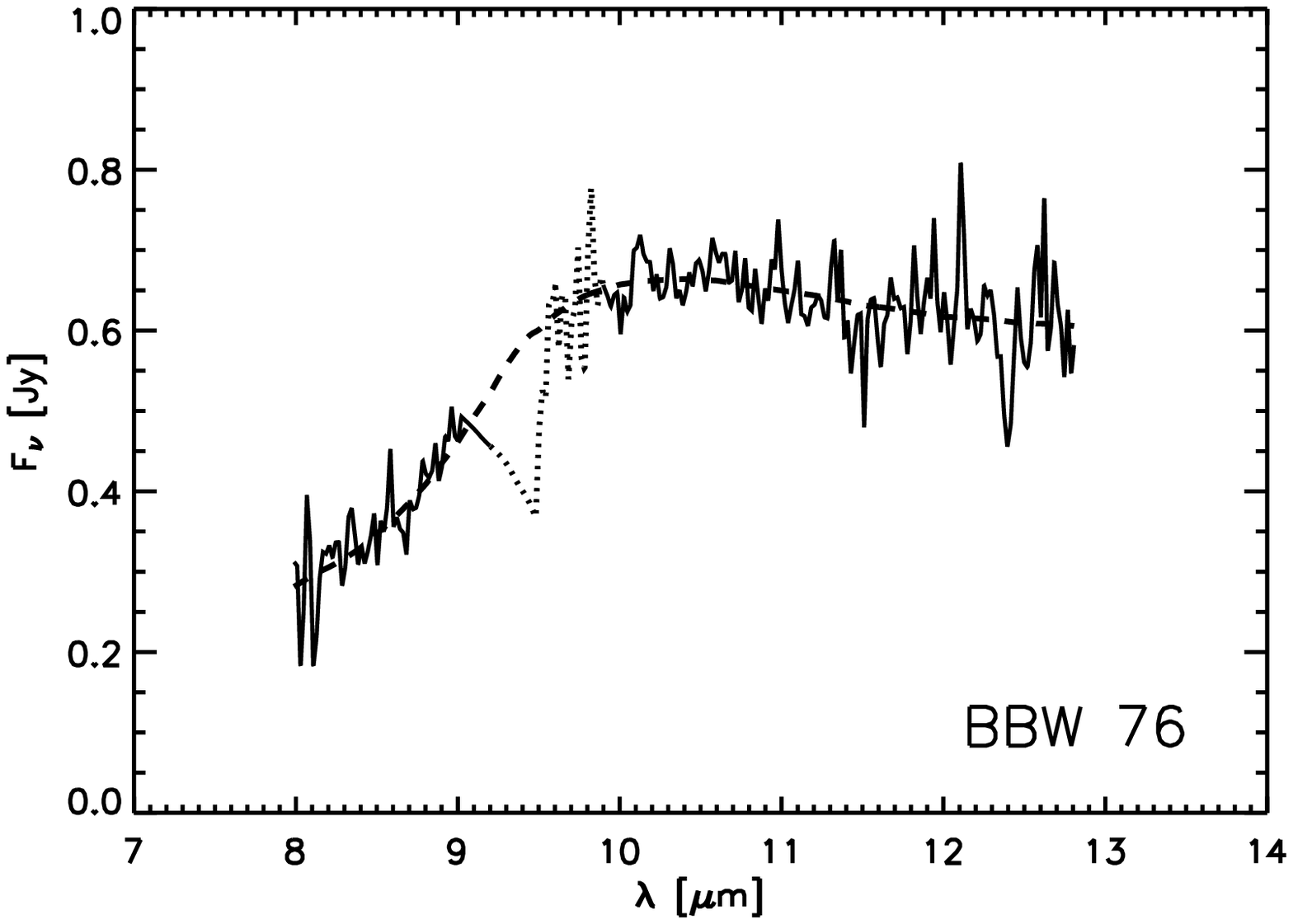}\includegraphics[%
  clip,
  scale=0.34]{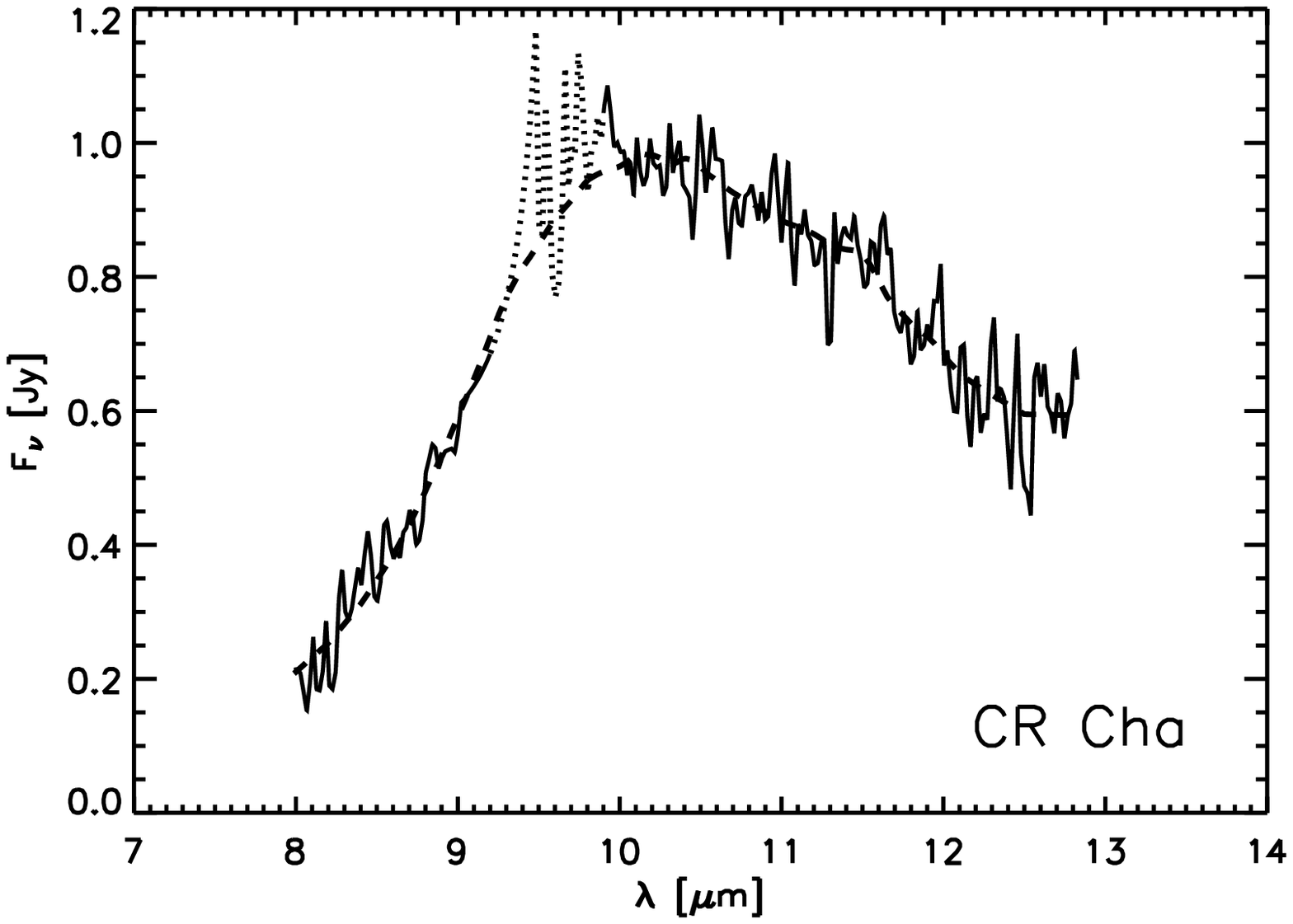}\includegraphics[%
  clip,
  scale=0.34]{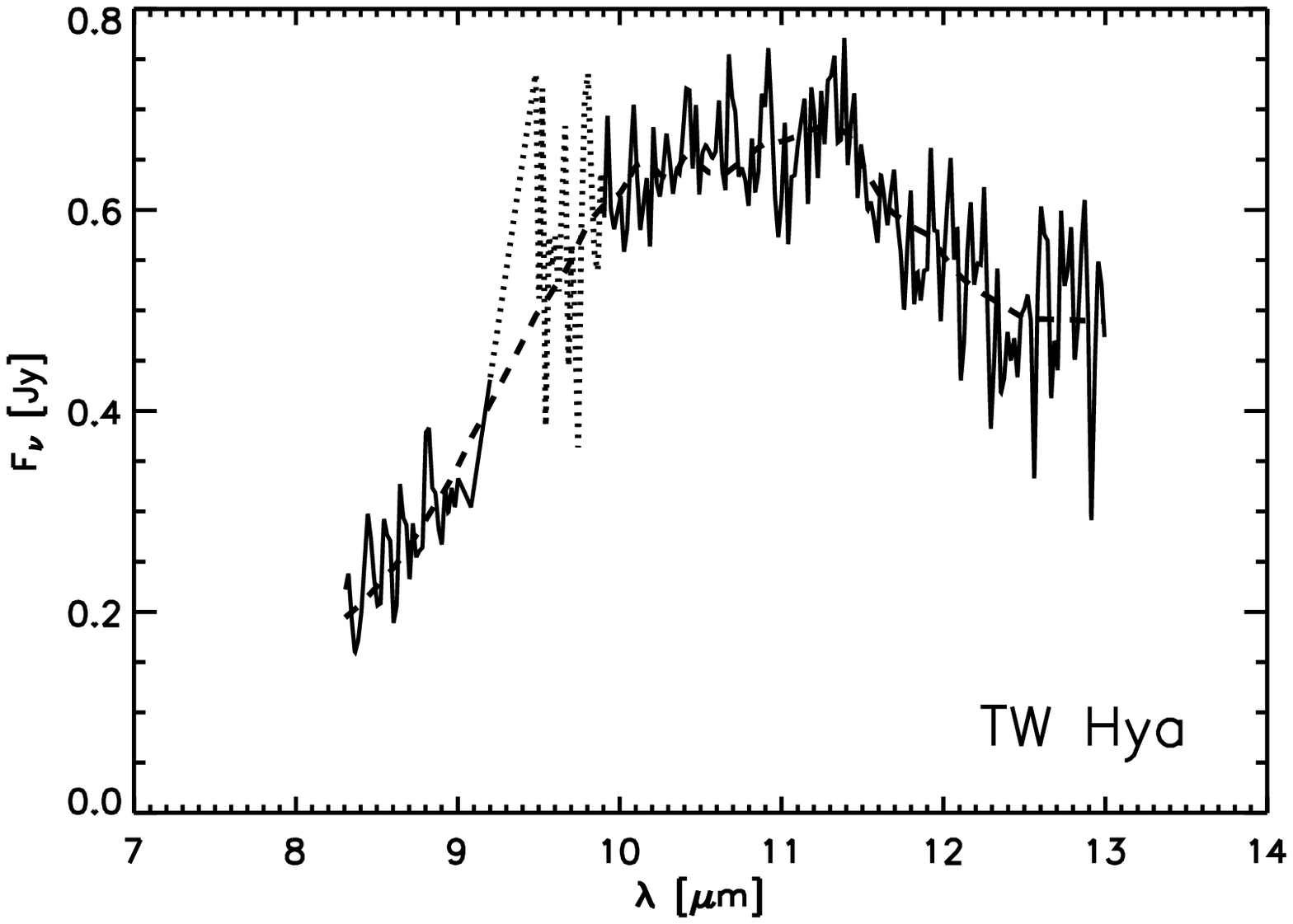}\hfill{}

\hfill{}\includegraphics[%
  scale=0.34]{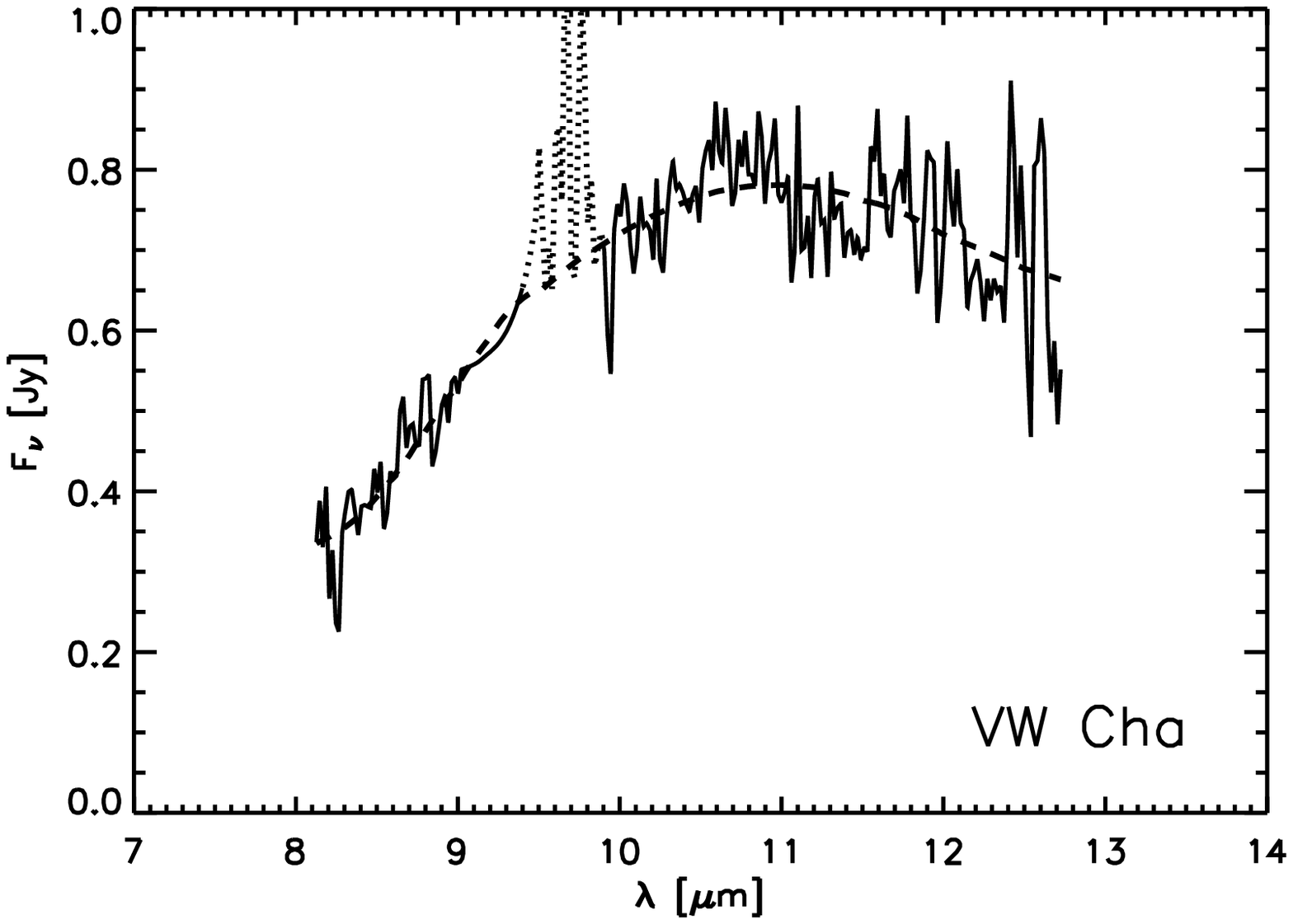}\includegraphics[%
  clip,
  scale=0.34]{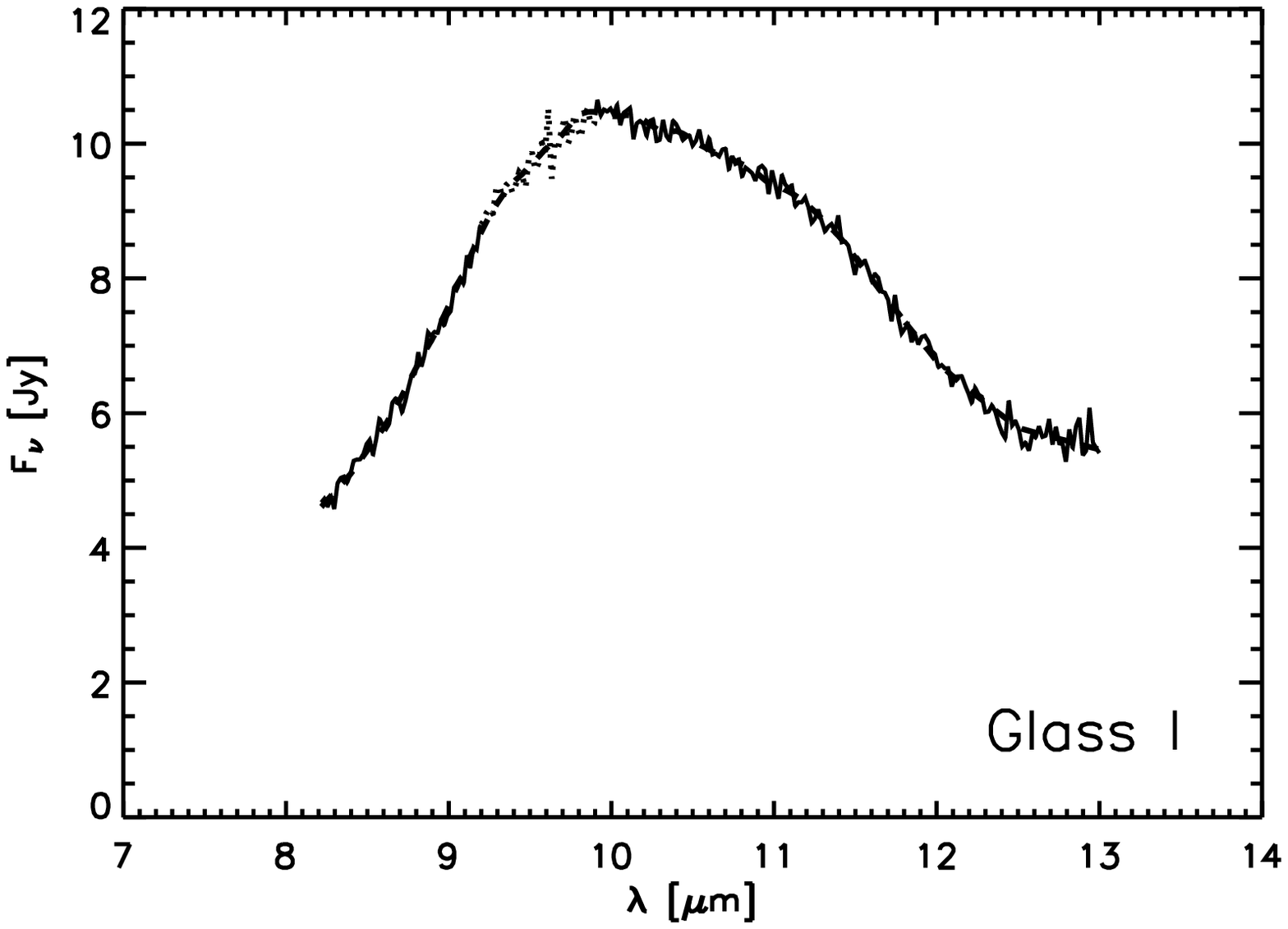}\includegraphics[%
  clip,
  scale=0.34]{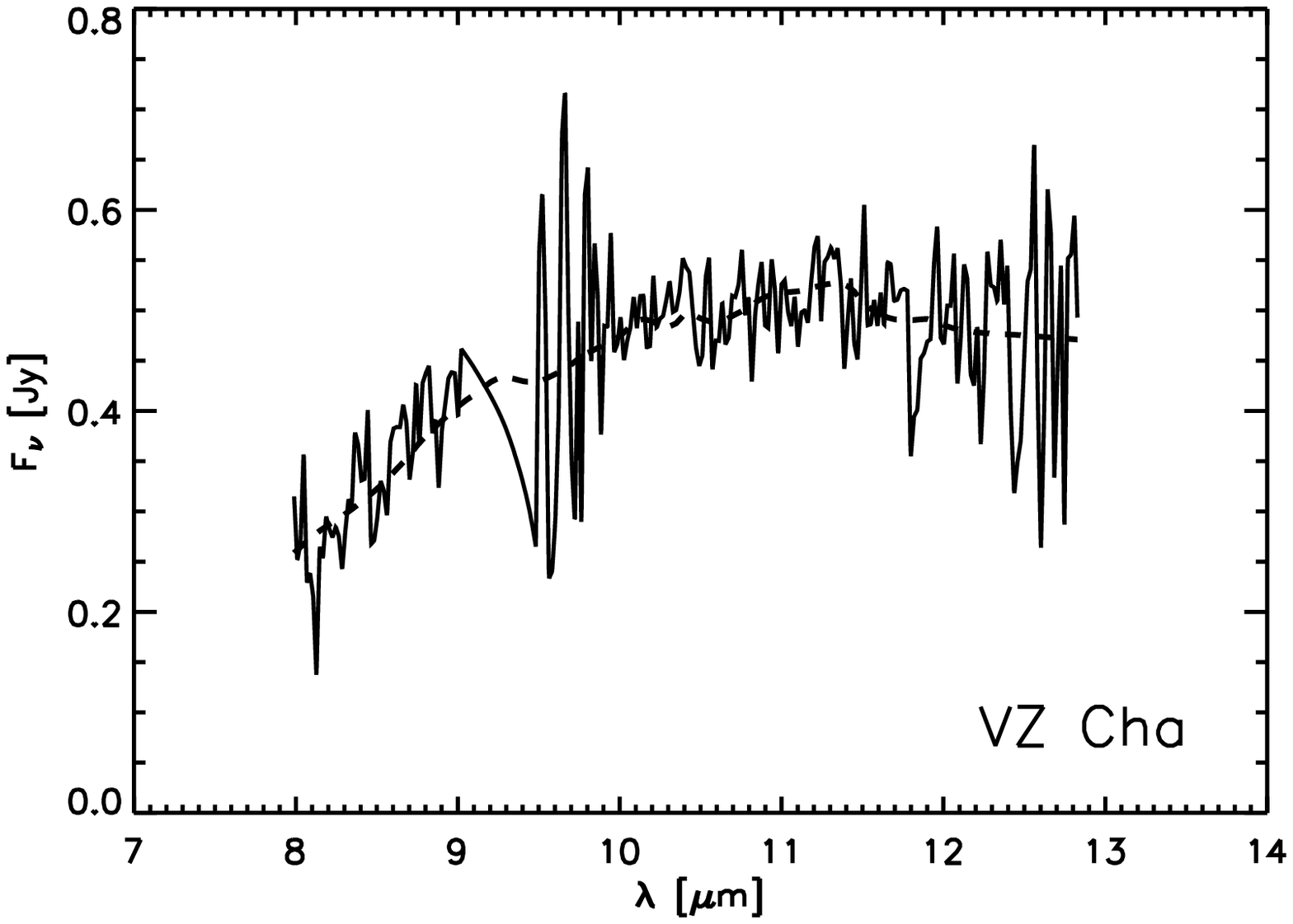}\hfill{}

\caption{\label{fig: fit-result}N band spectra of our sources (solid lines).
The best fits to our spectra are plotted, too (dashed lines).
We cut the wavelength interval $\sim 9.1\,\mu$m to $\sim 9.7\,\mu$m
from the spectra (dotted line) taking into account
the defect readout channel of the TIMMI2 detector and in order to
exclude remnants of the telluric ozone corrections.
}
\end{figure*}

The results of the fitting procedure are satisfactory 
(see Fig.~\ref{fig: fit-result}), in general. The quality of each fit, characterised
by $\chi^{2}$ (e.g., Press et al. 1986) with 
\begin{eqnarray} 
\chi^2=\frac{1}{N_{\rm \lambda}-f} \sum^{\rm N_{\rm \lambda}}_{\rm i=1}
\frac{({\cal F}^{\rm model}_{\rm \nu}(\lambda_{\rm i})-{\cal F}^{\rm data}_{\rm \nu}
(\lambda_{\rm i}))^2}{({\cal F}(\lambda_{\rm i})^2)_{\rm error}} 
\label{chi} 
\end{eqnarray}
is in the order of unity for most objects. The data with the largest $\chi^2$ have the worst 
signal-to-noise ratios (e.g., SZ\,82, VZ\,Cha). In Eq.~\ref{chi}, $N_{\rm \lambda}$
is the number of data points, $f$ is the degree of freedom, 
and ${\cal F}(\lambda_{\rm i})^2_{\rm error}$ the 
measurement error of each spectral data point. Following PR03 and KS05 the 
measurement error is the standard deviation of spectra that was observed in
subsequent nights. A root mean square (rms) of $\sim$9\% and $\sim$11\%
were determined for TIMMI2 and LWS observations, respectively. However, these errors 
were derived considering also parts of the spectra with a significantly lower 
signal-to-noise ratio such as the boundaries of the 
N band and the spectral range where the telluric ozone-band appears. In the following, we assume 
a rms of $\sim$6\% and $\sim$7\%, respectively, which we derive from the 
wavelength interval between 10\,$\rm \mu m$ and 12\,$\rm \mu m$. For several objects with higher  
signal-to-noise ratios, we determine a rms by using 
\begin{eqnarray} 
\sigma^2=\frac{1}{N_{\rm \lambda}-f}\sum^{\rm N_{\rm \lambda}}_{\rm i=1}
({\cal F}^{\rm model}_{\rm \nu}(\lambda_{\rm i})-{\cal F}^{\rm data}_{\rm \nu}
(\lambda_{\rm i}))^2.
\end{eqnarray}

The errors on the fit parameters are calculated using a Monte Carlo method. For 
every spectrum we generate 100 synthetic spectra, by randomly adding Gaussian noise 
to the spectrum at each wavelength point. The Gaussian noise has a rms of $\sim$6\% and $\sim$7\%, 
respectively. 
On each of these spectra, we perform the exact same compositional fit procedure, yielding 
different values of the fit parameters. From the resulting distribution of all fit 
parameters, we calculate the arithmetic mean value and standard deviation.

Approximately at $\rm 9.7\,\mu m$ a narrow ``feature'' appears
in some spectra (e.g., Haro\,1-16) which can not be fitted
by the available emission profiles. We ascribe this artefact to a
remnant of data reduction, i.e. sky subtraction, considering the spectral-adjacent,
broad atmospheric ozone line at $\rm \sim 9.5\,\mu m$. Additionally,
there is a defect readout channel of the TIMMI2 detector between
$\rm 9.14\,\mu m$ and $\rm 9.67\,\mu m$. Because of these two issues,
the fitting procedure does not account for the wavelength interval
between $\sim$9.2\,$\rm \mu m$ and $\sim$9.9\,$\rm \mu m$, which is
marked by a dotted line in the Fig. \ref{fig: fit-result}. 
In some spectra (e.g., DQ\,Tau) a second remnant of sky
subtraction appears at $\rm 12.7\,\mu m$, where several telluric OH-lines
accumulate. In order to improve the fit quality, we additionally cut
the spectral data in a narrow wavelength interval at $\rm \sim 12.7 \mu m$.

Although PAH molecules appear to be abundant throughout the 
universe (e.g., Lagache et al. 2004; Wu et al. 2005), our spectra can be 
sufficiently fitted without consi\-dering PAHs although they also show an 
emission feature near $\rm 11.2\,\mu m$ si\-milar to forsterite. The 
absence of its $\rm 11.3\,\mu m$ feature in the circumstellar disks of (our) 
T\,Tauri stars can be explained by the weak stellar UV radiation field compared 
to the higher-mass HAeBe stars. In fact, PAH molecules are thought to be excited 
by ultraviolet photons and cooled down by subsequent infrared emission. Furthermore, the 
circumstellar disk size and disk geometry affect the presence and strength of 
the PAH feature as Habart et al. (2004) and Acke \& van den Ancker (2004) showed.

Pyroxene is an important part of our dust mixture because it
is the dominant component at low temperatures (Gail 2003). BO05 included this
dust component considering its contribution to the short wavelength
range of the silicate feature at $\rm \sim 9\,\mu m$ 
(see Fig. \ref{fig:emission profiles}).
According to our results (see Table \ref{fitresults}) the large mass fraction of 
\emph{large} pyroxene grains (for ex\-ample, in VZ\,Cha) is suspicious. 
A corresponding richness of large
pyroxene grains was found in HAeBe stars (BO05), too. In 1994,
Pollack et al. determined that the initial composition of the silicate
dust mixture in circumstellar disks only consists of a maximum of
30\% pyroxene dust. Therefore, if the contribution of large
pyroxene grains is real, small pyroxene dust grains have already 
disappeared due to grain growth. 
If we do not consider pyroxene\footnote{large \emph{and} small grains} 
in our fitting function, the quantity $\chi^{2}$ increases slightly, 
about a few percents for the 
T\,Tauri stars but up to 100\% for the HAeBe stars 
HD\,163296, HD\,179218, and WW\,Vul.

Apart from the objects Hen\,3-600\,A and S\,CrA\,N
for {most of the} objects crystalline silicates contribute less than 10\% to the total fit. 
This result agrees with Pollack et al. (1994) who assumed a crystalline
initial contribution of only 10\% while notable amounts of crystalline
dust species like enstatite and forsterite are produced in annealing
reactions during disk evolution. 
Gail (2003) shows that crystalline dust dominates in the inner
parts of protoplanetary disks. However, the temperature is too
high in the inner regions of the disk in order to produce the prominent silicate feature
(see Fig.~\ref{cap:The-simulated-SED}). Apparently, the crystallisation
of dust is inhibited in the cool circumstellar disks around T\,Tauri
stars which is confirmed by the minor mass fraction of crystalline dust 
in contrast to HAeBe stars where a mean crystalline
contribution of $\sim 15.0 \%$ was determined (see BO05).

\onecolumn{
\begin{sidewaystable}
\begin{center}
\caption{\label{fitresults} Best results of our compositional fitting procedure.}
\begin{tabular}{clcccccccccccc}

\hline \hline

 &  &  & \multicolumn{2}{c}{Olivine} & \multicolumn{2}{c}{Pyroxene} & \multicolumn{2}{c}{Forsterite} & \multicolumn{2}{c}{Enstatite} & \multicolumn{2}{c}{Quartz} &  \\

\raisebox{1.5ex}[-1.5ex]{\#} & \raisebox{1.5ex}[-1.5ex]{Object} &  \raisebox{1.5ex}[-1.5ex]{$T[K]$} & Small [\%] & Large [\%] & Small [\%] & Large [\%] & Small [\%] & Large [\%] & Small [\%] & Large [\%]  & Small [\%]  & Large [\%] & \raisebox{1.5ex}[-1.5ex]{$\rm \chi^2 $}  \\[1.0ex] \hline

       1 & GG\,Tau &  384 $\rm \pm $   3 &  52.8 $\rm \pm $   2.3 &   3.7
 $\rm \pm $   4.9 &   1.2 $\rm \pm $   1.8 &  34.6 $\rm \pm $   5.1 &   3.3
 $\rm \pm $   0.2 &   3.7 $\rm \pm $   0.8 &   0.2 $\rm \pm $   0.3 &   0.4
 $\rm \pm $   0.6 & --  & --  &  0.88 $\rm \pm $  0.05
\\
       2 & AA\,Tau &  442 $\rm \pm $   6 &  44.3 $\rm \pm $   5.6 &   9.4
 $\rm \pm $  11.8 &  15.9 $\rm \pm $   5.5 &  23.9 $\rm \pm $  12.8 & -- &   
 5.6 $\rm \pm $   1.1 &   0.1 $\rm \pm $   0.3 &   0.2
 $\rm \pm $   0.5 &   0.1 $\rm \pm $   0.2 &   0.3 $\rm \pm $   0.5 &  0.49
 $\rm \pm $  0.02\\
       3 & LkCa\,15 & 1008 $\rm \pm $  95 &  52.6 $\rm \pm $   3.3 &  43.0
 $\rm \pm $   5.0 &   0.1 $\rm \pm $   0.5 &   2.3 $\rm \pm $   3.3 & -- &   
 0.4 $\rm \pm $   0.4 &   1.0 $\rm \pm $   0.4 &   0.4
 $\rm \pm $   0.6 & --  &   0.2 $\rm \pm $   0.3 &  0.70
 $\rm \pm $  0.03\\
       4 & DQ\,Tau &  300 $\rm \pm $   3 &  53.3 $\rm \pm $  15.8 &   3.1
 $\rm \pm $   5.3 &  19.9 $\rm \pm $  13.2 &   4.2 $\rm \pm $   6.8 &   0.5
 $\rm \pm $   0.4 &   1.5 $\rm \pm $   1.2 &   1.1 $\rm \pm $   1.2 &  11.2
 $\rm \pm $   2.3 &   2.0 $\rm \pm $   1.3 &   3.2 $\rm \pm $   2.7 &  0.62
 $\rm \pm $  0.03\\
       5 & DR\,Tau &  354 $\rm \pm $   1 &  46.7 $\rm \pm $   3.3 &  47.9
 $\rm \pm $   4.3 &   0.3 $\rm \pm $   0.9 &   1.5 $\rm \pm $   2.9 &   1.1
 $\rm \pm $   0.3 &   2.0 $\rm \pm $   0.8 & --  &   0.2
 $\rm \pm $   0.5 & --  &   0.2 $\rm \pm $   0.4 &  0.23
 $\rm \pm $  0.01\\
       6 & GM\,Aur &  403 $\rm \pm $   6 &  68.0 $\rm \pm $   5.2 &  27.5
 $\rm \pm $   6.2 & --  &   0.9 $\rm \pm $   2.1 &   1.2
 $\rm \pm $   0.3 &   0.5 $\rm \pm $   0.6 &   0.2 $\rm \pm $   0.3 &   1.6
 $\rm \pm $   1.1 & --  &   0.1 $\rm \pm $   0.3 &  0.74
 $\rm \pm $  0.03\\
       7 & SU\,Aur &  348 $\rm \pm $   2 &  62.0 $\rm \pm $   1.8 &  18.9
 $\rm \pm $   4.1 &   0.5 $\rm \pm $   1.0 &  12.6 $\rm \pm $   3.8 &   2.9
 $\rm \pm $   0.1 &   2.7 $\rm \pm $   0.5 & --  &   0.2
 $\rm \pm $   0.4 & --  & --  &  0.55 $\rm \pm $  0.03
\\
       8 & GW\,Ori &  334 $\rm \pm $   2 &  54.0 $\rm \pm $   1.1 &  21.9
 $\rm \pm $   2.9 &   0.4 $\rm \pm $   0.8 &  22.1 $\rm \pm $   2.8 &   1.0
 $\rm \pm $   0.1 &   0.3 $\rm \pm $   0.3 & --  &   0.1
 $\rm \pm $   0.2 & --  & --  &  0.37 $\rm \pm $  0.02
\\
       9 & FU\,Ori &  387 $\rm \pm $   1 &   8.1 $\rm \pm $   5.7 &  82.7
 $\rm \pm $   7.3 &   0.2 $\rm \pm $   0.8 &   2.4 $\rm \pm $   4.5 &   3.5
 $\rm \pm $   0.6 &   1.5 $\rm \pm $   1.5 &   0.3 $\rm \pm $   0.6 &   0.7
 $\rm \pm $   1.3 & --  &   0.6 $\rm \pm $   1.0 &  0.20
 $\rm \pm $  0.01\\
      10 & BBW76 &  324 $\rm \pm $   3 &  79.1 $\rm \pm $   8.6 &  13.3
 $\rm \pm $   8.0 &   0.6 $\rm \pm $   1.7 &   1.5 $\rm \pm $   3.2 &   0.2
 $\rm \pm $   0.2 &   1.1 $\rm \pm $   1.2 &   0.2 $\rm \pm $   0.4 &   0.5
 $\rm \pm $   0.9 & --  &   3.5 $\rm \pm $   1.7 &  1.51
 $\rm \pm $  0.09\\
      11 & CR\,Cha &  297 $\rm \pm $   3 &  59.7 $\rm \pm $   3.2 &   2.9
 $\rm \pm $   4.9 &   7.0 $\rm \pm $   2.9 &  23.9 $\rm \pm $   5.7 & -- &   
 5.0 $\rm \pm $   0.7 &   0.2 $\rm \pm $   0.3 &   1.0
 $\rm \pm $   0.8 & --  &   0.2 $\rm \pm $   0.4 &  2.31
 $\rm \pm $  0.42\\
      12 & TW\,Hya &  285 $\rm \pm $   3 &  37.0 $\rm \pm $   4.4 &  47.2
 $\rm \pm $   7.4 &   0.6 $\rm \pm $   1.5 &   3.5 $\rm \pm $   5.1 &   2.9
 $\rm \pm $   0.3 &   5.9 $\rm \pm $   1.1 & --  &   2.6
 $\rm \pm $   1.5 & --  &   0.2 $\rm \pm $   0.4 &  3.32
 $\rm \pm $  0.58\\
      13 & VW\,Cha &  332 $\rm \pm $   5 &   3.1 $\rm \pm $   4.2 &  80.9
 $\rm \pm $  11.3 &   2.1 $\rm \pm $   4.2 &   8.1 $\rm \pm $  11.7 &   0.2
 $\rm \pm $   0.3 &   0.7 $\rm \pm $   1.0 &   0.5 $\rm \pm $   0.8 &   0.5
 $\rm \pm $   0.9 &   0.6 $\rm \pm $   0.4 &   3.3 $\rm \pm $   1.4 &  2.24
 $\rm \pm $  0.15\\
      14 & Glass\,I &  453 $\rm \pm $   2 &  37.0 $\rm \pm $   0.7 &  37.0
 $\rm \pm $   2.0 &   0.2 $\rm \pm $   0.4 &  21.1 $\rm \pm $   1.9 &   0.6
 $\rm \pm $   0.1 &   1.4 $\rm \pm $   0.2 &   0.6 $\rm \pm $   0.2 &   2.1
 $\rm \pm $   0.4 & --  & --  &  0.05 $\rm \pm $  0.00
\\
      15 & VZ\,Cha &  348 $\rm \pm $   5 &   4.0 $\rm \pm $   8.0 &  13.8
 $\rm \pm $  21.9 &   7.4 $\rm \pm $  11.7 &  53.2 $\rm \pm $  24.3 &   6.2
 $\rm \pm $   1.7 &   9.2 $\rm \pm $   4.6 &   1.5 $\rm \pm $   2.1 &   1.6
 $\rm \pm $   2.7 &   2.7 $\rm \pm $   1.0 &   0.4 $\rm \pm $   0.8 &  5.51
 $\rm \pm $  0.49\\
      16 & WW\,Cha &  355 $\rm \pm $   2 &  14.2 $\rm \pm $   1.6 &  35.9
 $\rm \pm $   5.2 &   1.2 $\rm \pm $   2.0 &  45.4 $\rm \pm $   5.6 &   0.4
 $\rm \pm $   0.1 &   1.4 $\rm \pm $   0.4 & --  &   0.1
 $\rm \pm $   0.2 & --  &   1.4 $\rm \pm $   0.3 &  0.23
 $\rm \pm $  0.01\\
      17 & Hen3-600\,A &  301 $\rm \pm $   4 &   2.3 $\rm \pm $   4.7 &  15.8
 $\rm \pm $  19.6 &   1.0 $\rm \pm $   2.6 &  36.6 $\rm \pm $  21.2 &   7.5
 $\rm \pm $   1.2 &   1.0 $\rm \pm $   1.5 &   7.3 $\rm \pm $   3.0 &  27.8
 $\rm \pm $   3.7 &   0.2 $\rm \pm $   0.3 &   0.4 $\rm \pm $   0.9 &  0.70
 $\rm \pm $  0.03\\
      18 & IRAS\,14050 &  406 $\rm \pm $   6 &  74.9 $\rm \pm $   2.5 &   2.4
 $\rm \pm $   3.6 &   0.5 $\rm \pm $   1.0 &  19.1 $\rm \pm $   4.1 &   0.3
 $\rm \pm $   0.2 &   0.3 $\rm \pm $   0.4 &   1.8 $\rm \pm $   0.5 &   0.6
 $\rm \pm $   0.7 & --  &   0.2 $\rm \pm $   0.4 &  0.70
 $\rm \pm $  0.03\\
      19 & SZ\,82 &  411 $\rm \pm $   7 &   2.3 $\rm \pm $   3.6 &  16.5
 $\rm \pm $  15.8 &   2.2 $\rm \pm $   4.4 &  58.9 $\rm \pm $  16.8 &   0.4
 $\rm \pm $   0.4 &  12.2 $\rm \pm $   2.4 &   0.5 $\rm \pm $   0.9 &   6.8
 $\rm \pm $   2.5 &   0.1 $\rm \pm $   0.2 &   0.2 $\rm \pm $   0.6 &  1.52
 $\rm \pm $  0.08\\
      20 & RU\,Lup &  346 $\rm \pm $   2 &   1.1 $\rm \pm $   2.1 &   5.5
 $\rm \pm $   8.2 &   2.7 $\rm \pm $   3.2 &  75.0 $\rm \pm $  10.2 &   6.9
 $\rm \pm $   0.4 &   0.3 $\rm \pm $   0.6 &   3.5 $\rm \pm $   0.9 &   1.8
 $\rm \pm $   1.4 &   0.7 $\rm \pm $   0.4 &   2.5 $\rm \pm $   1.1 &  0.14
 $\rm \pm $  0.01\\
      21 & AS\,205\,N &  329 $\rm \pm $   1 &  38.9 $\rm \pm $   3.3 &   5.7
 $\rm \pm $   6.7 &   2.0 $\rm \pm $   2.7 &  47.5 $\rm \pm $   8.6 &   1.9
 $\rm \pm $   0.3 &   3.3 $\rm \pm $   0.9 &   0.2 $\rm \pm $   0.3 &   0.5
 $\rm \pm $   0.8 & --  & --  &  0.13 $\rm \pm $  0.01
\\
      22 & AS\,205\,S &  507 $\rm \pm $   4 &  20.4 $\rm \pm $   1.3 &  68.1
 $\rm \pm $   3.2 &   0.5 $\rm \pm $   0.9 &   9.5 $\rm \pm $   3.1 &   0.4
 $\rm \pm $   0.1 &   0.9 $\rm \pm $   0.4 & --  &   0.2
 $\rm \pm $   0.3 & --  & --  &  0.15 $\rm \pm $  0.01
\\
      23 & HBC\,639 &  366 $\rm \pm $   3 &   2.3 $\rm \pm $   3.0 &  69.8
 $\rm \pm $   8.3 &   0.5 $\rm \pm $   1.3 &  13.5 $\rm \pm $   9.3 &   1.9
 $\rm \pm $   0.4 &   0.4 $\rm \pm $   0.5 &   5.9 $\rm \pm $   1.0 &   4.1
 $\rm \pm $   1.7 & --  &   1.4 $\rm \pm $   0.7 &  0.37
 $\rm \pm $  0.01\\
      24 & Haro\,1-16 &  373 $\rm \pm $   4 &  73.4 $\rm \pm $   2.5 &   2.7
 $\rm \pm $   3.8 &   1.0 $\rm \pm $   1.6 &  18.9 $\rm \pm $   5.9 &   1.7
 $\rm \pm $   0.3 &   0.2 $\rm \pm $   0.4 &   0.1 $\rm \pm $   0.2 &   1.8
 $\rm \pm $   1.1 & --  & --  &  0.69 $\rm \pm $  0.03
\\
      25 & AK\,Sco &  559 $\rm \pm $   5 &  63.8 $\rm \pm $   1.1 &  31.6
 $\rm \pm $   1.7 & --  &   0.5 $\rm \pm $   0.9 &   1.7
 $\rm \pm $   0.1 &   2.1 $\rm \pm $   0.4 & --  &   0.2
 $\rm \pm $   0.3 & --  & --  &  0.58 $\rm \pm $  0.03
\\
      26 & S\,CrA\,N &  427 $\rm \pm $   1 &   2.5 $\rm \pm $   4.1 &  42.2
 $\rm \pm $   7.9 & --  &   1.8 $\rm \pm $   4.1 &  18.4
 $\rm \pm $   1.0 &  11.6 $\rm \pm $   2.3 &   0.4 $\rm \pm $   0.7 &  22.9
 $\rm \pm $   2.6 & --  &   0.3 $\rm \pm $   0.8 &  0.15
 $\rm \pm $  0.01\\
      27 & S\,CrA\,S &  375 $\rm \pm $   1 &   1.0 $\rm \pm $   1.8 &   2.7
 $\rm \pm $   3.9 &   1.0 $\rm \pm $   1.8 &  89.1 $\rm \pm $   5.4 &   3.8
 $\rm \pm $   0.3 &   0.3 $\rm \pm $   0.5 &   0.6 $\rm \pm $   0.6 &   1.4
 $\rm \pm $   1.1 & --  &   0.1 $\rm \pm $   0.2 &  0.25
 $\rm \pm $  0.01\\ \hline 
      28 & MWC\,480 &  801 $\rm \pm $  44 &   0.7 $\rm \pm $   1.3 &  59.2
 $\rm \pm $   7.4 &   0.1 $\rm \pm $   0.5 &  33.9 $\rm \pm $   6.8 &   1.2
 $\rm \pm $   0.3 &   3.0 $\rm \pm $   0.9 &   1.2 $\rm \pm $   0.5 &   0.5
 $\rm \pm $   0.7 & --  & --  &  0.69 $\rm \pm $  0.03
\\
      29 & HD\,163296 &  431 $\rm \pm $  11 &  20.2 $\rm \pm $   5.6 &  24.7
 $\rm \pm $   8.9 &   0.8 $\rm \pm $   1.9 &  44.5 $\rm \pm $   5.1 &   2.5
 $\rm \pm $   0.3 &   0.2 $\rm \pm $   0.4 &   0.2 $\rm \pm $   0.3 &   5.3
 $\rm \pm $   1.2 & --  &   1.6 $\rm \pm $   0.6 &  0.69
 $\rm \pm $  0.02\\
      30 & HD\,179218 &  255 $\rm \pm $   8 &   1.2 $\rm \pm $   2.1 &   4.1
 $\rm \pm $   7.1 &   1.7 $\rm \pm $   3.7 &  83.9 $\rm \pm $   7.6 &   1.4
 $\rm \pm $   0.2 &   0.9 $\rm \pm $   0.7 &   2.5 $\rm \pm $   1.0 &   0.8
 $\rm \pm $   0.9 &   3.3 $\rm \pm $   0.3 &   0.2 $\rm \pm $   0.3 &  0.69
 $\rm \pm $  0.02\\
      31 & WW\,Vul &  399 $\rm \pm $   9 &  52.7 $\rm \pm $   5.1 &   6.9
 $\rm \pm $   7.2 &   0.6 $\rm \pm $   1.3 &  33.8 $\rm \pm $   3.1 &   1.6
 $\rm \pm $   0.2 &   0.4 $\rm \pm $   0.5 &   0.2 $\rm \pm $   0.3 &   3.5
 $\rm \pm $   1.0 & --  &   0.3 $\rm \pm $   0.5 &  0.69
 $\rm \pm $  0.02\\
      32 & HD\,184761 &  519 $\rm \pm $  10 &  39.6 $\rm \pm $   3.4 &  57.4
 $\rm \pm $   4.0 & --  &   0.4 $\rm \pm $   0.9 &   0.7
 $\rm \pm $   0.2 &   1.6 $\rm \pm $   0.7 & --  &   0.1
 $\rm \pm $   0.3 & --  & --  &  0.69 $\rm \pm $  0.02
\\

\end{tabular}
\end{center}
{\scriptsize
Note. The first two columns correspond to number
(see Table \ref{fig:emission profiles}) and name of each object.
The value $T$ represents the temperature of the underlying continuum
which is determined assuming a black body function (see Eq.~(\ref{eq:planck})).
The abundance of small (0.1\,$\mu$m) and
large (1.5\,$\mu$m) grains of the various dust species are given as relative
fractions of the total mass of silicate dust.
We use the symbol '--' if a fitting parameter was not found
in the corresponding spectra.
The value $\chi^2$ refers to the quality of each fit (see Eq.~(\ref{chi})).
}
\end{sidewaystable}}
\twocolumn

\begin{figure*}[!ht]

\hfill{}\includegraphics[%
  clip,
  scale=0.34]{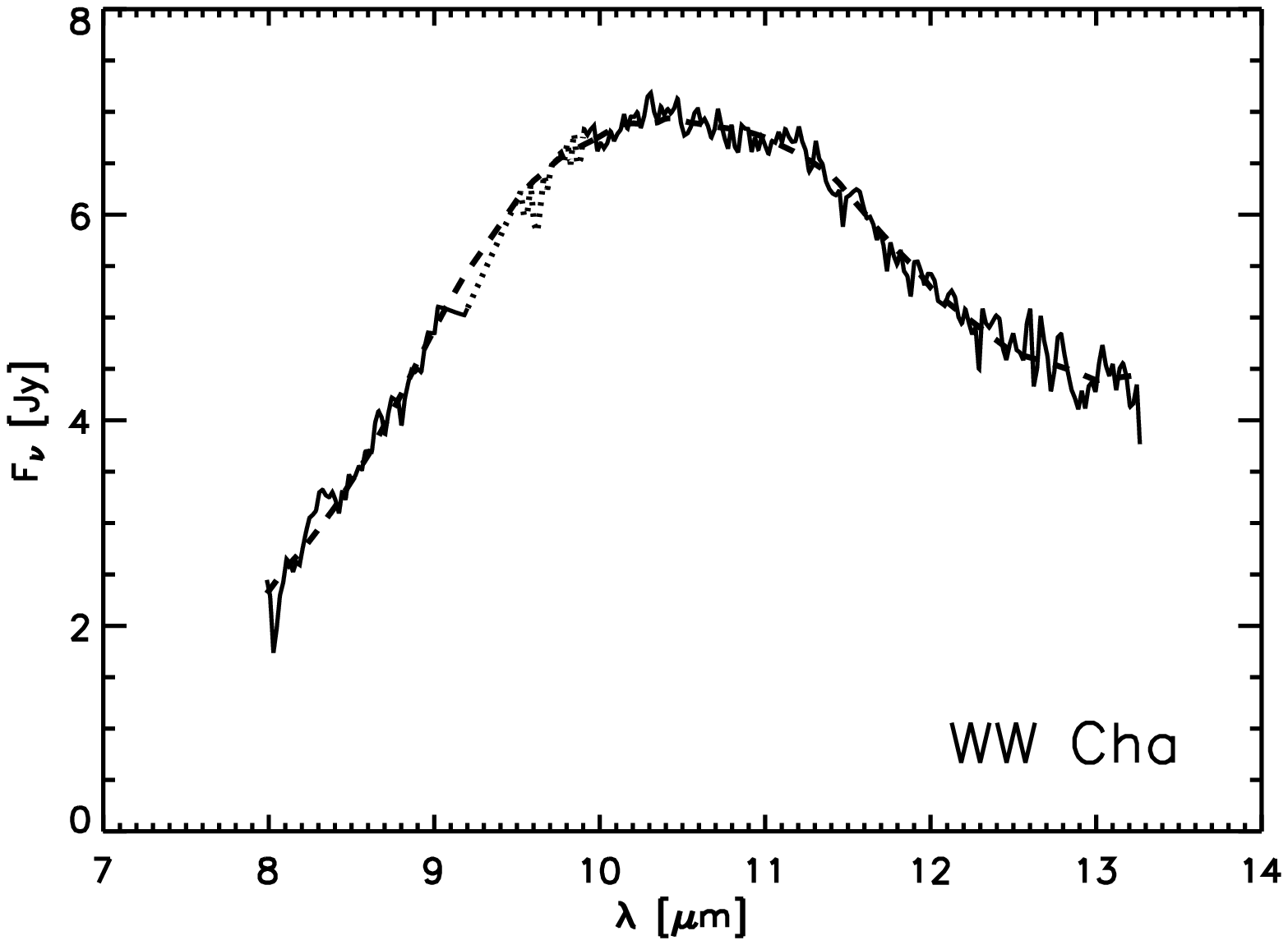}\includegraphics[%
  clip,
  scale=0.34]{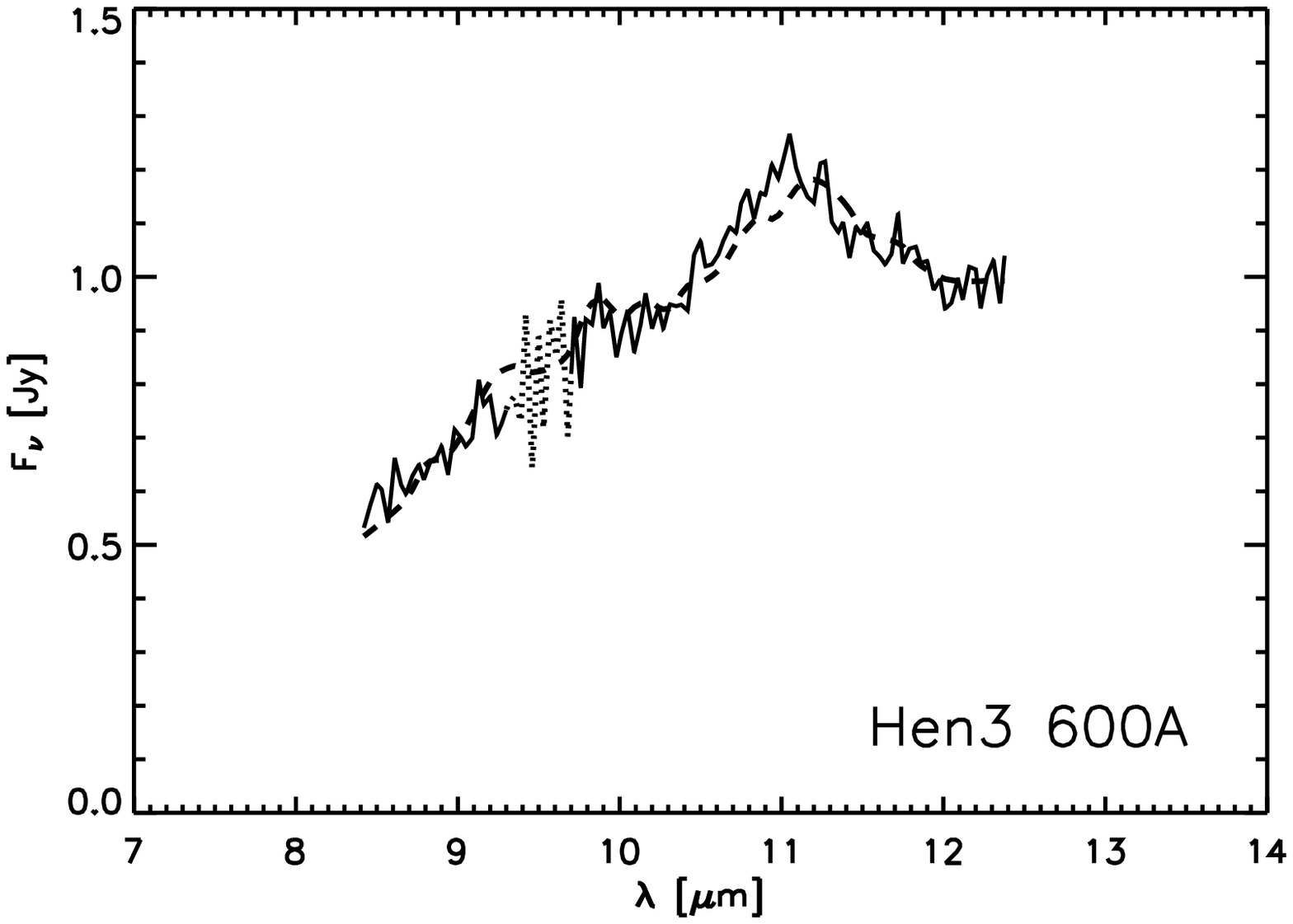}\includegraphics[%
  clip,
  scale=0.34]{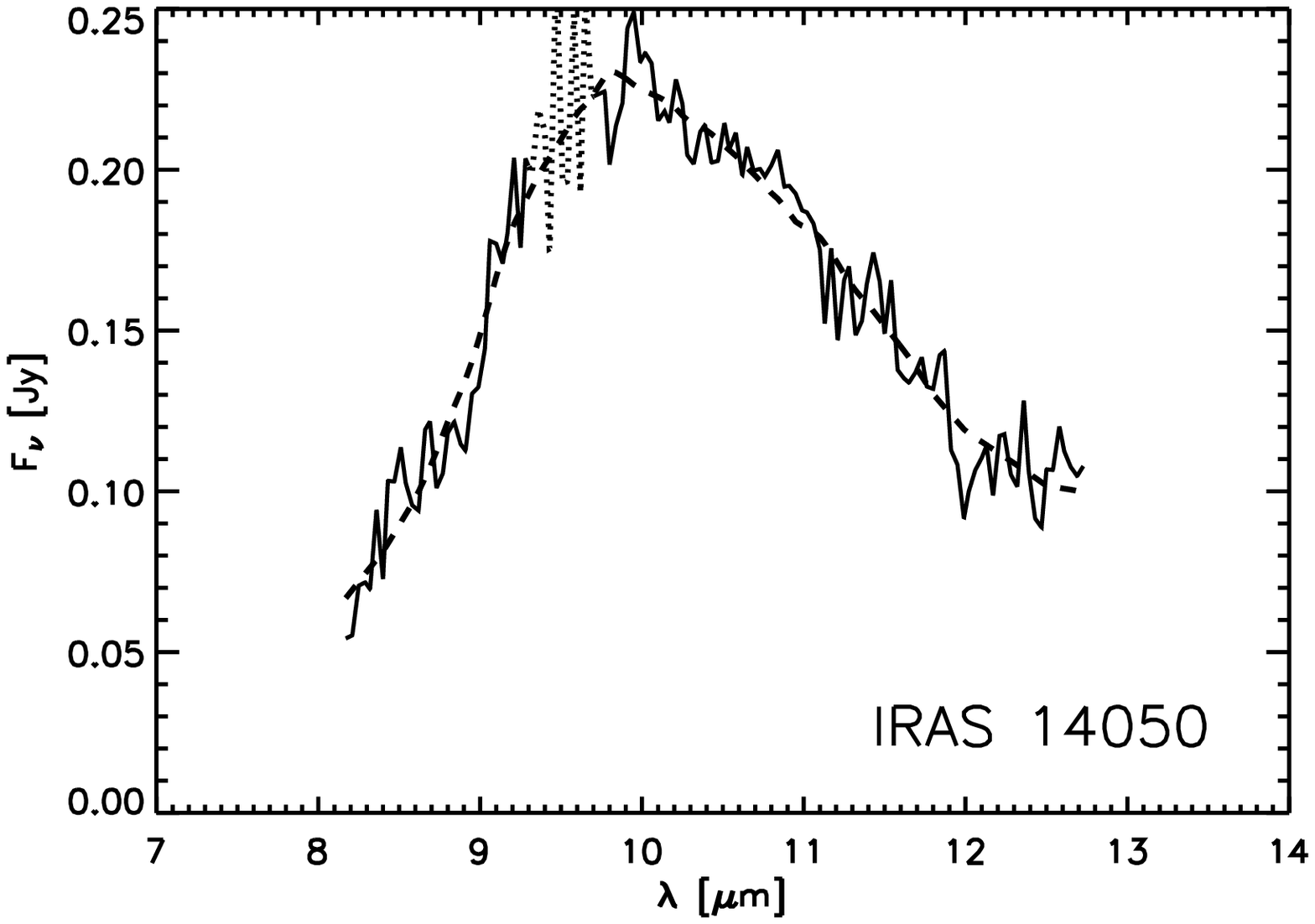}\hfill{}

\hfill{}\includegraphics[%
  clip,
  scale=0.34]{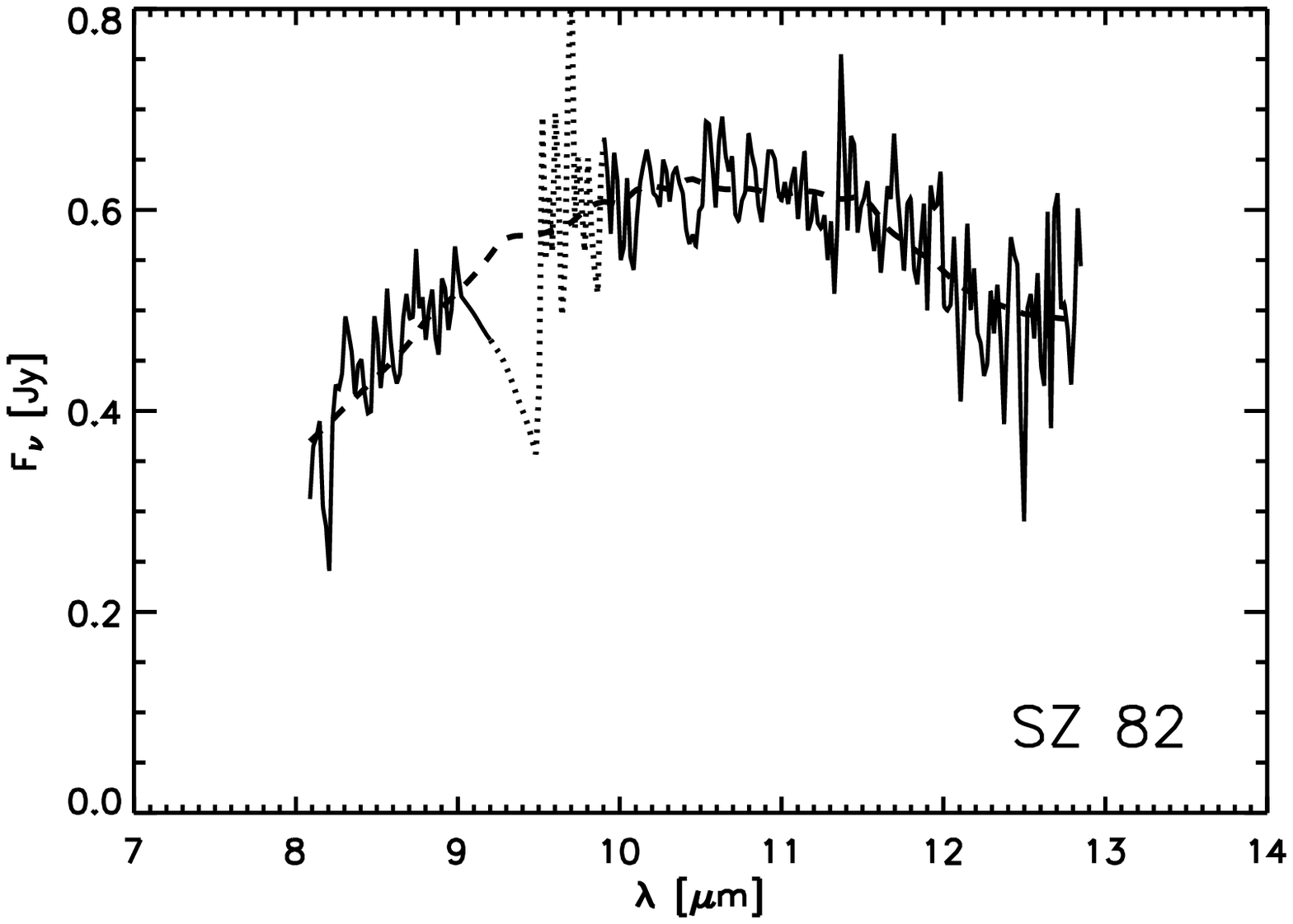}\includegraphics[%
  clip,
  scale=0.34]{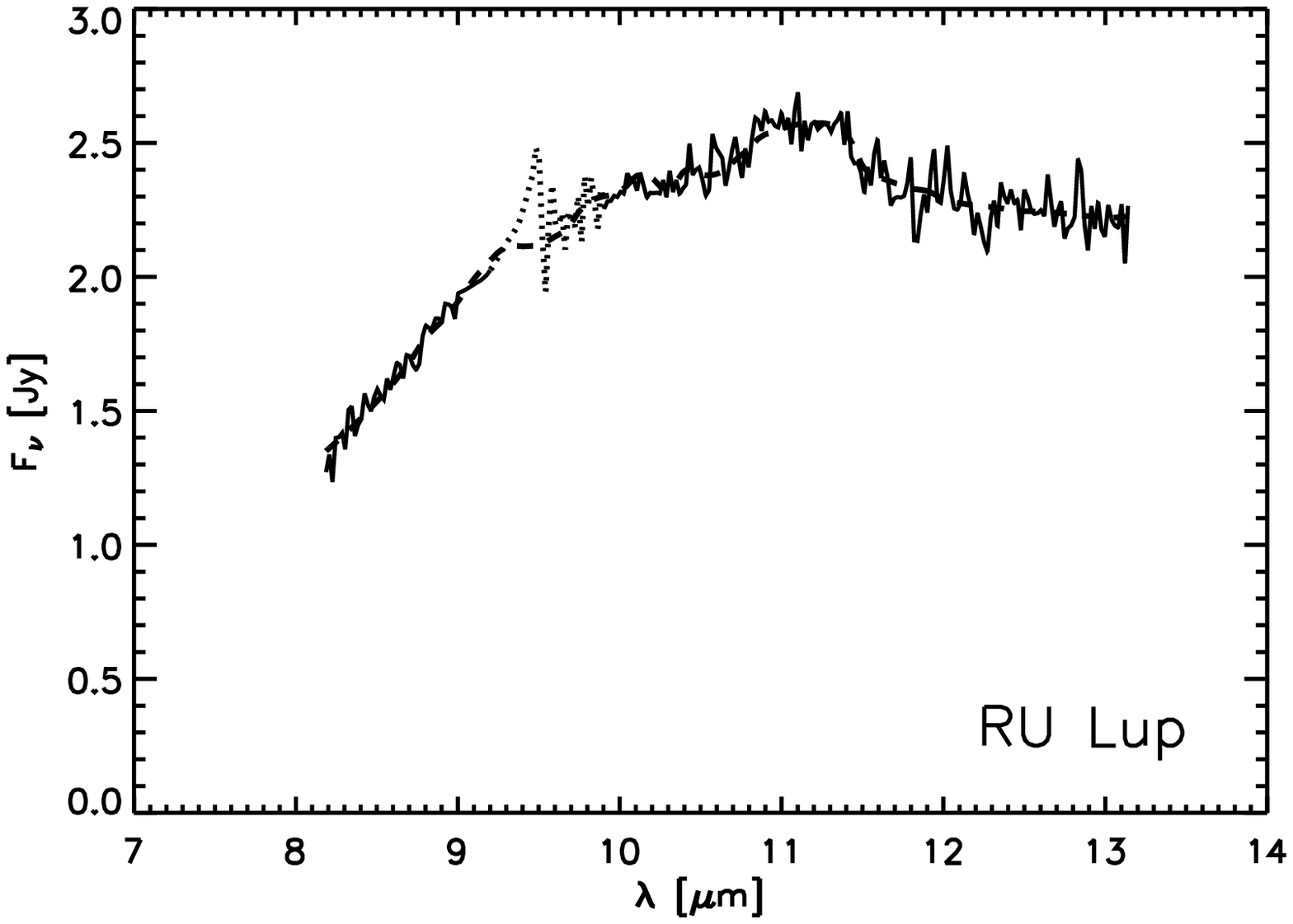}\includegraphics[%
  clip,
  scale=0.34]{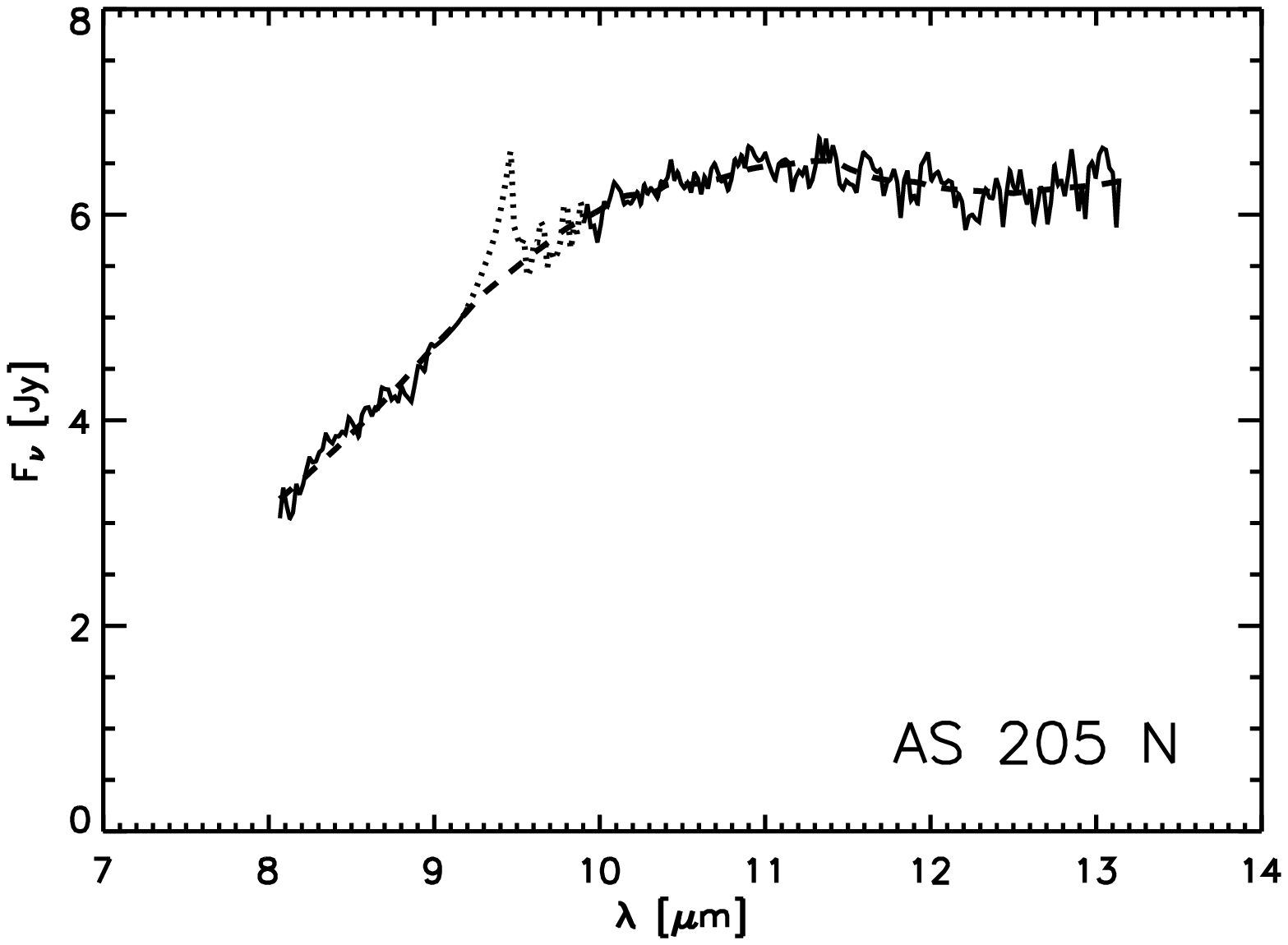}\hfill{}

\hfill{}\includegraphics[%
  clip,
  scale=0.34]{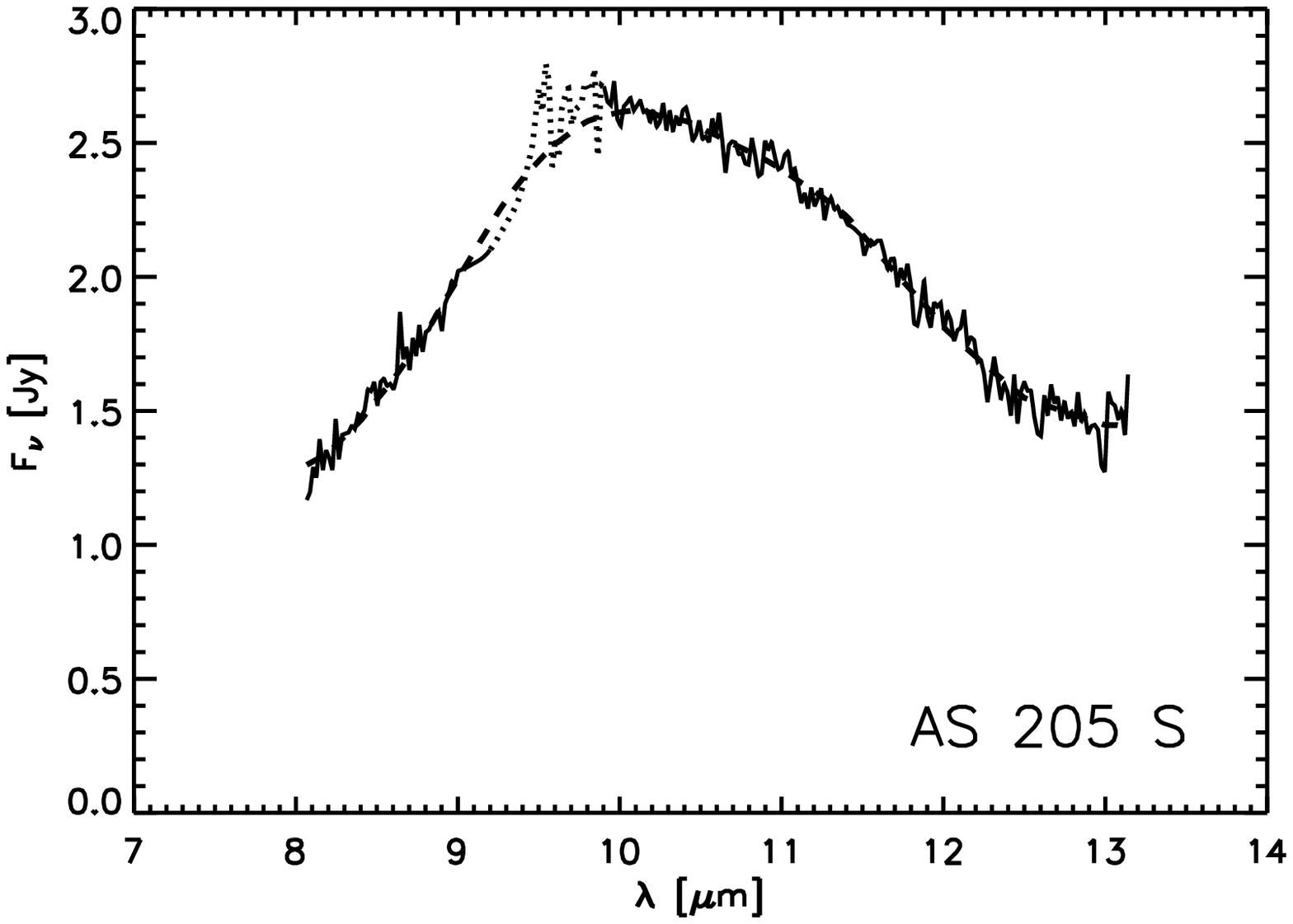}\includegraphics[%
  clip,
  scale=0.34]{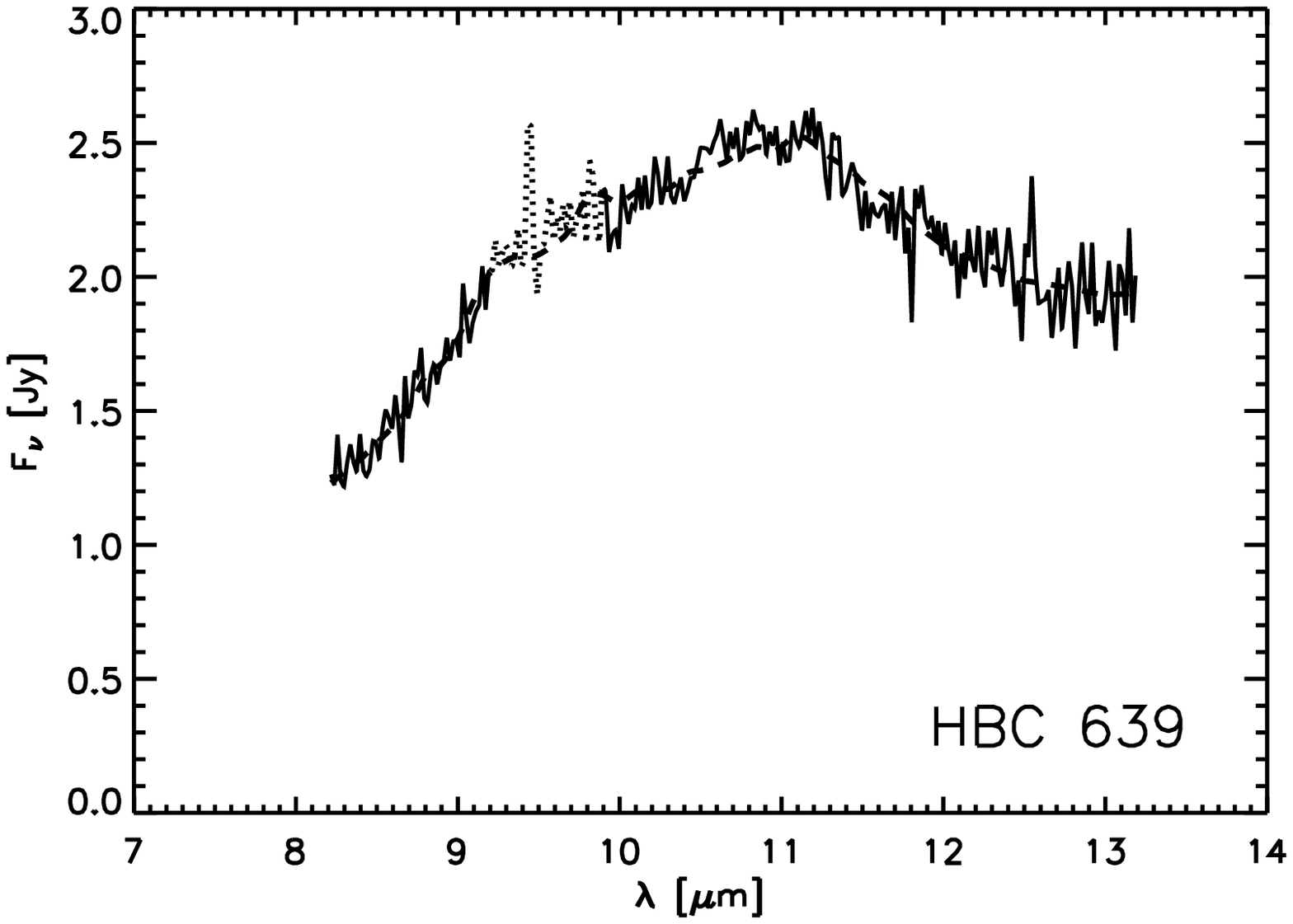}\includegraphics[%
  clip,
  scale=0.34]{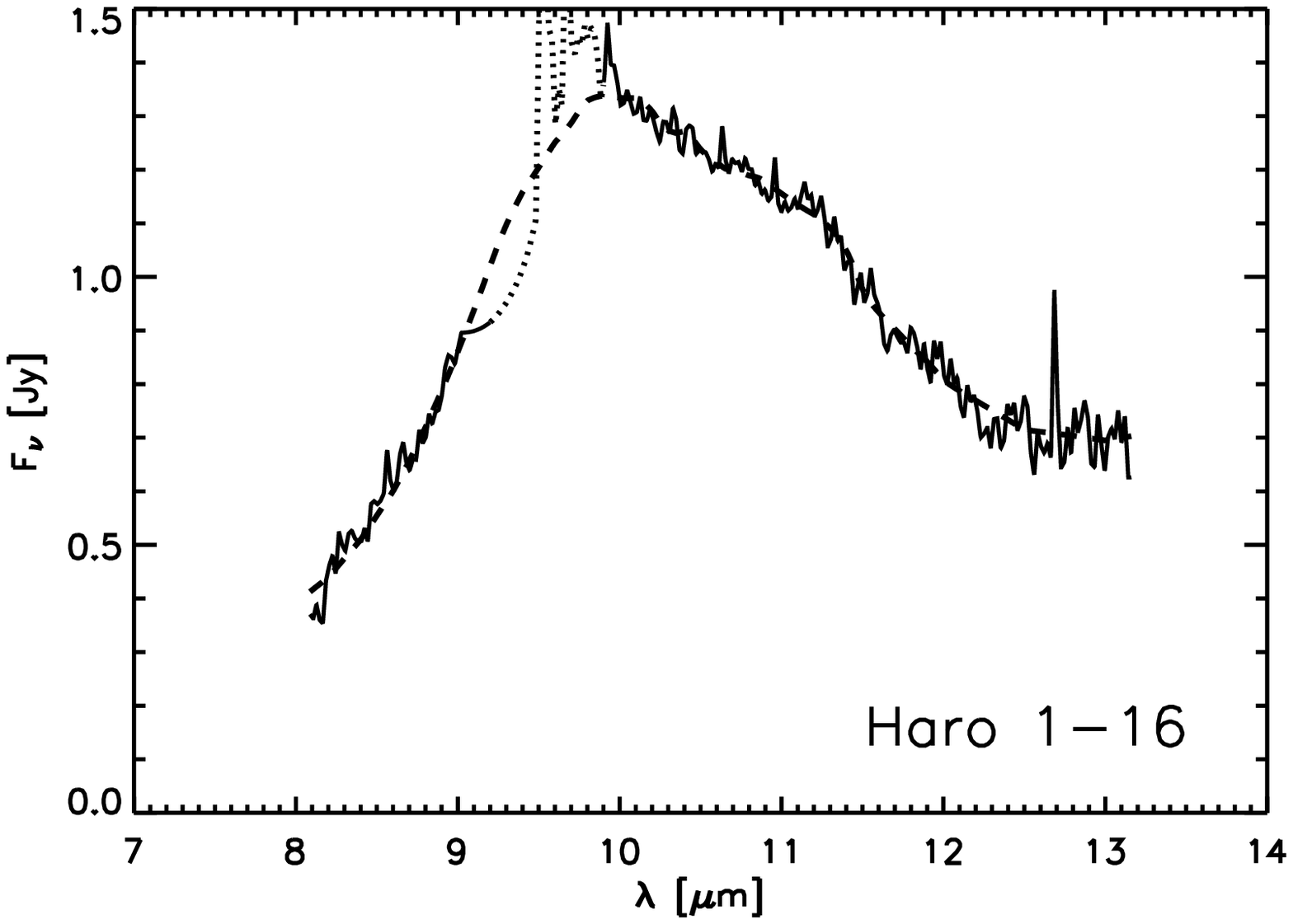}\hfill{}

\hfill{}\includegraphics[%
  scale=0.34]{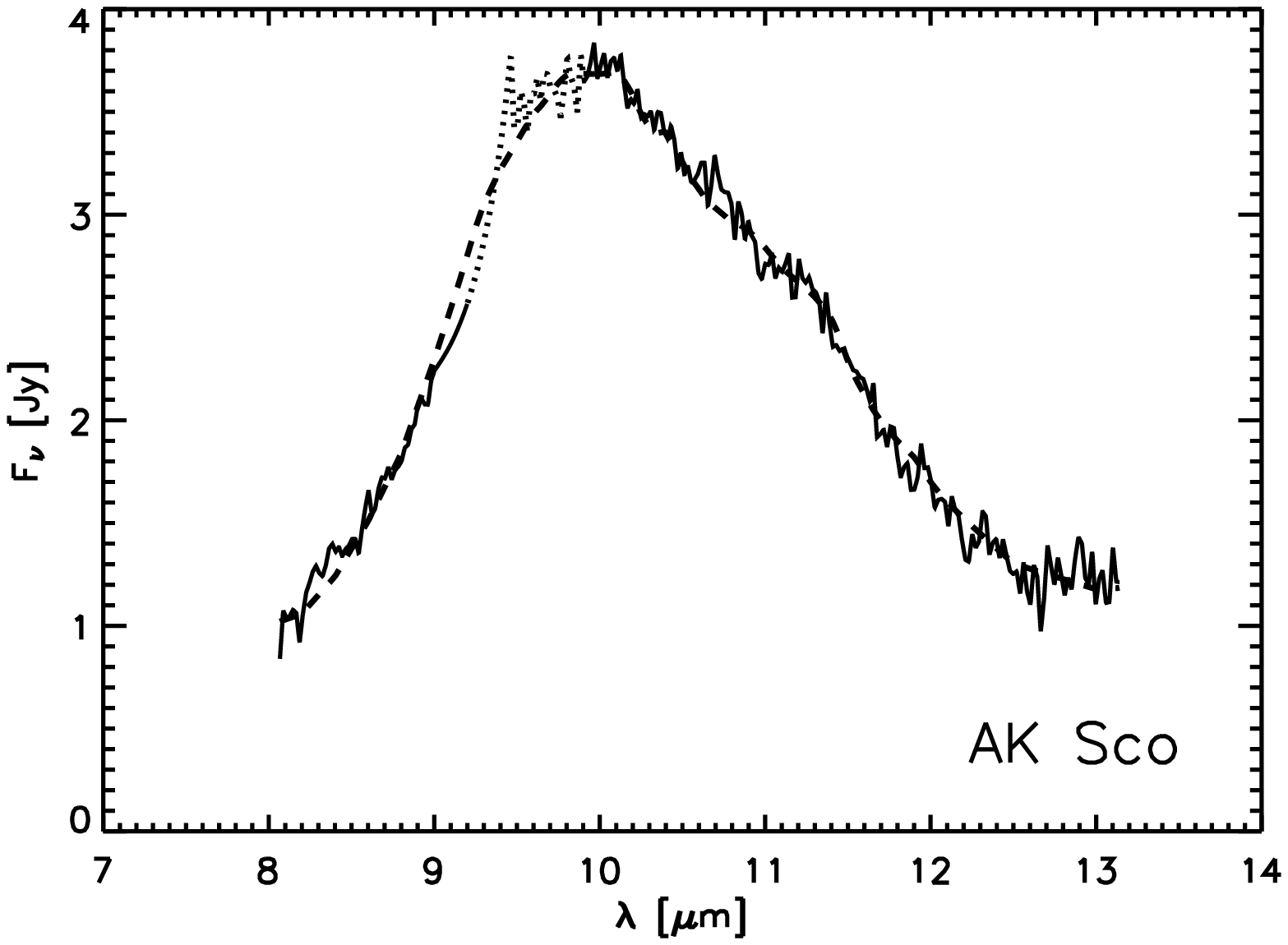}\includegraphics[%
  clip,
  scale=0.34]{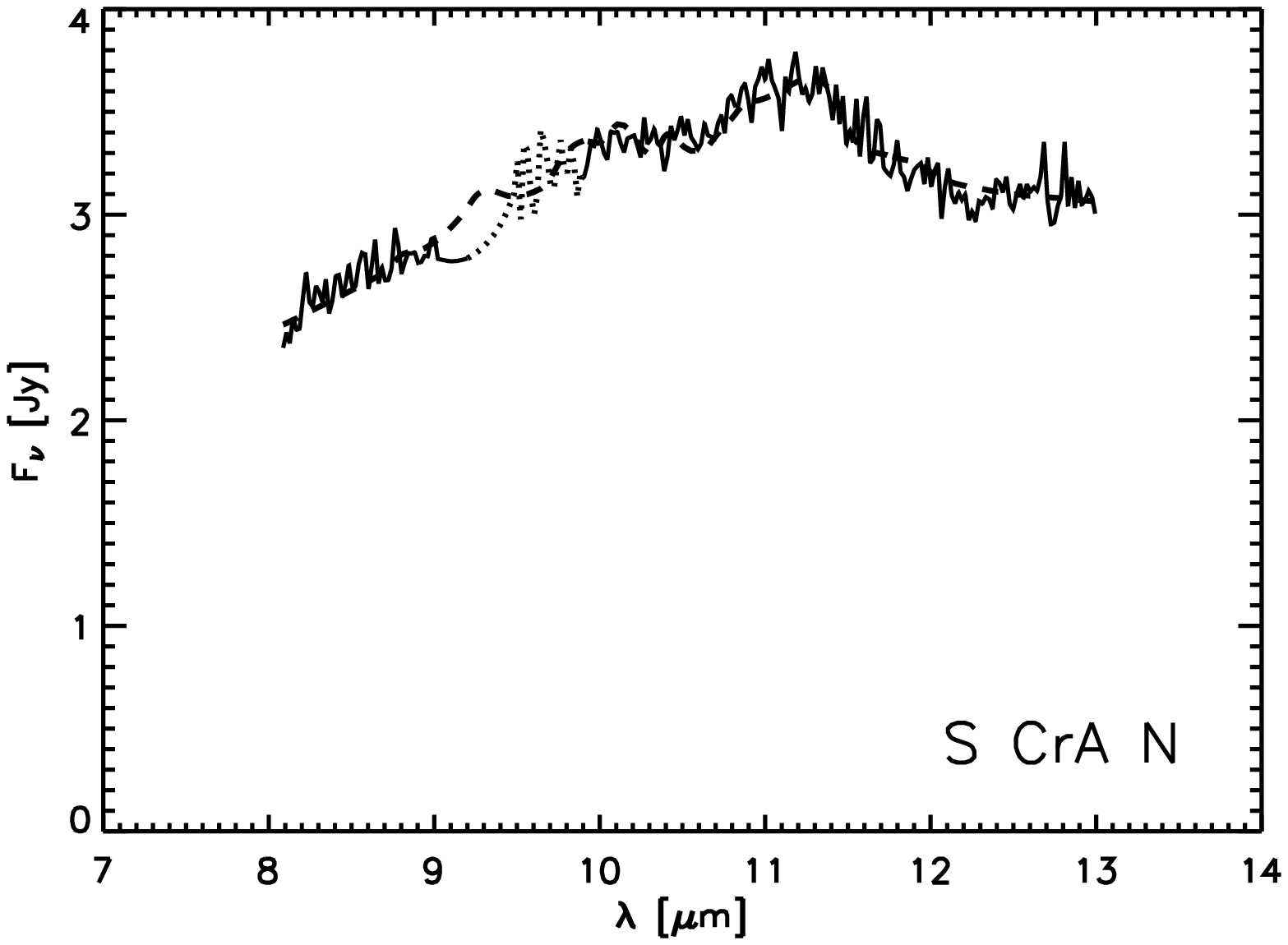}\includegraphics[%
  clip,
  scale=0.34]{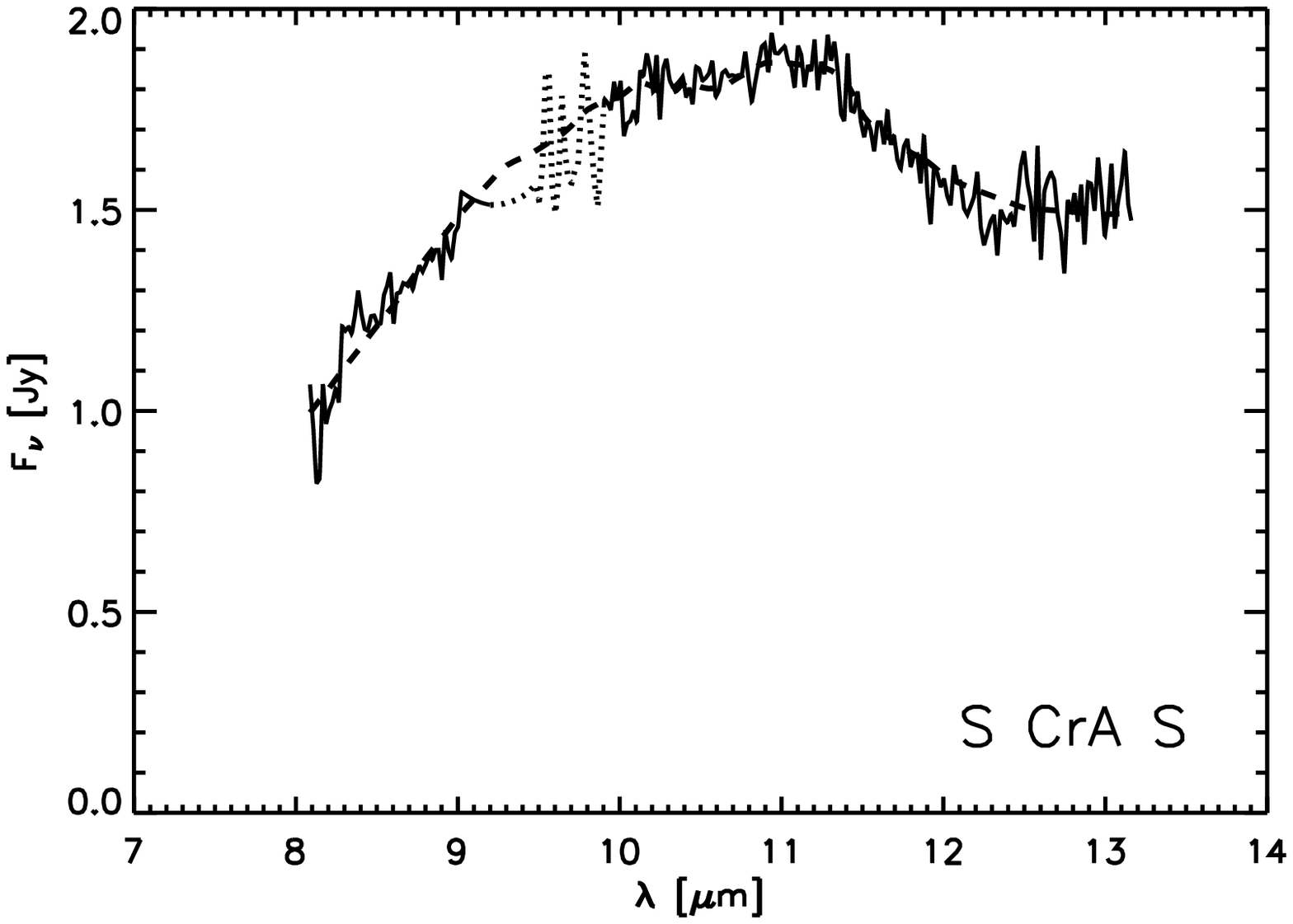}\hfill{}

\caption{\label{cap:Continuation-of-figure}Continuation of figure \ref{fig: fit-result}.}
\end{figure*}

\begin{figure*}[!ht]

\hfill{}\includegraphics[%
  clip,
  scale=0.34]{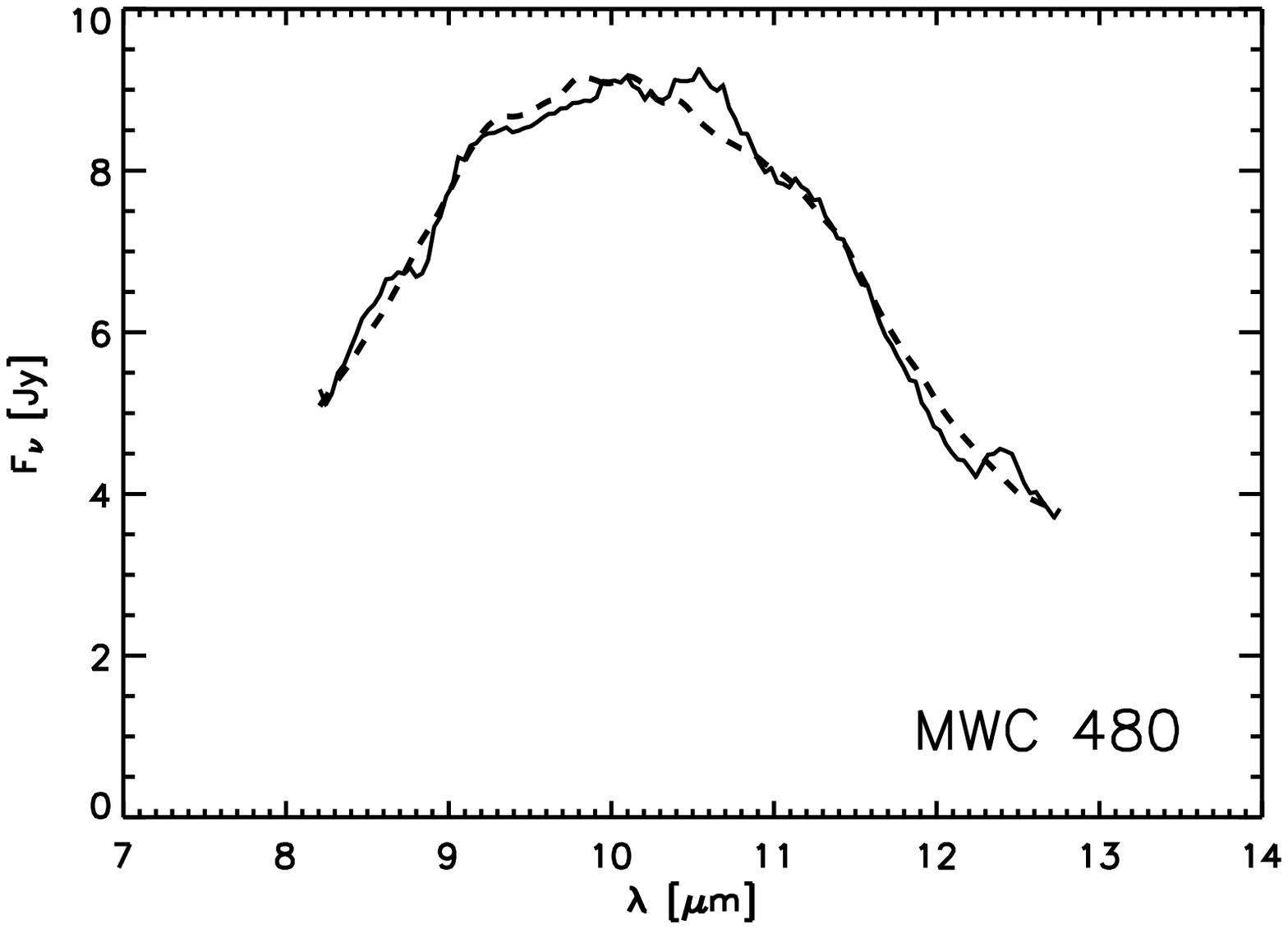}\includegraphics[%
  clip,
  scale=0.34]{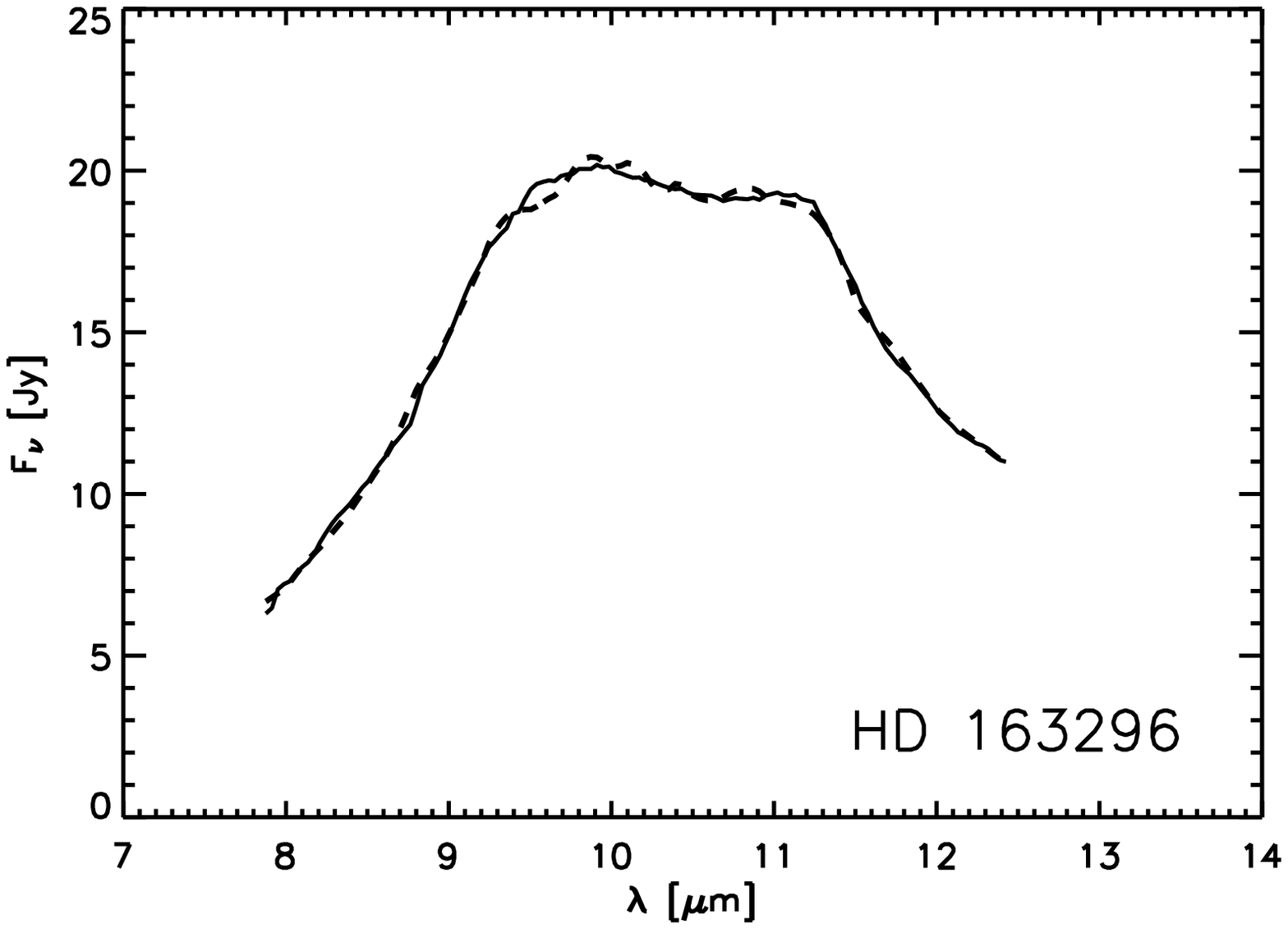}\includegraphics[%
  clip,
  scale=0.34]{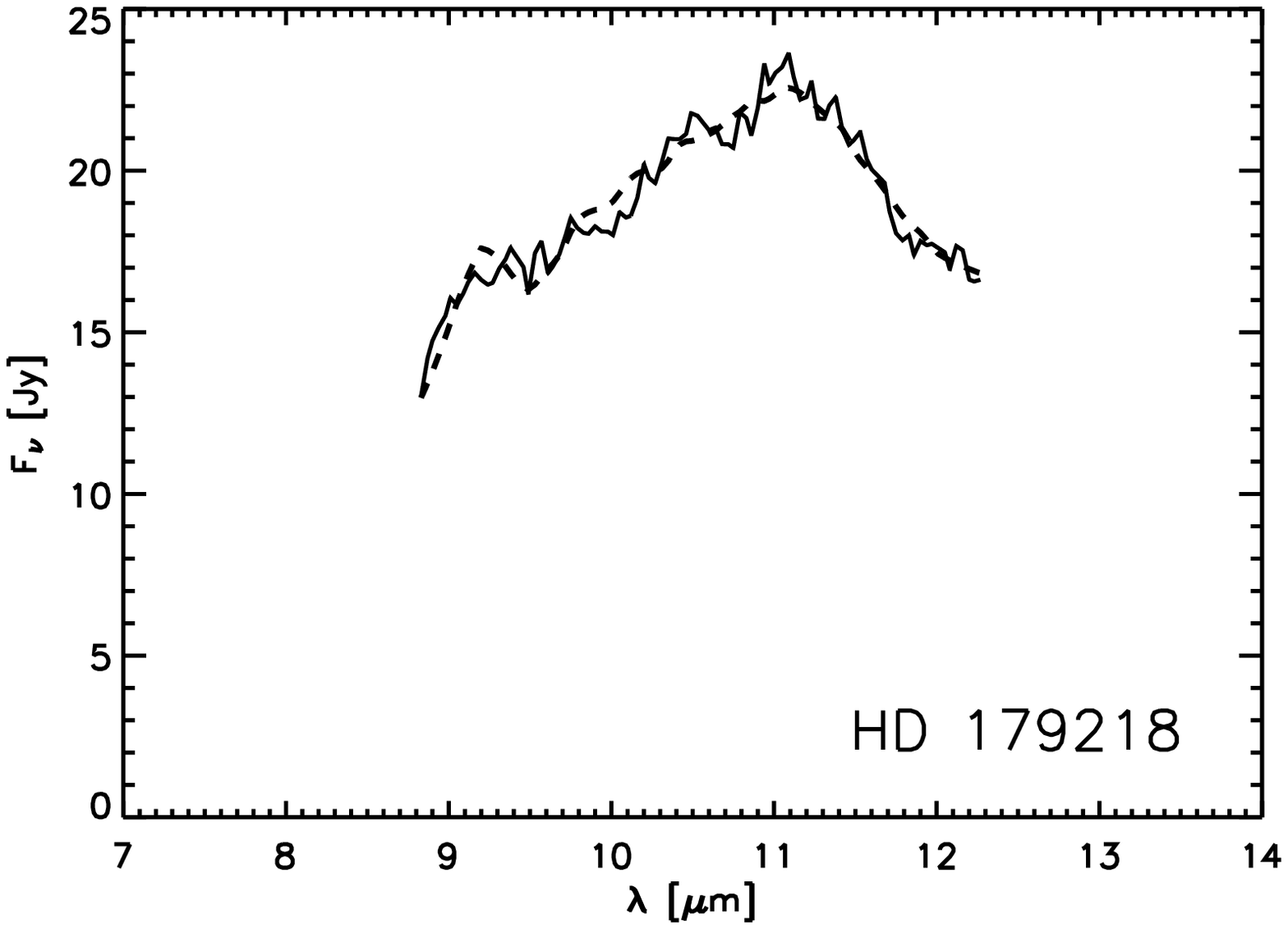}\hfill{}

\hfill{}\includegraphics[%
  clip,
  scale=0.34]{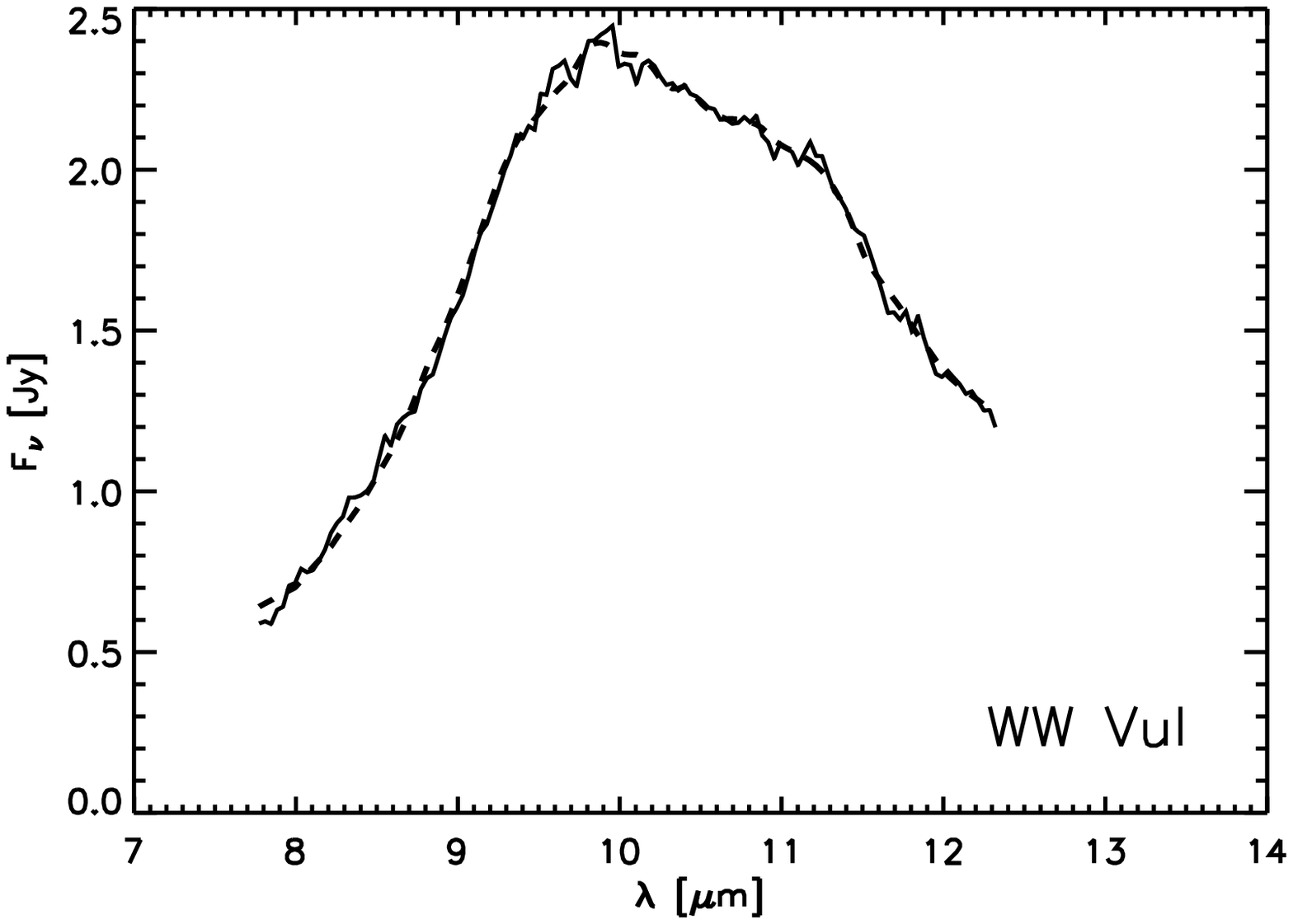}\includegraphics[%
  clip,
  scale=0.34]{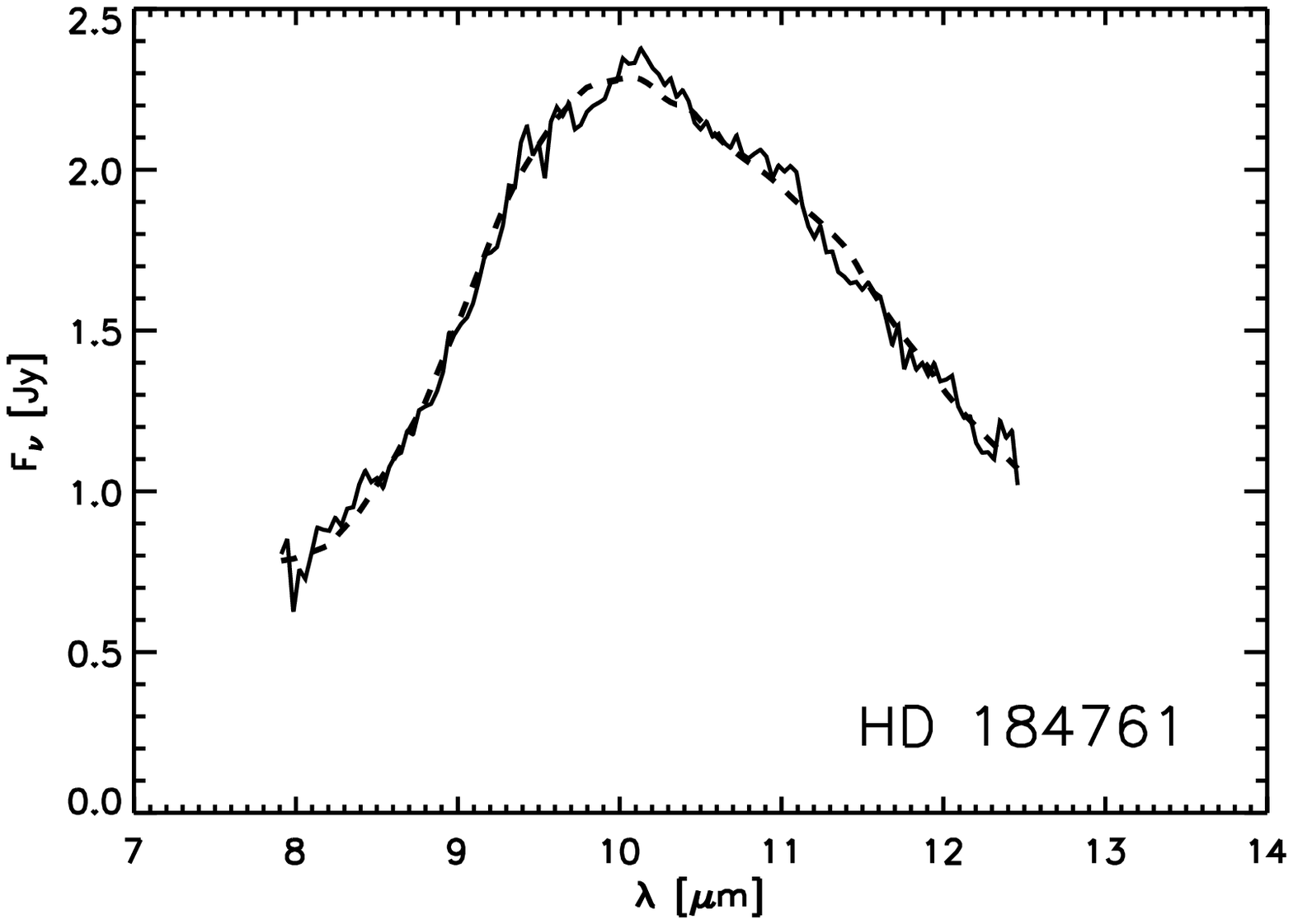}\hfill{}

\caption{\label{cap:Continuation-of-figure}Continuation of figure \ref{fig: fit-result}.
The spectra of HAeBe stars are shown.}
\end{figure*}

Analysing the results in Table \ref{fitresults}, forsterite is the most 
abundant crystalline dust component
in the circumstellar matter of T\,Tauri systems in 19 of 32 objects. According to Gail
(1998) forsterite is the thermodynamically most stable silicate compound
and therefore the final product of the annealing procedure: amorphous
grains $\rightarrow$ polycrystalline grains $\rightarrow$ enstatite
$\rightarrow$ forsterite. We point out that a back-reaction to enstatite
is also possible if quartz is present: 
Mg$_{\rm 2}$SiO$_{\rm 4}$ + SiO$_{\rm 2} \rightarrow$ 2\,MgSiO$_{\rm 3}$ 
(Rietmeijer et al. 1986). Furthermore, the last step in this cascade
can also be inhibited, e.g. by an O- or Mg-deficiency (Fabian et al.
2000) or by too low temperatures (Gail 2003). Thus, the presence of enstatite in circumstellar 
disks of T\,Tauri stars can not be excluded although it appears less frequently than in
HAeBe stars (BO05).
\footnote{We want to mention that recent mineralogical examinations 
have shown that enstatite, which could be developed from forsterite, is a final stage of 
silicate evolution in interplanetary medium and meteorites, in particular.
(see Fig. \ref{fig: enstatite}; private communication with Dominik Hezel, 
Institute for Mineralogy, Cologne).} 

Quartz is less abundant (see Table \ref{fitresults}). This 
result is in agreement with the conclusion about the composition
of the silicate mixture in stationary protoplanetary disks (see Table~8 in 
Gail 2003). Gail (2003) determined a fraction of quartz which is $\rm \le 5\%$.  
Furthermore, the small abundance of quartz -in contrast to HAeBe stars- could 
be a result of the quartz emission profile that we implement in our fitting procedure: 
our profile shows a more discontinous, less smooth curve in contrast to
the emission profile which BO05 used. 

We conclude by comparing our analysis with previous
investigations. 
Honda et al. (2003) succeeded to fit the mid-infrared spectra of
the 10\,Myr old T\,Tauri star Hen\,3-600\,A with larger amounts
of quartz and less enstatite. The absence of quartz in our fitting results 
could be a result of ``observational selection'':
the lower resolution of LWS (R=150) in comparison with R=250
at the 8.2m Subaru Telescope equipped by
the Cooled Mid-Infrared Camera and Spectrometer
(Honda et al. 2003). But the distinction can be also the consequence of
the dif\-ferent fitting functions and the implementation of pyroxene
and large crystalline grains in our fitting procedure.
It is suspicuous that our spectra
tend to rise to longer wavelengths which indicates the negligible
contribution of small amorphous grains in contrast to data of Honda et al.
The modified shape of Hen\,3-600\,A could have its origin
in a variable N band emission.

N band spectra of the HAeBe stars HD\,179218 and
HD\,163296 have been previously analysed by BO05, where
a higher fraction of large enstatite grains was found for the object
HD\,179218. This distinction may be due to different spectral
resolutions of the instruments (LWS versus TIMMI2) and differences
in the fitting procedure. Our fitting results for HD\,163296 
confirm the results of BO05.

\subsection{Correlations\label{sub:Interdependence-of-grain}}\label{sec:correlation}

While a correlation between feature strength and feature
shape has already been observed in HAeBe and T\,Tauri stars indicating grain 
growth (PR03; Acke \& van den Ancker 2004; BO05), we examine the question 
if there is any correlation between grain growth, crystallinity and stellar 
properties like stellar age and luminosity. Here, we include only T\,Tauri systems. 
\begin{enumerate}
\item The \emph{degree of crystallinity} in T\,Tauri systems cor\-relates
with neither stellar age nor luminosity (Fig.~\ref{fig:age-crys}, 
\ref{fig:lum-crys}). It is not clear if the lack of such correlations
is a consequence of the uncertainty which adheres to the measurements
in Table \ref{object-properties}. 
However, we suppose that a correlation 'crystallinity -- 
stellar age' and 'crystallinity -- stellar luminosity', respectively,
exists in the \emph{inner} circumstellar regions, where dust is more
exposed to stellar irradiation. Interferometric observations with
the Mid-Infrared Interferometer (MIDI) show that the degree of crystallinity
increases with decreasing radial distances to the star (van Boekel
et al. 2004). But the dust emission from these inner and hotter regions 
of the circumstellar disk ($\rm >$700\,K) contributes only marginally to our
spectra (Fig. \ref{fig: fit-result} - 
\ref{cap:Continuation-of-figure}) 
as it is illustrated in Fig. \ref{cap:The-simulated-SED}.

With respect of HAeBe stars, BO05 did not inves\-tigate the correlations
between crystallinity versus the stellar age and the stellar lumi\-no\-sity,
respectively. Nevertheless, their ana\-lysis of the correlation between
the crystalline mass fraction and the stellar mass yielded a weak
dependence: objects with larger stellar masses tend to have larger
mass fractions of crystalline grains. Assuming a correlation between
stellar mass and stellar luminosity this result disagrees with our
study at first glance. But the correlation that was found by BO05
appears only for the more massive HAeBe stars ($ \ga 3.0M_{\rm \odot}$).
Those HAeBe stars which are less massive are more comparable
to T\,Tauri stars.
\begin{figure}[!h]
\hfill{}\includegraphics[%
  scale=0.48]{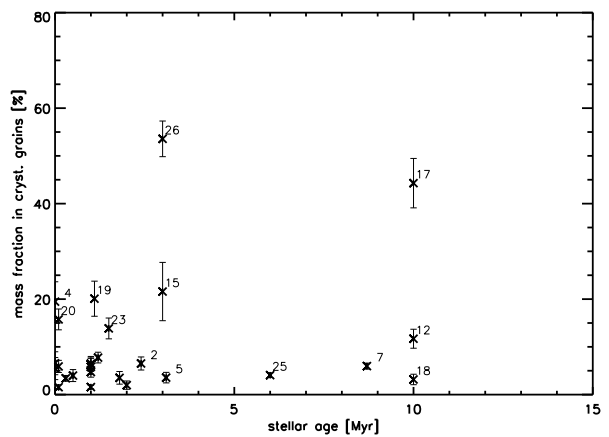}\hfill{}
\caption{\label{fig:age-crys}Mass fraction in crystalline grains vs. 
stellar age according to Table~\ref{fitresults}.
Numbers of objects are indicated.}
\end{figure}
\begin{figure}[!h]
\hfill{}\includegraphics[%
  scale=0.48]{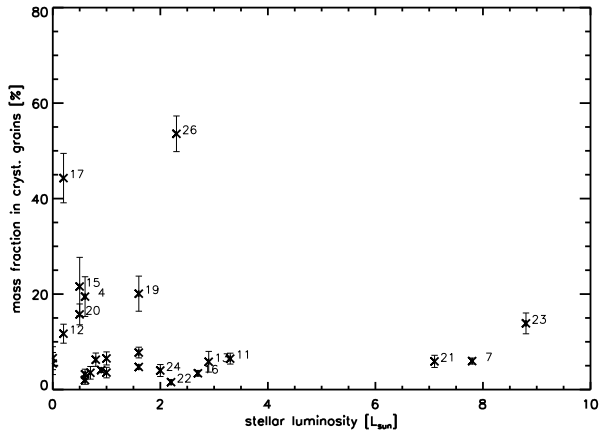}\hfill{}
\caption{\label{fig:lum-crys}Mass fraction in crystalline grains vs. 
stellar luminosity.}
\end{figure}

\item A clear correlation between the mass fraction of \emph{large
grains} (including large crystalline grains) and stellar lumi\-no\-sity
does not exist in our sample of T\,Tauri stars (Fig. \ref{fig:lum-large}). 
If we exclude
objects with the worst signal-to-noise ratio ($\chi^2$, TW\,Hya, 
VZ\,Cha and Sz\,82)
we find that most data points in Fig.~\ref{fig:lum-large} are located
in the area which is limited by the dashed lines.
The resulting, potentially exis\-tent correlation is weak because of
the lack of objects with larger and smaller stellar luminosities 
($> 4 L_{\rm \odot}$ and $< 0.5 L_{\rm \odot}$).
\begin{figure}[!h]
\hfill{}\includegraphics[%
  scale=0.48]{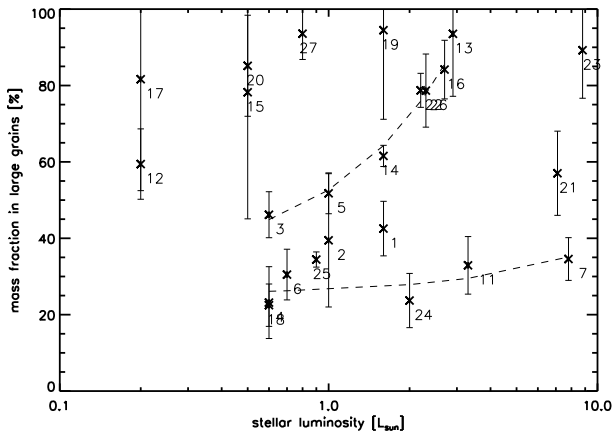}\hfill{}
\caption{\label{fig:lum-large}Mass fraction in large grains versus stellar
luminosity. The stellar luminosity is plotted in logarithmic scale.
Excluding objects with the worst signal-to-noise ratio most objects
are located in the area which is limited by the dashed lines. The
number of each data point corresponds to the object.}
\end{figure}
BO05 analysed the correlation between stellar mass and mass fraction
in large grains. Again, a correlation between both parameters exists
only for more massive objects: HAeBe stars with a mass $\rm M \ga 2.5 M_{\odot}$
have a mass fraction of large grains above 85\%. Less massive HAeBe 
stars do not show any dependence on the mass of the central star. 

We have to point out, that the stellar age does not affect the existence
of large grains, which a further study shows (see Fig. \ref{fig:age-large}). 
\begin{figure}[!h]
\hfill{}\includegraphics[%
  scale=0.48]{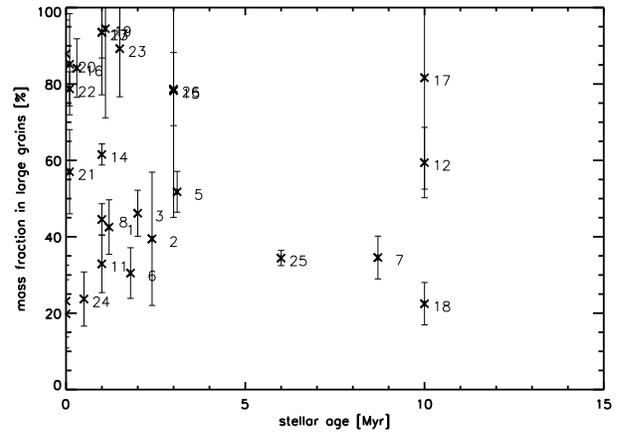}\hfill{}
\caption{\label{fig:age-large}Mass fraction in large grains vs. 
stellar age.}
\end{figure}

\item The Fig. \ref{fig:large-crys} shows the relative mass fraction in \emph{large
crystalline} grains versus the relative mass fraction in \emph{large}
grains. Apart from the objects Hen\,3-600\,A, S\,CrA\,N 
and DQ\,Tau
most objects show a low mass fraction of crystalline grains ($\rm <$20\%).
Such a correlation could also be derived
in HAeBe stars (BO05) where crystalline dust was \emph{exclusively}
found in objects with bigger amounts ($\rm >$80\%) of large grains. %
\begin{figure}[!h]
\hfill{}\includegraphics[%
  scale=0.5]{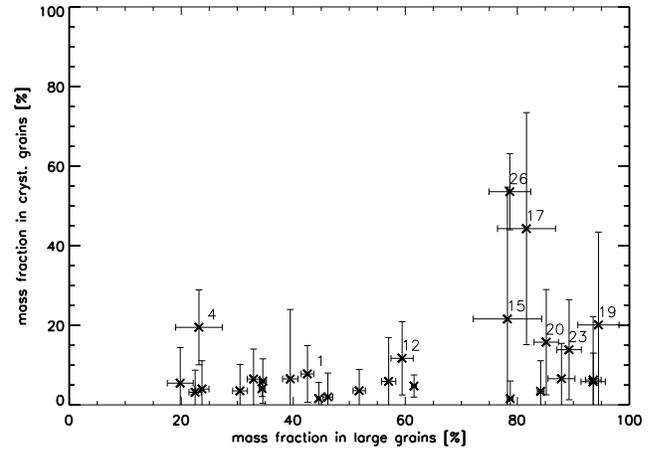}\hfill{}

\caption{\label{fig:large-crys} Relative mass fraction in crystalline grains
plotted versus the relative mass fraction in large grains. The number
of each data point corresponds to the object in Table \ref{fitresults}. }
\end{figure}
 Around half of our T\,Tauri objects have only a negligible degree
of crystallisation in spite of a high mass fraction in large grains. 

A correlation between grain growth of amorphous grains and grain growth
of crystalline grains can not be found.%
\footnote{\label{Both-mass-fractions}Both mass fractions are normalised by
the mass fraction of crystalline and the mass fraction of amorphous
grains, respectively, considering small and large grains, simultaneously. %
} The existence of such a correlation strongly depends on the rate
of crystallisation and grain growth, respectively, and the ratio between
both quantities. Note also that in some conditions the amorphisation
of crystalline silicate occurs after ion irradiation
(see, e.g., Djouadi et al. 2005).
\item An agreement between T\,Tauri and HAeBe systems is
apparent if the \emph{enstatite} fraction of crystalline grains is plotted
versus the mass fraction of \emph{crystalline} grains. BO05 found increasing
enstatite fractions with increasing degrees of crystallinity. A similar
tendency can be found in our sample (see Fig. \ref{fig: enstatite}).
The explanation for this weak correlation is a delayed production of
forsterite or the reproduction of enstatite from forsterite (see Sect.
\ref{sec:Results}, BO05 and Gail 2003).%
\begin{figure}[!h]
\hfill{}\includegraphics[%
  scale=0.5]{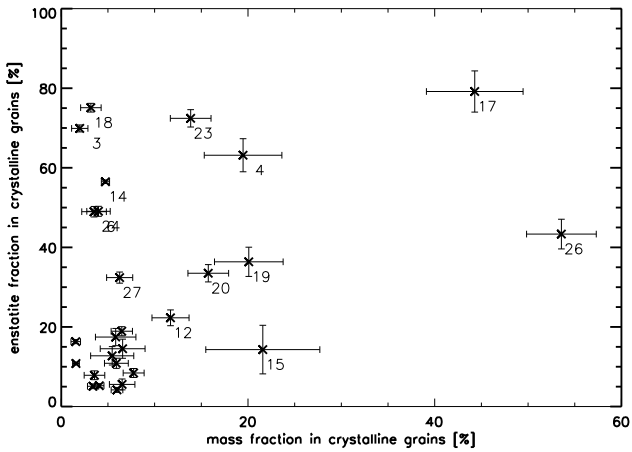}\hfill{}

\caption{\label{fig: enstatite} Enstatite fraction, normalised by the mass
fraction of crystalline grains plotted vs. the mass fraction of crystalline
grains. }
\end{figure}

\end{enumerate}
Summarising our study, the analysis of the silicate feature of T\,Tauri
systems is complicated by their generally lower brightness in contrast to HAeBe stars (BO05). 
However, we find 
correlations which partly agree with the analysis of BO05, for instance 
if their correlations are extrapolated to less massive systems.

\section{Composite grains\label{sec:Fit-results-with}}

The opacities $\kappa_{{\rm abs;} i}(\lambda)$ are conventionally calculated
by Mie theo\-ry assuming homogeneous dust particles. In reality,  collisions
with gas, other grains and the interaction with
radiation can lead to grain coagulation, partial grain evaporation,
crystallisation, annealing, amorphisation, etc.
As a result, compo\-site, partially flaky and inhomogeneous dust grains with 
inclusions of vacuum evolve (Dominik et al.\,1993; Poppe et al. 2000a,b). 
The calculations of the optical properties of such grains is a rather difficult
problem which can be solved using special algorithms like
the discrete dipole approximation (DDA; e.g., Andersen et al.\,2002,\,2003).

The most often used approximation includes the con\-sideration of homogeneous
particles with an average (effective) refractive index derived 
from effective medium theory (EMT; see Bruggeman 1935; Ch\'ylek et al.\,2000).
Comparing the DDA and EMT-Mie calculations,
Voshchinnikov et al.~(2005a) have recently discovered that
the EMT-Mie approach can give relatively
accurate results only for porous particles with ``Rayleigh''
inclusions. \footnote{inclusions which are small in comparison with the wavelength of incident radiation}
Otherwise, the approach becomes unacceptable
when the porosity exceeds $\sim$50\% of the grain volume. 
At the same time, the optical properties of heterogeneous
spherical particles having inclusions of various sizes
(Rayleigh and non-Rayleigh) and very large porosity
were found to resemble well those of
spheres with a large number ($\ga 15-20$) of concentric layers 
with different compositions.

It was also found that
the layered-sphere model predicts a broadening of silicate bands and
a shift of the peak position to larger wavelengths as porosity
grows while in the case of the EMT-Mie model variations of the feature profile
are less significant (see Voshchinnikov et al. 2005b for details).

Since the cosmic particles can have 
inclusions of different sizes, here, we apply the  model of
multi-layered spheres for explanation of the $10\,\mu$m silicate feature.
For simpli\-city, we consider particles consisting of two materials
(carbon and silicate) and 18 concentric
layers\footnote{Voshchinnikov et al. (2005a)  have shown that this was enough
to preclude influence of the order of materials on the results.}. 
Va\-cuum can be one of the components, too.
In contrast to pre\-vious modelling with compact grains,
the model parameters are grain size, composition and porosity. 

In the context of the described layered sphere model, pre\-liminary calculations showed that variations of
the volume fractions of va\-cuum, silicate and carbon produce diverse profiles
of the 10~$\mu$m feature from strongly peaked to very flat (Voshchinnikov et al. 2005b).  
It is important to note that different profiles are obtained for
small grains with the same compact grain radius $r_{\rm compact}=0.1\,\mu$m.\footnote{ 
The quantity $r_{\rm compact}$ is the radius of the corresponding dust grain without voids.}
This means that the variations of the shape of the silicate band can be
related to the change of grain porosity and composition but not necessarily to
the grain growth.

\begin{figure}[!h]
\hfill{}\includegraphics[%
  scale=0.5]{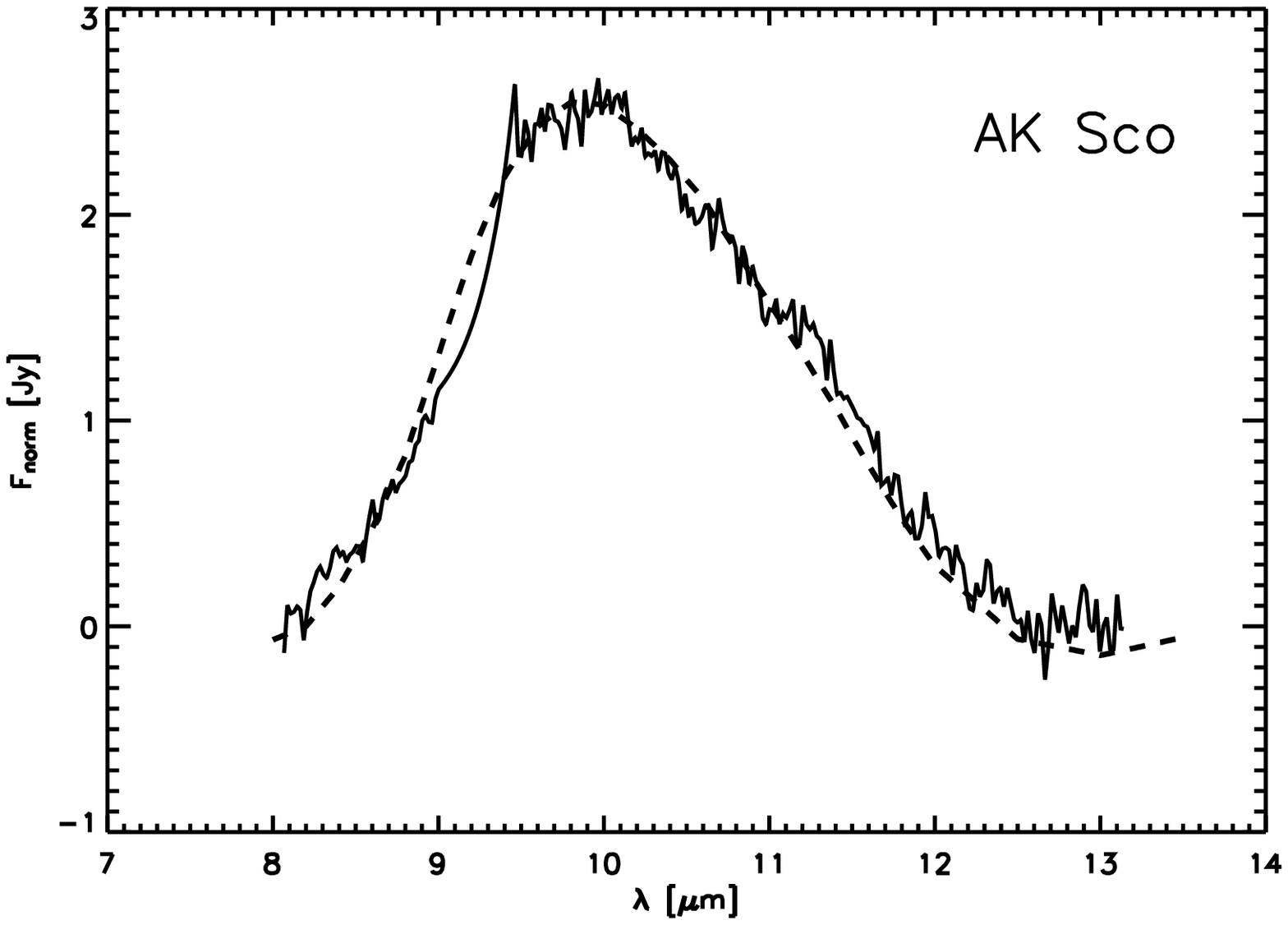}\hfill{}
\hfill{}\includegraphics[%
  scale=0.5]{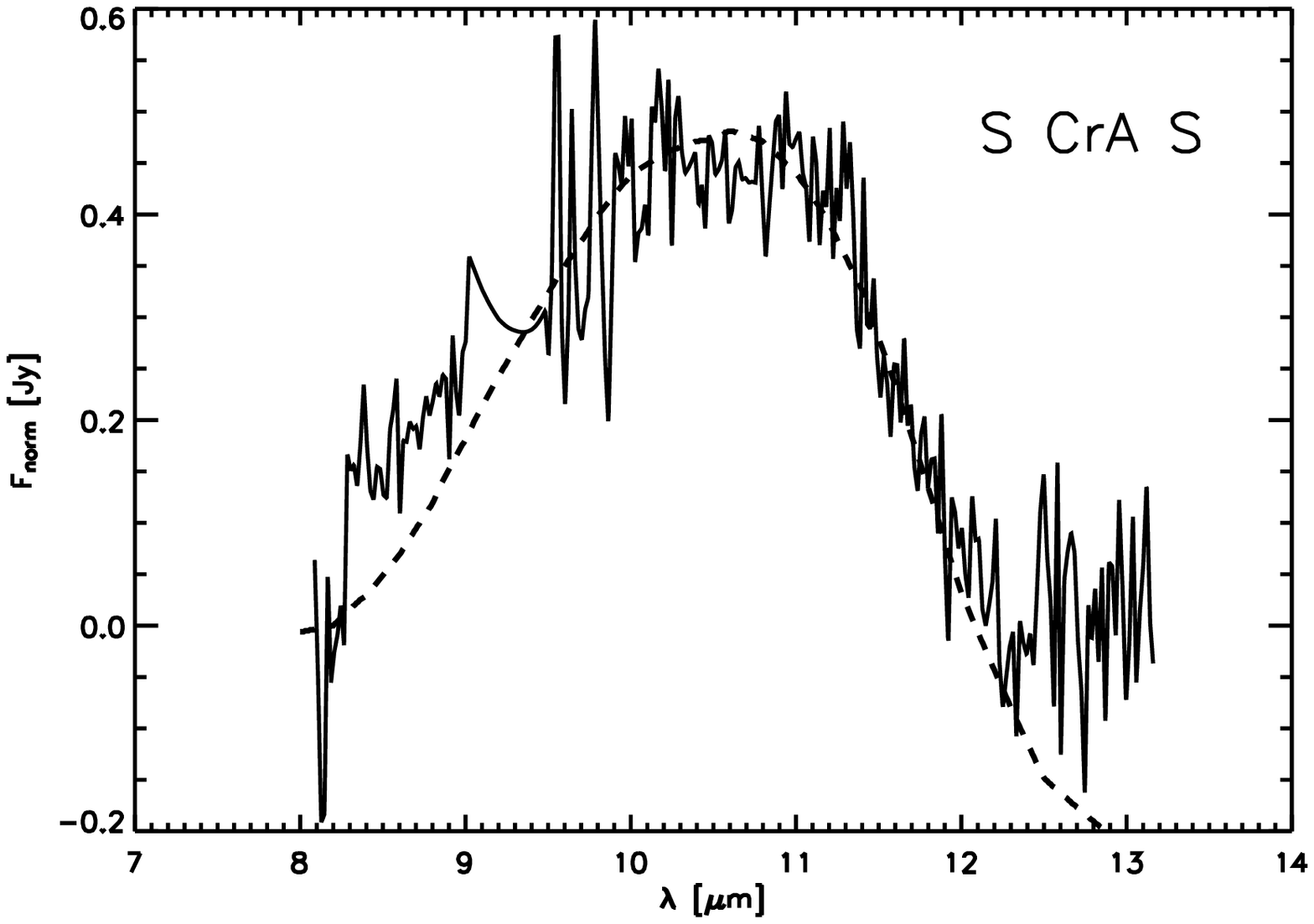}\hfill{}
\caption{\label{fig:nikolai}Normalised, continuum subtracted fluxes of AK\,Sco and S\,CrA\,S. The
dashed curves represent the results using the multi-layered model for porous grains. The relative
\emph{volume} fraction and the grain radius for AK\,Sco and S\,CrA\,S
are presented in the Table \ref{tab:fitresults-nikolai}.}
\end{figure}

As illustration Fig.~\ref{fig:nikolai} shows the result of our modelling
for the objects AK\,Sco and S\,CrA\,S having very 
different shapes of silicate band in their spectra. We used glassy 
olivine, va\-cuum and amorphous carbon Be1 (Rouleau \& Martin\,1991) and 
grains with a compact radius $r_{\rm compact}=0.1\,\mu$m in both cases. 
Crystalline components are not considered. Figure \ref{fig:nikolai}%
and Table \ref{tab:fitresults-nikolai} %
\begin{table}[!htb]
\caption{\label{tab:fitresults-nikolai} Parameters
of composite grains used for modelling silicate band in spectra
AK\,Sco and S\,CrA\,S (Fig. \ref{fig:nikolai}).}
\hfill{}\begin{tabular}{cccccc}
\hline 
\hline 
object& $r_{\rm porous}$& olivine& Be1& vacuum& $\chi^{2}$\tabularnewline
\hline 
AK Sco& 0.12\,$\mu$m& 49.5\%&5.5\%&45\%&0.57\\
S CrA S& 0.27\,$\mu$m&3.35\%&1.65\%&95\%&7.81\\
\hline
\end{tabular}\hfill{}

{\scriptsize 
Note. The quantity $r_{\rm porous}$ is the total radius of the porous grains (with vacuum).
In col. 3-5 volume fractions of materials are given. The parameter $\chi^{2}$ is 
explained in Eq.~\ref{chi}. }
\end{table}
show our best results after varying the relative \emph{volume} fractions
of the three dust components.
If the longer wavelength regime ($\rm \lambda \ge 12.3\, \mu m$)
of S\,CrA\,S is excluded
the fits are satisfying according to Eq.~(\ref{chi}).
Considering this result and previous fitting with compact grains,
the question about the unam\-biguous interpretation certainly arises.
An answer on it can be found in the use of additional observations
capable to discriminate between these two models (see Sect.~\ref{sec:Conclusion-and-future}).

There are previous attempts to compare the silicate feature with the
mid-infrared emission behaviour of composite, complex dust grains.
Bradley et al. (1999) found that the sub-micron components of interplanetary
dust particles, glass with embedded metals and sulfides (GEMS), show
features that are very similar to those in YSOs. The structure of
GEMS and the properties of their iron-nickel inclusions
refer to their origin: Westphal et al. (2004) suggested that
GEMS were probably formed from crystalline grains by the exposure
of hard, ioni\-zing radiation in the stellar outflows from massive stars.
Finally, the particles are accelerated in the shock waves of subsequent
supernova explosions and distributed over space. The study of the importance of
GEMS in circumstellar disks of YSOs and their actual influence on
the 10\,$\rm \mu$m-silicate-feature is out of the scope of this study.

\section{Conclusion \label{sec:Conclusion-and-future}}

In this study we investigated the mid-infrared silicate feature of 27
T\,Tauri and 5 HAeBe stars. The relative mass fractions of
amorphous and crystalline, large and small dust grains were derived by fitting 
a linear combination of the corresponding dust opacities. We assume that the
emission feature has its origin in the optically thin surface layer
of the circumstellar disk while optical thick disk layers only contribute
to the underlying continuum of the SED. Because of magnetorotational 
turbulences and turbulences which arise from Kelvin-Helmholtz instability 
dust grains with radii $r$ $\rm <$ 1cm are assumed to be 
vertically well mixed in the disk (e.g., Johansen et al. 
2005). The relative mass fractions obtained are a copy of the mean composition 
of micron-sized silicate grains in the whole disk, therefore.

A central focus of this study was the question which phy\-sical quantities can
be derived from the observation profile of the silicate feature, in general. 
In terms of an impending vio\-lation of the uniqueness of the fitting results 
the {\it narrow} spectral interval has a very restrictive
effect on the number of fitting parameters. Therefore, the number
of the selected fitting parameters should be as low as possible. Apart
from the temperature $T$ and a constant term $C_{\rm 0}$ we chose
the opacities $\kappa_{\rm abs}$ of following dust species as fitting parameters:
0.1\,$\rm \mu$m- and 1.5\,$\rm \mu$m-sized olivine, pyroxene,
forsterite, enstatite and quartz. PAH emissivities were not implemented. 
The satisfying fits (see Table \ref{fitresults}) 
underline the negligible abundance of excited PAHs in T\,Tauri systems in contrast 
to HAeBe systems. However, forsterite could overlay the PAH emission as it 
has a feature maxima at $11.3 \mu m$, as well.   

As a subsequent result we found hints that both, grain growth and
crystallisation appear simultaneously in (our sample of) T\,Tauri
systems while grain growth is more dominant. This result has to been
seen in the context of previous investigations: Mathis et al. (1977)
found a particle size distribution of several silicates in interstellar
dust that can be reproduced by a power law of a strongly decreasing
grain number with increa\-sing grains size. In fact, Bouwman et al. (2001)
found negligible mass fractions of large silicate ($\rm \ge 1.5\, \mu m$)
grains in interstellar dust as parent material of dust in circumstellar
(HAeBe) disks. 
However, the conclusion of grain growth is valid only for the case of 
compact dust grains which consist of one silicate component. Assuming inhomogeneous 
and flaky dust particles, which can be described by the multi-layered sphere model, 
for instance, the increase of porosity of composite grains up to 50\% and higher
can be considered as an alternative to grain growth as it generates similar, 
flat spectra (Voshchinnikov et al. 2005b). 

There are indications that 
all particles larger than $\rm 0.1\,\mu m$ are porous. But it is still 
a central question how large the degree of porosity is or if there is even a porosity distribution for grains of certain sizes. 
Furthermore, the EMT-model and the layered sphere model for porous grains are still
competing models and partly provide different results. Although the layered sphere 
model has its preferences (e.g., Rayleigh and non-Rayleigh inclusion can be handled) it 
focuses only on sphe\-rical grains. Grains with crystalline inclusion have not been 
considered, so far. 

In respect of our fitting results, it is not clear, at all, how strongly porosity 
affects our fitting results. But according to the layered sphere model the contribution 
of large grains would be overestimated as a flat $\rm 10\,\mu m$ feature can arise from
large grains of any porosity or from very porous, small grains while a narrow feature 
results from small grains, exclusively.

Considering the fact that T\,Tauri systems are less massive than
HAeBe stars our analysis partly agrees with previous results.
In fact, our analysis showed that -- in terms of the pro\-perties of the circumstellar dust -- 
T\,Tauri systems are a continuation 
of HAeBe systems at their lower mass end. Further observations -- in particular,
the si\-licate feature at 20\,$\mu$m (Kessler-Silacci et al. 2006) --  
and polarization in silicate bands accompanying with more comprehensive 
modelling could provide stronger constraints for existent silicate species in the
circumstellar disks of young stars.

Certainly, the determination of the individual silicate composition
of the disk material is an important result, by its own, but it has also 
a further consequence for radiative transfer si\-mulations. While
recent radiative transfer simu\-lations of circumstellar disks used
to be based on the extinction coefficients of an averaged mixture
of silicates (e.g., ``astronomical silicate'', Draine \& Lee 1984),
the consideration of the results of compositional fits such as those
presented here are meant to significantly improve disk modelling. 
Radiative transfer models finally provide further constraints with
respect of the fundamental characteristics of the objects like geometry
and structure, sky position (inclination), (stellar) luminosity and
object mass (star and disk). Together with the chemical composition
of the grains, models of YSOs become quite individual. Furthermore,
such complete disk models open up new perspectives, e.g. the ``compositional''
analysis of \emph{absorption} features which are strongly affected
by disk structure and inclination. 

\begin{acknowledgements}
We thank J. Bouwman for discussions and for the opacity data. Furthermore, 
we thank the referee for his helpful suggestions.
A. Schegerer and S. Wolf were supported by the German Research Foundation
(DFG) through the Emmy-Noether grant WO 857/2-1.
\end{acknowledgements}

\end{document}